\documentclass[11pt]{article}

\usepackage[final]{acl}

\usepackage{times}
\usepackage{latexsym}
\usepackage[T1]{fontenc}
\usepackage[utf8]{inputenc}
\usepackage{microtype}
\usepackage{inconsolata}
\usepackage{natbib}

\usepackage{xspace}
\usepackage{xcolor}
\usepackage{graphicx}
\usepackage[normalem]{ulem} %
\usepackage{amsmath}
\usepackage{enumitem}
\usepackage{amssymb}

\usepackage{hyperref}
\hypersetup{linkcolor=black,citecolor=black,anchorcolor=black,filecolor=black,menucolor=black,runcolor=black,urlcolor=black,hidelinks}
\usepackage{breakurl}

\usepackage{booktabs}
\usepackage{multirow}
\usepackage{makecell}
\usepackage{ragged2e}

\usepackage{caption}
\usepackage{subcaption}
\usepackage{placeins} %

\usepackage{listings}

\usepackage{tikz}
\usetikzlibrary{calc}
\usetikzlibrary{shapes.geometric}
\usetikzlibrary{decorations.pathreplacing}
\usetikzlibrary{positioning}
\usetikzlibrary{fit}
\usetikzlibrary{arrows.meta}
\usetikzlibrary{backgrounds}

\usepackage{pifont}  %
\usepackage[most]{tcolorbox}  %

\usepackage{datetime} %

\makeatletter
\newcommand{\DefMacro}{\@ifstar\@DefMacroAllowRedefine\@DefMacro}
\newcommand{\@DefMacro}[2]{\expandafter\newcommand\csname rmk-#1\endcsname{#2}}
\newcommand{\@DefMacroAllowRedefine}[2]{\expandafter\providecommand\csname rmk-#1\endcsname{} \expandafter\renewcommand\csname rmk-#1\endcsname{#2}}
\makeatother
\newcommand{\UseMacro}[1]{\csname rmk-#1\endcsname}

\newcommand{\XSpace}[1]{}
\newcommand{\XComment}[1]{}

\newcommand{\Code}[1]{{\ifmmode{\mathtt{#1}}\else$\mathtt{#1}$\fi}}
\newcommand{\CodeIn}[1]{{\ifmmode{\mathtt{#1}}\else$\mathtt{#1}$\fi}}

\newcommand{\specialcell}[2][c]{%
\begin{tabular}[#1]{@{}c@{}}#2\end{tabular}}
\newcolumntype{R}[1]{>{\RaggedLeft\arraybackslash}p{#1}}
\newcolumntype{L}[1]{>{\RaggedRight\arraybackslash}p{#1}}

\definecolor{gray}{RGB}{211,211,211}
\newcommand{\jbasicstyle}{\small\sffamily} %

\newcommand{\jnumberstyle}{\scriptsize}

\lstdefinelanguage{pseudo}
{
morekeywords={},
keywordstyle=\bfseries,
lineskip=-0.1em,
numbers=left, %
numberstyle=\jnumberstyle,
numbersep=4pt,
basicstyle=\jbasicstyle,
breaklines=true,
breakautoindent=true,
tabsize=2,
columns=fullflexible,
morecomment=*[l][\textsl]{//},
mathescape=true,
xleftmargin=10pt,
}

\lstdefinelanguage{todo-comment}
{
morekeywords={},
keywordstyle=\bfseries,
lineskip=-0.1em,
numbers=none,
basicstyle=\jbasicstyle,
breaklines=true,
breakautoindent=true,
tabsize=2,
columns=fullflexible,
morecomment=*[l][\textsl]{//},
mathescape=true,
xleftmargin=-10pt,
}

\lstdefinelanguage{java-pretty}
{
language=java,
numbers=left,
basicstyle=\scriptsize\ttfamily,
numberstyle=\scriptsize,
breaklines=true,
columns=fullflexible,
xleftmargin=16pt,
showstringspaces=false,
}

\newcommand{\codelora}{Code2LoRA\xspace}
\newcommand{\repopeftbench}{RepoPeftBench\xspace}
\newcommand{\Title}{\codelora{}: Hypernetwork-Generated Adapters for Code Language Models under Software Evolution}

\newcommand{\URLRepo}{\url{https://anonymous.4open.science/r/code2lora-6857}}
\newcommand{\URLData}{\url{https://huggingface.co/code2lora}}

\newcommand{\codelorastatic}{\codelora-Static\xspace}
\newcommand{\codeloraevo}{\codelora-Evo\xspace}

\newcommand{\textlora}{Text2LoRA\xspace}
\newcommand{\doclora}{Doc2LoRA\xspace}

\DefMacro{TH-em}{EM (\%)}
\DefMacro{TH-editsim}{EditSim}
\DefMacro{TH-codebleu}{CodeBLEU}
\DefMacro{TH-method}{Method}
\DefMacro{TH-cr}{Cross-Repo (CR Test)}
\DefMacro{TH-ir}{In-Repo (IR Test)}
\DefMacro{TH-ood}{Post-Cutoff OOD}
\DefMacro{TH-tasks}{Tasks}
\DefMacro{TH-tasks-per-repo}{Tasks / repo}

\DefMacro{method-pretrained}{Pretrained}
\DefMacro{method-rag}{RAG ($k$=3)}
\DefMacro{method-drc}{Dep.-Resolved Context}
\DefMacro{method-fft}{FFT}
\DefMacro{method-fft-rag}{FFT + RAG}
\DefMacro{method-fft-drc}{FFT + DRC}
\DefMacro{method-slora}{Single LoRA} %
\DefMacro{method-slora-drc}{Single LoRA + DRC}
\DefMacro{method-plora}{Per-repo LoRA}
\DefMacro{method-codelorastatic-drc}{\codelorastatic{} + DRC}

\DefMacro{TH-group-inference}{Inference-only (no fine-tuning)}
\DefMacro{TH-group-finetuned}{Fine-tuned}
\DefMacro{TH-group-hypernet}{Hypernetwork-based}
\DefMacro{note-plora}{\UseMacro{method-plora} is an in-repo upper bound and is not applicable to the cross-repo setting.}

\DefMacro{cr-em-pretrained}{45.7}
\DefMacro{cr-em-rag}{47.6}
\DefMacro{cr-em-drc}{49.1}
\DefMacro{cr-em-fft}{51.4}
\DefMacro{cr-em-fft-rag}{54.1}
\DefMacro{cr-em-delta-codelorastatic}{9.7}
\DefMacro{cr-em-slora}{47.4}
\DefMacro{cr-em-textlora}{45.8}
\DefMacro{cr-em-codelorastatic}{63.8}

\DefMacro{ir-em-pretrained}{46.8}
\DefMacro{ir-em-rag}{48.0}
\DefMacro{ir-em-drc}{50.0}
\DefMacro{ir-em-fft}{55.9}
\DefMacro{ir-em-slora}{50.4}
\DefMacro{ir-em-plora}{64.0}
\DefMacro{ir-em-codelorastatic}{66.2}

\DefMacro{cd-cr-em-pretrained}{31.5}
\DefMacro{cd-cr-em-rag}{36.8}
\DefMacro{cd-cr-em-drc}{36.5}
\DefMacro{cd-cr-em-slora}{55.1}
\DefMacro{cd-cr-em-codelorastatic}{55.7}
\DefMacro{cd-cr-em-codeloraevo}{60.3}
\DefMacro{cd-cr-em-delta-codeloraevo}{5.2}

\DefMacro{cd-ir-em-pretrained}{29.3}
\DefMacro{cd-ir-em-rag}{37.4}
\DefMacro{cd-ir-em-drc}{36.9}
\DefMacro{cd-ir-em-slora}{61.3}
\DefMacro{cd-ir-em-codelorastatic}{60.6}
\DefMacro{cd-ir-em-codeloraevo}{64.5}

\DefMacro{ood-em-pretrained}{44.6}
\DefMacro{ood-em-rag}{55.6}
\DefMacro{ood-em-drc}{56.7}
\DefMacro{ood-em-textlora}{60.4}
\DefMacro{ood-em-slora}{72.3}
\DefMacro{ood-em-codelorastatic}{72.2}
\DefMacro{ood-em-codeloraevo}{74.1}

\DefMacro{cr-locally-solvable}{45.7}
\DefMacro{ir-locally-solvable}{46.8}
\DefMacro{cr-reporeq-codelorastatic}{38.1}
\DefMacro{ir-reporeq-codelorastatic}{41.5}
\DefMacro{cr-reporeq-rag}{10.1}
\DefMacro{ir-reporeq-rag}{9.7}
\DefMacro{cr-reporeq-drc}{9.6}
\DefMacro{ir-reporeq-drc}{9.5}

\DefMacro{num-repos-total}{604}
\DefMacro{num-repos-in-dist}{512}
\DefMacro{num-repos-train}{409}
\DefMacro{num-repos-cr-test}{52}
\DefMacro{num-repos-ood}{92}

\DefMacro{num-static-qnas}{62{,}294}
\DefMacro{num-static-qnas-short}{62K}
\DefMacro{num-static-train}{39{,}612}
\DefMacro{num-static-train-short}{40K}
\DefMacro{num-static-val}{11{,}046}
\DefMacro{num-static-val-short}{11K}
\DefMacro{num-static-test}{11{,}636}
\DefMacro{num-static-test-short}{12K}

\DefMacro{num-commit-qnas}{399{,}649}
\DefMacro{num-commit-qnas-short}{0.40M}
\DefMacro{num-commit-train}{215{,}129}
\DefMacro{num-commit-train-short}{215K}
\DefMacro{num-commit-val}{97{,}727}
\DefMacro{num-commit-val-short}{98K}
\DefMacro{num-commit-test}{86{,}793}
\DefMacro{num-commit-test-short}{87K}
\DefMacro{repo-tok-mean}{285K}
\DefMacro{repo-tok-max}{3.0M}

\title{\Title}

\author{Liliana Hotsko, Yinxi Li, Yuntian Deng, Pengyu Nie \\
University of Waterloo \\
\texttt{\{lhotsko, yinxi.li, yuntian, pynie\}@uwaterloo.ca}}

\begin{document}
\maketitle

\begin{abstract}

Code language models need repository-level context to resolve imports, APIs, and project conventions.
Existing methods inject this knowledge as long inputs (retrieved through RAG or dependency analysis) or through per-repository fine-tuning and LoRA---costly at repository scale and brittle to evolving codebases.
We introduce \codelora{}, a hypernetwork framework that generates repository-specific LoRA adapters, effectively injecting repository knowledge with zero inference-time token overhead.
\codelora{} supports two usage scenarios: \codelorastatic{} converts a single repository snapshot into an adapter, suitable for comprehension of stable codebases;
while \codeloraevo{} maintains an adapter backed by a GRU hidden state updated per code diff, suitable for active development of evolving codebases.
To evaluate \codelora{} against parameter-efficient fine-tuning baselines, we build \repopeftbench{}, a benchmark of \UseMacro{num-repos-total} Python repositories with two tracks: a static track with \UseMacro{num-static-train-short} training and \UseMacro{num-static-test-short} test assertion-completion tasks, and an evolution track with \UseMacro{num-commit-train-short} commit-derived training and \UseMacro{num-commit-test-short} commit-derived test tasks.
On the static track, \codelorastatic{} achieves \UseMacro{cr-em-codelorastatic}\% cross-repo and \UseMacro{ir-em-codelorastatic}\% in-repo exact match, matching the per-repository LoRA upper bound;
on the evolution track, \codeloraevo{} achieves \UseMacro{cd-cr-em-codeloraevo}\% cross-repo exact match (+\UseMacro{cd-cr-em-delta-codeloraevo}\,pp over a single shared LoRA).\footnote{\codelora{}'s code can be found at \URLRepo{}; the model checkpoints and \repopeftbench{} datasets can be found at \URLData{}.}

\end{abstract}

\section{Introduction}
\label{sec:intro}

Real codebases span thousands of files whose imports, APIs, and conventions a code language model must know to complete assertions, fix bugs, or navigate a project.
Today's LLM-based coding assistants typically inject this repository knowledge as long inputs, in the form of retrieved relevant files through RAG (retrieval-augmented generation) or dependency analysis, and pay for the retrieved context at every query.
This is costly because repository-level context can be massive, stressing the LLM's context window and RAG's retrieval capability.
Another approach is to fine-tune the model or LoRA adapters~\cite{hu2022lora} for one repository or a group of related repositories, pushing knowledge into parameters.
These methods also require costly training, and even worse, are brittle to \emph{evolving} codebases, where every commit can invalidate the adapter and require retraining.

Recent work on hypernetwork-generated LoRA adapters~\cite{ha2017hypernetworks,charakorn2025text2lora,charakorn2026doctolora} is promising: a single forward pass over a conditioning input produces task- or document-specific weights for a frozen LLM.
These methods, however, are built for short natural-language task descriptions or single documents, not the long context a repository typically carries, and they assume a static conditioning input with no mechanism for tracking a codebase as it evolves.

We propose \codelora{}, a hypernetwork framework that generates repository-specific LoRA adapters, effectively injecting repository knowledge with zero inference-time token overhead.
We design around two orthogonal axes---\emph{how} knowledge enters the parameters and \emph{when} it is updated---and instantiate them as two usage scenarios:
\codelorastatic{} converts a single repository snapshot into an adapter, suitable for comprehension of stable codebases;
\codeloraevo{} maintains an adapter backed by a GRU hidden state updated per code diff, so the recurrence augments (rather than replaces) the snapshot prior, suitable for active development of evolving codebases.

We evaluate \codelora{} on \repopeftbench{}, a benchmark of \UseMacro{num-repos-total} Python repositories (\UseMacro{num-repos-in-dist} in-distribution and a \UseMacro{num-repos-ood}-repository temporal holdout created after the scrape cutoff).
\repopeftbench{} divides each repository into non-test and test portions: the model may use non-test code as repository context and must complete assertion-completion tasks in the test portion, which is a task that requires complex reasoning capabilities~\cite{jain2025livecodebench}.
Two tracks instantiate our usage scenarios: a static track with \UseMacro{num-static-train} training and \UseMacro{num-static-test} test tasks on a single repository snapshot, and an evolution track with \UseMacro{num-commit-train} training and \UseMacro{num-commit-test} test tasks drawn from commit history.
Evaluation uses in-repo (IR) and cross-repo (CR) splits on the in-distribution corpus, plus a temporal out-of-distribution (OOD) test split on the post-cutoff holdout (\S\ref{sec:results:ood}).

On the static track, \codelorastatic{} achieves \UseMacro{cr-em-codelorastatic}\% cross-repo exact match, well above context-injection methods such as RAG and dependency-resolved context; without any per-repository training it also reaches \UseMacro{ir-em-codelorastatic}\% in-repo exact match, matching the per-repository LoRA upper bound.
On the evolution track, snapshot-based adaptation goes stale once evaluation uses commit-derived tasks; \codeloraevo{} reaches \UseMacro{cd-cr-em-codeloraevo}\% cross-repo exact match, +\UseMacro{cd-cr-em-delta-codeloraevo}\,pp over a shared LoRA.

\vspace{3pt}
\noindent
The main contributions of this work include:

\begin{itemize}[topsep=3pt,itemsep=1ex,partopsep=0ex,parsep=0ex,leftmargin=*]
\item \textbf{Idea.}
We propose using hypernetworks to effectively inject repository knowledge into code language models, and frame the problem along \emph{how} knowledge enters model parameters and \emph{when} it is refreshed.
\item \textbf{Framework.}
We design and implement \codelora{}, a hypernetwork that maps repository code to LoRA adapters with zero inference-time token overhead, instantiated as \codelorastatic{} (mapping one repository snapshot) and \codeloraevo{} (maintaining an adapter from sequential code diffs).
\item \textbf{Benchmark.}
We curate \repopeftbench{}, a benchmark of \UseMacro{num-repos-total} Python repositories, including a \UseMacro{num-repos-ood}-repository temporal holdout for out-of-distribution evaluation.
\item \textbf{Evaluation.}
\codelora{} outperforms the strongest baselines on \repopeftbench{} by +\UseMacro{cr-em-delta-codelorastatic}\,pp on the static track and +\UseMacro{cd-cr-em-delta-codeloraevo}\,pp on the evolution track, with consistent gains on the temporal OOD holdout (\S\ref{sec:results:ood}).
\end{itemize}

\section{Related Work}
\label{sec:related}

\paragraph{Parameter-efficient fine-tuning.}
LoRA~\cite{hu2022lora} enables efficient adaptation through low-rank decomposition of weight updates; extensions include QLoRA~\cite{dettmers2023qlora}, DoRA~\cite{liu2024dora}, weight merging~\cite{yadav2024ties}, multi-LoRA routing~\cite{huang2024lorahub}, LoRACode~\cite{chaturvedi2025loracodeloraadapterscode}, and MoLE~\cite{zong2025mixlanguageexperts}, which trains a separate LoRA module per programming language.
These methods treat adapters as static artifacts, trained per task, per language, or per repository; \codelora{} instead \emph{generates} adapters conditioned on repository context, enabling adaptation to unseen codebases without retraining.

\paragraph{Hypernetworks for LoRA generation.}
Hypernetworks~\cite{ha2017hypernetworks} generate the parameters of a target network from a conditioning signal.
Recent applications to language models include HyperTuning~\cite{phang2023hypertuning} and HyperLoRA~\cite{von2024language} for cross-task generalization, Generative Adapter~\cite{chen2025generativeadapter} for single-pass contextualization, and Zhyper~\cite{abdalla2025zhyper} for factorized conditioned LoRA generation.
Closest to our framework are \textlora{}~\cite{charakorn2025text2lora} and \doclora{}~\cite{charakorn2026doctolora}, which both map a whole input (a task description and a document, respectively) to a LoRA in one forward pass.
\textlora{} conditions on a short task description via an external text encoder and targets only Q/V projections; \doclora{} conditions on a document via per-layer activations of the frozen target LLM (Perceiver~\cite{jaegle2021perceiver} encoder, MLP \texttt{down\_proj} only) and is built for document QA, not code.
SHINE~\cite{liu2026shine} likewise maps a natural-language context to a LoRA in a single pass, reusing the frozen LLM's own parameters as an in-context hypernetwork.
\codelorastatic{} extends this family to a new input modality, an entire code repository (\S\ref{sec:method:static}).
\codeloraevo{} adds a second hypernetwork design: a GRU aggregates sequential code diffs into a hidden state that conditions adapter generation at each commit, yielding an adapter trajectory over a repository's lifetime; no analogue exists in the \textlora{}/\doclora{} line of work, which only model a single static input.

\paragraph{Software evolution and continual code adaptation.}
Software evolution and mining software repository---tracking how code changes commit by commit, file by file---is a well-established line of software engineering research~\cite{kagdi2007survey,hassan2008road}, underpinning analyses of change impact, bug introduction~\cite{sliwerski2005when}, and refactoring detection~\cite{tsantalis2018refdiff} over long version histories.
In NLP, a parallel line investigates \emph{when} a deployed model should be refreshed: continual pretraining and online fine-tuning aim to keep language models current under temporal drift~\cite{lazaridou2021mind,jang2022continual}, but typically maintain a single global checkpoint and have no notion of \emph{which} repository is being adapted to.
\codeloraevo{} sits at the intersection of these two lines: it treats sequential code diffs as the unit of update and refreshes a repository-specific adapter as the commit history unfolds.

\paragraph{Repository-level code understanding and generation.}
Prior work on incorporating repository context typically routes information through the \emph{input}: RepoFusion~\cite{shrivastava2023repofusion} trains with cross-file context, RepoCoder~\cite{zhang2023repocoder} iteratively retrieves and generates, RepoFormer~\cite{wu2024repoformer} uses selective retrieval, CoCoMIC~\cite{ding2024cocomic} jointly models in-file and cross-file context, R2C2-Coder~\cite{deng2024r2c2coder} enhances repo-level completion with repository-context-aware methods, and RepoHyper~\cite{phan2024repohypersearchexpandrefinesemanticgraphs} uses semantic-graph retrieval.
Evaluation benchmarks include CrossCodeEval~\cite{ding2024crosscodeeval} and RepoBench~\cite{liu2023repobenchbenchmarkingrepositorylevelcode}.
\codelora{} instead distills repository knowledge into model \emph{parameters}, avoiding context-window limits and per-query retrieval cost, and---through \codeloraevo{}---tracks how that knowledge changes as code evolves commit by commit.
We base our experiments on Qwen2.5-Coder-1.5B~\cite{hui2024qwen25coder}; other recent code LLMs include CodeLlama~\cite{roziere2023code}, StarCoder~\cite{li2023starcoder}, and DeepSeek-Coder~\cite{guo2024deepseekcoderlargelanguagemodel}.

\section{Method}
\label{sec:method}

\begin{figure*}[t]
\centering
\includegraphics[width=.9\textwidth, trim=3 3 3 3, clip]{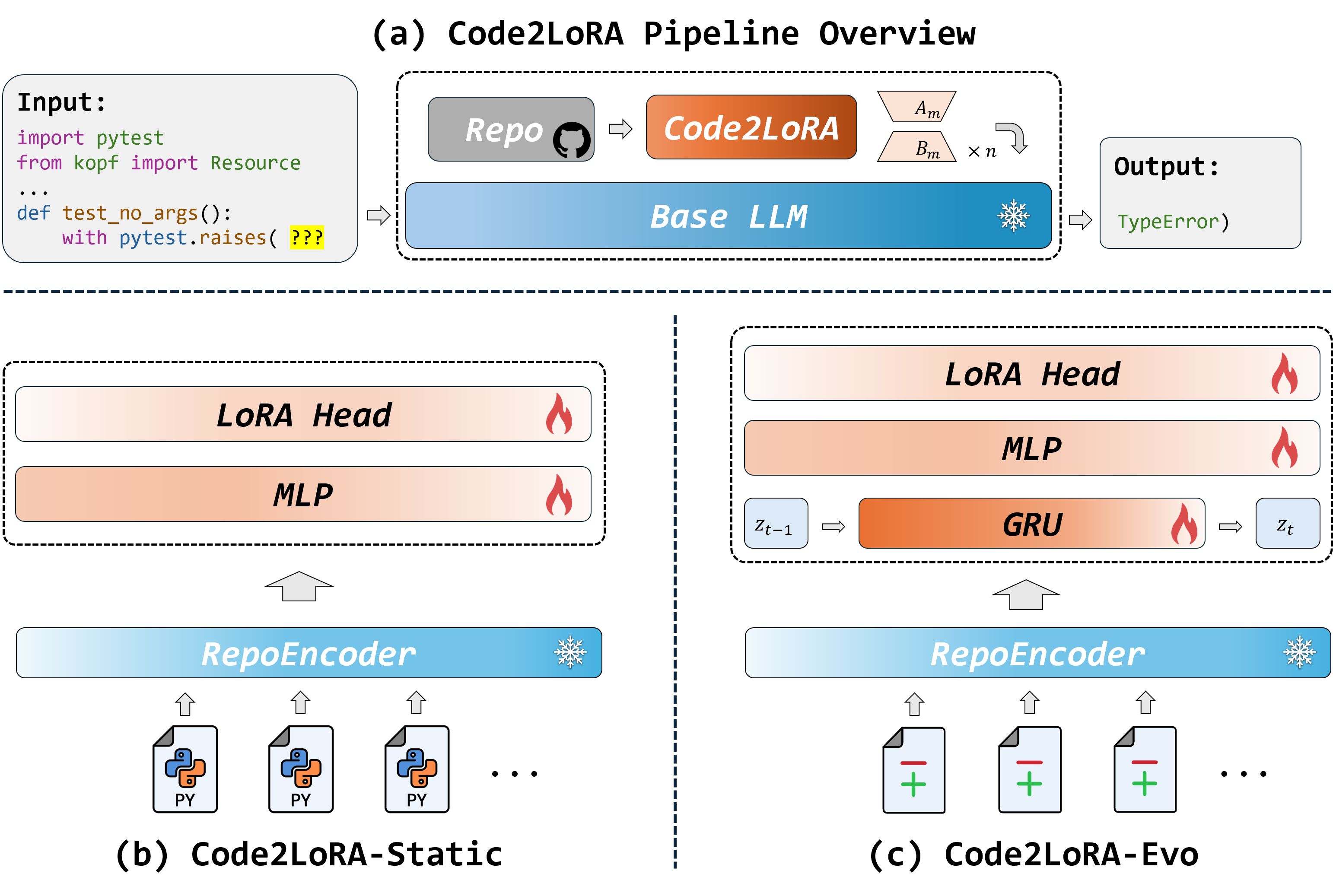}
\caption{\codelora{} architecture.
\textbf{(a)}~Overall pipeline: repository context is encoded and mapped to LoRA adapters, which are injected into a frozen LLM to support inference (example task: assertion completion).
\textbf{(b)}~\codelorastatic{}'s static hypernetwork.
\textbf{(c)}~\codeloraevo{}'s recurrent hypernetwork.}
\label{fig:architecture}
\end{figure*}

\codelora{} is a hypernetwork framework that generates repository-specific LoRA adapters for a frozen code LM, effectively injecting repository knowledge with zero inference-time token overhead.
As illustrated in Figure~\ref{fig:architecture}a, the framework has three components:
a shared \textbf{repository encoder} (\S\ref{sec:method:encoding}) that maps repository-level context to dense embeddings,
a \textbf{hypernetwork} that maps those embeddings to LoRA weights,
a \textbf{base LLM} that receives the generated adapter and performs inference.
Only the hypernetwork is trained, via the standard language-modeling loss; the repository encoder and base LLM are frozen.
The two usage scenarios differ in hypernetwork design:
\codelorastatic{} (\S\ref{sec:method:static}) directly projects the repository embedding into LoRA weights;
\codeloraevo{} (\S\ref{sec:method:evo}) inserts a GRU before the projection head to aggregate a sequence of diff embeddings.

\subsection{Repository Encoder}
\label{sec:method:encoding}

Repository-level context must be compressed into a fixed-size vector before the hypernetwork can consume it (we analyze how this scales with repository size in Appendix~\ref{app:repo_size}).
We adopt a training-free two-step embedding approach that works effectively in practice using a frozen Qwen3-Embedding-0.6B model. %

\paragraph{Step 1: file-level embedding.}
Each file $f_i$ in the repository context (or its diff $\Delta f_i$) is divided into 4096-token chunks with 512-token overlap, embedded by the frozen model, and mean-pooled over chunks to produce a file vector $\mathbf{f}_i \in \mathbb{R}^{d}$ ($d{=}1024$).

\paragraph{Step 2: repository-level aggregation.}
For a full repository snapshot, each file vector receives an importance weight $w_i$ based on a combination of content distinctiveness, file size, and path importance.
The repository embedding is the concatenation of a weighted mean and a max pool,
\begin{equation*}
\mathbf{e} = \big[\textstyle\sum_i w_i \mathbf{f}_i \,;\, \max_i \mathbf{f}_i\big] \in \mathbb{R}^{2d},
\end{equation*}
capturing both the average character and the most distinctive features of the codebase.
The embeddings are pre-computed at training time.

\subsection{\codelorastatic{}}
\label{sec:method:static}

The static hypernetwork in \codelorastatic{} maps a single embedding $\mathbf{e}$ to a LoRA adapter in one forward pass.
For each module type $m \in \{\texttt{q,k,v,o,gate,up,down}\}$, its LoRA matrices $\mathbf{A}_m$ and $\mathbf{B}_m$ are generated by a shared 2-layer MLP with GELU activation followed by dedicated output heads:
\begin{align*}
\mathbf{h} &= \sqrt{d_h}\,\mathrm{L2Norm}(\mathrm{MLP}(\mathbf{e})), \\
\mathbf{A}_m &= \tanh(\mathrm{Head}^A_m(\mathbf{h})) \cdot \exp(s^A_m),\\
\mathbf{B}_m &= \tanh(\mathrm{Head}^B_m(\mathbf{h})) \cdot \exp(s^B_m),
\end{align*}
where
learnable log-scales $s^{A/B}_m$ control adapter magnitudes (initialized to $-3.5$).
LoRA matrices are shared across all layers of base LLM and injected via $\mathbf{W}' = \mathbf{W} + \tfrac{\alpha}{r}\mathbf{B}_m\mathbf{A}_m$.
With hidden dimension $d_h{=}512$ and LoRA rank $r{=}16$, the \codelorastatic{} hypernetwork has ${\sim}$340M trainable parameters, almost all of them in the output heads, which emit the ${\approx}0.66$M entries of one shared $(\mathbf{A}_m,\mathbf{B}_m)$ set per projection type.
\codelorastatic{}'s hypernetwork architecture is similar to that of \textlora{}~\cite{charakorn2025text2lora} and \doclora{}~\cite{charakorn2026doctolora}, but (1)~is driven by a whole-repository embedding summarized from millions of tokens rather than a task description, and (2)~injects LoRA to all seven module types rather than just Q/V or down-projection to be more flexible.

\subsection{\codeloraevo{}}
\label{sec:method:evo}

The recurrent hypernetwork in \codeloraevo{} maintains a repository-specific adapter over a chronological stream of diff embeddings~$\{\mathbf{e}_t\}$.
The diff embeddings are aggregated by a GRU recurrent neural network:
at step~$t$ the encoder (\S\ref{sec:method:encoding}) supplies~$\mathbf{e}_t$, which is linearly projected and combined with the previous state,
\begin{equation*}
\mathbf{z}_t = \mathrm{GRU}(\mathrm{LayerNorm}(\mathrm{Linear}(\mathbf{e}_t)),\, \mathbf{z}_{t-1}).
\end{equation*}
The initial GRU state $\mathbf{z}_0$ is initialized by a small projector given the initial repository embedding (e.g., on the first commit).
At each step~$t$, the LoRA adapter is generated by a projection head of the same design as \codelorastatic{}'s (\S\ref{sec:method:static}) with $\mathbf{z}_t$ substituted for~$\mathbf{e}$, yielding an \emph{adapter trajectory} over the repository's lifetime.
Each update requires only one GRU step on the stored diff embedding~$\mathbf{e}_t$, far less model compute than re-encoding the full repository (end-to-end costs in Appendix~\ref{app:efficiency}).
\codeloraevo{} uses this head with a wider trunk ($d_h{=}1024$, ${\sim}$679M parameters), as does the \codelorastatic{} reference we report on the evolution track, so the two differ only by the recurrence;
the GRU and initial-state projector add ${\sim}$34M, for ${\sim}$713M trainable parameters in total.

\subsection{Training}
\label{sec:method:training}

We train the hypernetwork end-to-end by minimizing cross-entropy on assertion-completion pairs from the frozen base LLM, with LoRA weights supplied by $\mathrm{Hypernetwork}_\theta$:
\begin{equation*}
\mathcal{L}(\theta) = -\!\!\!\sum_{(x,y) \in \mathcal{D}}\!\!\! \log p\!\left(y \mid x; \mathrm{Hypernetwork}_\theta(\mathbf{u})\right),
\end{equation*}
where $x$ is the input prefix, $y$ the output target, and $\mathbf{u}=\mathbf{e}$ for \codelorastatic{} or $\mathbf{u}=\mathbf{z}_t$ for \codeloraevo{}.
For \codeloraevo{}, we optimize with truncated backpropagation through time, detaching $\mathbf{z}_t$ every $K{=}16$ steps (App.~\ref{app:implementation}).
Batches are formed by first sampling a repository, then a pair of input-output from it, so that the hypernetwork sees diverse repositories and does not overfit to data-rich ones.

\section{\repopeftbench{}: A Repository-Level PEFT Benchmark}
\label{sec:benchmark}

We construct \repopeftbench{}, a repository-level benchmark for parameter-efficient fine-tuning of code language models.
The corpus comprises \UseMacro{num-repos-total} Python repositories drawn from GitHub under shared quality filters---each uses \texttt{pytest} or \texttt{unittest}, carries a permissive license, and shows recent activity---partitioned along a fixed temporal cutoff (\texttt{2025-04-01}) into \UseMacro{num-repos-in-dist} in-distribution repositories and a \UseMacro{num-repos-ood} out-of-distribution (OOD) repositories.
The in-distribution set was collected before the cutoff date, and requires an additional filter of having at least 300 stars (to ensure high-quality), which supplies all training and validation splits; commit histories are truncated at the cutoff date. %
The OOD set comprises repositories created strictly after the cutoff date and is reserved for held-out test-time evaluation only (\S\ref{sec:results:ood}).
We collect both the last snapshot as well as the full commit histories of each repository.

Two evaluation tracks share the same task, metrics, and CR/IR repository partitions but differ in how instances are indexed in history (\S\ref{sec:benchmark:tracks}).
Table~\ref{tab:dataset_actual} summarizes the split sizes used in all reported results.
Benchmark construction details are in Appendix~\ref{app:dataset_construction}.

\paragraph{Task.}
Each instance is an assertion-completion input-output pair: the model receives a structured prefix from a test file and must predict the expected value of an assertion.
The task follows the code-execution probe of LiveCodeBench~\cite{jain2025livecodebench}, but replaces hand-curated single-function snippets with assertions mined at scale from real test suites.
Assertion completion is well suited to repository-level evaluation because all instances in a repository share the same non-test source as conditioning context. %
Repository-level code completion, as in RepoBench~\cite{liu2023repobenchbenchmarkingrepositorylevelcode}, is not suitable because each target file requires excluding that file from context to prevent leakage and thus a different repository slice per instance.
CrossCodeEval~\cite{ding2024crosscodeeval}, RepoHyper~\cite{phan2024repohypersearchexpandrefinesemanticgraphs}, and R2C2-Coder~\cite{deng2024r2c2coder} likewise ship only retrieval-selected slices; \repopeftbench{} releases full information of each repository to evaluate methods that ingest the full codebase.

We extract instances from five assertion families: bare \texttt{assert}, \texttt{self.assert*}, \texttt{pytest.raises}, \texttt{pytest.approx}, and NumPy-style \texttt{assert\_*}.
The \emph{input} concatenates imports, the enclosing class (if any), helper methods, and the test-function body up to the assertion cut point;
the \emph{output} is the expected value of the assertion, namely the right-hand side of the binary comparison operator, or the last argument of the assertion function call.

\paragraph{Repository splits.}
\label{sec:benchmark:splits}
We partition the \UseMacro{num-repos-in-dist} in-distribution repositories into cross-repo (CR) and in-repo (IR) sets, shared by both evaluation tracks.
\textbf{Cross-Repo (CR)} holds out 103 repositories entirely at training time (51 validation, \UseMacro{num-repos-cr-test} test) to measure generalization to unseen codebases.
\textbf{In-Repo (IR)} uses the remaining \UseMacro{num-repos-train} repositories for training and is the only setting in which per-repository LoRA is defined; held-out instances within each training repository are assigned by the track-specific protocol below.

\paragraph{Evaluation tracks.}
\label{sec:benchmark:tracks}
\label{sec:benchmark:evolution}
The \textbf{Static} track draws every instance from a single snapshot per repository (\UseMacro{num-static-qnas} tasks) and corresponds to \codelorastatic{}:
on CR splits, tasks are extracted from each held-out repository's last commit snapshot;
on IR splits, tasks are also extracted from last commits, and are randomly split into training, validation, and test sets in a ratio of 8:1:1.
The \textbf{Evolution} track replays each repository's commit history and emits a task whenever a commit adds or modifies an assertion, storing the input-output pair together with the production-code diff~$\Delta_t$; it corresponds to \codeloraevo{}.
On CR splits, evaluation uses all commit-derived tasks from held-out repositories;
on IR splits, following the time-segmented methodology of~\citet{nie2022impact}, commits within each training repository are partitioned chronologically so that training examples strictly precede validation and test.
Evolution-track training and evaluation each retain at most eight tasks per commit; \codeloraevo{} training further caps at four tasks per test file so that no commit dominates a backpropagation window.
Table~\ref{tab:dataset_actual} reports the number of tasks for every split used in our experiments.
Commit histories are bursty: repositories accumulate hundreds of test-touching commits in irregular clusters (Appendix Figure~\ref{fig:dataset_construction}), which motivates streaming adaptation via \codeloraevo{} rather than a single frozen snapshot.

\begin{table*}[t]
\begin{small}
\begin{center}
\caption{Dataset statistics for \repopeftbench{}, divided into static and evolution tracks (sharing the same set of \UseMacro{num-repos-in-dist} in-distribution repositories) and \UseMacro{num-repos-ood} out-of-distribution repositories, split into train/val/test sets under cross-repo (CR) and in-repo (IR) settings.}
\label{tab:dataset_actual}

\setlength{\tabcolsep}{4pt}
\begin{tabular}{l r r r r}
\toprule
Split & Repos & Commits & \UseMacro{TH-tasks} & \UseMacro{TH-tasks-per-repo} \\
\midrule
\multicolumn{5}{l}{\emph{Static track}} \\
\midrule
Train             & 409 & 409 & 39{,}612 & 96.9 \\
CR Val \,/\, Test &  51 \,/\,  52 &  51 \,/\,  52 & 6{,}213 \,/\, 6{,}414 & 121.8 \,/\, 123.3 \\
IR Val \,/\, Test & 409 \,/\, 409 & 409 \,/\, 409 & 4{,}833 \,/\, 5{,}222 &  11.8 \,/\,  12.8 \\
\midrule
\multicolumn{5}{l}{\emph{Evolution track}} \\
\midrule
Train (\codelorastatic{} and baselines) & 400$^\dagger$ & 400 & 44{,}149 & 110.4 \\
Train (\codeloraevo{})     & 400$^\dagger$ & 45{,}516 & 215{,}129 & 537.8 \\
CR Val \,/\, Test &  49 \,/\,  51 & 8{,}614 \,/\, 6{,}618 & 58{,}944 \,/\, 44{,}732 & 1{,}203 \,/\,    877 \\
IR Val \,/\, Test & 389 \,/\, 389 & 5{,}710 \,/\, 6{,}179 & 38{,}783 \,/\, 42{,}061 &     99.7 \,/\, 108.1 \\
\midrule
\multicolumn{5}{l}{\emph{Out-of-distribution holdout}} \\
\midrule
OOD Test          & 92 & 1{,}950 & 14{,}813 & 161.0 \\
\bottomrule
\multicolumn{5}{l}{\footnotesize $^\dagger$ 9 repositories lack sufficient commit histories and are excluded from \codeloraevo{} training.}
\end{tabular}

\end{center}
\end{small}
\end{table*}

\begin{table*}[t]
\begin{small}
\begin{center}
\caption{Results on \repopeftbench{} static track. \UseMacro{method-rag} and \UseMacro{method-drc} use the calibrated context-injection configuration of Appendix~\ref{app:context_calibration}.}\label{tab:main_results}

\begin{tabular}{l@{\hskip 8pt}ccc@{\hskip 12pt}ccc}
\toprule
& \multicolumn{3}{c}{\textbf{Cross-Repo (CR Test)}} & \multicolumn{3}{c}{\textbf{In-Repo (IR Test)}} \\
\cmidrule(lr){2-4} \cmidrule(lr){5-7}
Method & EM (\%) & EditSim & CodeBLEU & EM (\%) & EditSim & CodeBLEU \\
\midrule
\multicolumn{7}{l}{\emph{\UseMacro{TH-group-inference}}} \\
\midrule
\UseMacro{method-pretrained} & 45.7 & 0.605 & 0.646 & 46.8 & 0.624 & 0.655 \\
\UseMacro{method-rag} & 47.6 & 0.624 & 0.655 & 48.0 & 0.632 & 0.661 \\
\UseMacro{method-drc} & 49.1 & 0.631 & 0.662 & 50.0 & 0.645 & 0.670 \\
\midrule
\multicolumn{7}{l}{\emph{\UseMacro{TH-group-finetuned}}} \\
\midrule
\UseMacro{method-fft} & 51.4 & 0.695 & 0.678 & 55.9 & 0.727 & 0.714 \\
\UseMacro{method-fft-rag} & 54.1 & 0.703 & 0.689 & 56.8 & 0.730 & 0.713 \\
\UseMacro{method-slora} & 47.4 & 0.663 & 0.649 & 50.4 & 0.687 & 0.675 \\
\UseMacro{method-plora}$^\dag$ & --- & --- & --- & 64.0 & 0.801 & 0.788 \\
\midrule
\multicolumn{7}{l}{\emph{\UseMacro{TH-group-hypernet}}} \\
\midrule
\textlora{} & 45.8 & 0.606 & 0.647 & 46.7 & 0.625 & 0.655 \\
\textbf{\codelorastatic{}} & \textbf{63.8} & \textbf{0.784} & \textbf{0.778} & \textbf{66.2} & \textbf{0.806} & \textbf{0.797} \\
\bottomrule
\multicolumn{7}{l}{\footnotesize $^\dag$ \UseMacro{note-plora}} \\
\end{tabular}

\end{center}
\end{small}
\end{table*}

\begin{table*}[t]
\begin{small}
\begin{center}
\caption{Results on \repopeftbench{} evolution track, with calibrated context-injection baselines as in Table~\ref{tab:main_results}.}\label{tab:per_commit_results}

\begin{tabular}{l@{\hskip 8pt}ccc@{\hskip 12pt}ccc}
\toprule
& \multicolumn{3}{c}{\textbf{Cross-Repo (CR Test)}} & \multicolumn{3}{c}{\textbf{In-Repo (IR Test)}} \\
\cmidrule(lr){2-4} \cmidrule(lr){5-7}
Method & EM (\%) & EditSim & CodeBLEU & EM (\%) & EditSim & CodeBLEU \\
\midrule
\multicolumn{7}{l}{\emph{\UseMacro{TH-group-inference}}} \\
\midrule
\UseMacro{method-pretrained} & 31.5 & 0.490 & 0.515 & 29.3 & 0.469 & 0.501 \\
\UseMacro{method-rag} & 36.8 & 0.528 & 0.546 & 37.4 & 0.532 & 0.551 \\
\UseMacro{method-drc} & 36.5 & 0.524 & 0.544 & 36.9 & 0.523 & 0.544 \\
\midrule
\multicolumn{7}{l}{\emph{\UseMacro{TH-group-finetuned}}} \\
\midrule
\UseMacro{method-slora} & 55.1 & 0.749 & 0.709 & 61.3 & 0.787 & 0.753 \\
\UseMacro{method-plora}$^\dag$ & --- & --- & --- & 64.2 & 0.803 & 0.788 \\
\midrule
\multicolumn{7}{l}{\emph{\UseMacro{TH-group-hypernet}}} \\
\midrule
\textlora{} & 41.7 & 0.596 & 0.600 & 43.5 & 0.612 & 0.613 \\
\codelorastatic{} & 55.7 & 0.760 & 0.716 & 60.6 & 0.787 & 0.749 \\
\textbf{\codeloraevo{}} & \textbf{60.3} & \textbf{0.810} & \textbf{0.763} & \textbf{64.5} & \textbf{0.828} & \textbf{0.790} \\
\bottomrule
\multicolumn{7}{l}{\footnotesize $^\dag$ \UseMacro{note-plora}} \\
\end{tabular}

\end{center}
\end{small}
\end{table*}

\section{Experimental Setup}
\label{sec:setup}

\paragraph{Models.}
The base LLM is Qwen2.5-Coder-1.5B~\cite{hui2024qwen25coder}, loaded in \texttt{bfloat16}; all baselines and both \codelora{} usage scenarios share this backbone.
Repository encoder uses Qwen3-Embedding-0.6B~\cite{zhang2025qwen3embedding}. %
Both models are released under the Apache 2.0 license and our research use is consistent with their model cards.

\paragraph{Hyperparameters.}
\codelora{} generate rank-$r{=}16$ LoRA adapters with $\alpha{=}32$ for all seven attention/MLP projection types, with each $(\mathbf{A}_m, \mathbf{B}_m)$ pair shared across all $28$ transformer layers (\S\ref{sec:method}).
\codelorastatic{} has ${\sim}$340M trainable parameters, while \codeloraevo{} has ${\sim}$713M trainable parameters.
We train both for 3 epochs with AdamW (cosine schedule) on a single H100 80\,GB GPU using TRL~\cite{vonwerra2022trl}; full hyperparameters, schedules, and sequence-length budgets are in Appendix~\ref{app:implementation}.

\paragraph{Baselines.}
We evaluate against various baselines:
\begin{itemize}[topsep=3pt,itemsep=1ex,partopsep=0ex,parsep=0ex,leftmargin=*]
\item \UseMacro{method-pretrained}: base LLM (Qwen2.5-Coder-1.5B).
\item \UseMacro{method-rag}: non-test source files pre-chunked into 512-token segments, embedded with Qwen3-Embedding-0.6B; top-$k$ retrieved chunks prepended to the prefix at inference. The main tables use the budget-matched $k{=}3$ (at most $1{,}536$ retrieved tokens, comparable to the context \UseMacro{method-drc} adds); larger budgets ($k\!\in\!\{5,10\}$, chunk size $256$) are reported in Appendix~\ref{app:rag_k}.
\item \UseMacro{method-drc}\label{sec:setup:drc}\label{sec:benchmark:oracle}: prepends function and class definitions reachable from each prefix's imports via dependency analysis, with relevance-aware compression under an adaptive token budget (Appendix~\ref{app:drc_construction}).
Both context-injection baselines are calibrated to the backbone before comparison: each uses the stronger of plain prepending and Qwen2.5-Coder's native repository-completion layout (native for \UseMacro{method-rag}, plain for \UseMacro{method-drc}; Appendix~\ref{app:context_calibration}).
\item \UseMacro{method-fft}: all model parameters are made trainable. %
\item \UseMacro{method-fft-rag}: the \UseMacro{method-fft} model with retrieval at inference; the top-$5$ retrieved $256$-token chunks (at most $1{,}280$ tokens) are prepended to the prefix (static track; Appendix~\ref{app:rag_k}).
\item \UseMacro{method-slora}: one rank-16 adapter trained on \emph{all} repositories.
\item \UseMacro{method-plora}~\cite{zong2025mixlanguageexperts}: one rank-16 adapter trained \emph{per} repository (IR splits only), serving as an upper bound on repository-level adaptation.
\item \textlora{}~\cite{charakorn2025text2lora}: a hypernetwork that emits a LoRA from an external task embedding.
To control for input modality and target-module coverage, we strengthen the upstream baseline along both axes: the natural-language task description is replaced with the same repository encoder that \codelora{} uses (mean$+$max-pooled Qwen3-Embedding-0.6B), and the output heads are extended from $\{\mathbf{Q},\mathbf{V}\}$ to all seven attention and MLP projections.
Training data, loss, and budget match \codelora{}, so only the LoRA-generation head differs (details in Appendix~\ref{app:implementation}).
\end{itemize}

\paragraph{Evaluation metrics.}
\label{sec:setup:metrics}
We report \textbf{Exact Match}
(\textbf{EM};
with post-processing to discard whitespace and over-generation parts; Appendix~\ref{app:em_rule} shows the EM results without this post-processing which yields the same conclusions);
\textbf{Edit Similarity} (\texttt{difflib}~\cite{python_difflib} \texttt{SequenceMatcher} ratio);
and \textbf{CodeBLEU}~\cite{ren2020codebleu}, which incorporates AST and data-flow structure in addition to n-gram overlap.

\section{Results}
\label{sec:results}

\codeloraevo{} requires commit history and is therefore evaluated on the evolution track and the OOD holdout but not on the static track.

\subsection{Static Track}
\label{sec:results:static}

Table~\ref{tab:main_results} shows the results on \repopeftbench{}'s static track.
On CR evaluation, \codelorastatic{} reaches \UseMacro{cr-em-codelorastatic}\% EM, +\UseMacro{cr-em-delta-codelorastatic}\,pp over the strongest baseline (\UseMacro{method-fft-rag}, \UseMacro{cr-em-fft-rag}\%) and above every context-injection method (\UseMacro{method-rag} \UseMacro{cr-em-rag}\%, \UseMacro{method-drc} \UseMacro{cr-em-drc}\%) and other fine-tuned baselines (\UseMacro{method-fft} \UseMacro{cr-em-fft}\%, \UseMacro{method-slora} \UseMacro{cr-em-slora}\%).
The margin over the strongest context baseline is $+$14.7\,pp (CR) and $+$16.2\,pp (IR), and all gaps to non-upper-bound baselines are significant (Appendix~\ref{app:significance}).
This margin concentrates on instances the no-context backbone cannot solve, where context injection recovers only ${\sim}$10\% of instances versus \UseMacro{cr-reporeq-codelorastatic}\% / \UseMacro{ir-reporeq-codelorastatic}\% for \codelorastatic{} (Appendix~\ref{app:local_solvability}).
The strengthened \textlora{} baseline, matched with \codelora{} on input modality and target modules, reaches only \UseMacro{cr-em-textlora}\% EM, which points to the LoRA-generation head as the bottleneck for repository-level adaptation.
On IR evaluation, \codelorastatic{} reaches \UseMacro{ir-em-codelorastatic}\% EM, matching the \UseMacro{method-plora} upper bound (\UseMacro{ir-em-plora}\%) without any per-repository training, suggesting that cross-repository transfer can substitute for fitting one adapter per repository.

\begin{table}[tb]
\begin{small}
\begin{center}
\caption{Results on \repopeftbench{} OOD set.}\label{tab:ood_results}

\begin{tabular}{@{}l@{\hspace{4pt}}c@{\hspace{8pt}}c@{\hspace{8pt}}c@{}}
\toprule
Method & EM (\%) & EditSim & CodeBLEU \\
\midrule
\multicolumn{4}{l}{\emph{\UseMacro{TH-group-inference}}} \\
\midrule
\UseMacro{method-pretrained} & 44.6 & 0.568 & 0.630 \\
\UseMacro{method-rag} & 55.6 & 0.649 & 0.680 \\
\UseMacro{method-drc} & 56.7 & 0.662 & 0.696 \\
\midrule
\multicolumn{4}{l}{\emph{\UseMacro{TH-group-finetuned}}} \\
\midrule
\UseMacro{method-slora} & 72.3 & 0.836 & 0.817 \\
\midrule
\multicolumn{4}{l}{\emph{\UseMacro{TH-group-hypernet}}} \\
\midrule
\textlora{} & 60.4 & 0.720 & 0.740 \\
\codelorastatic{} & 72.2 & 0.842 & 0.818 \\
\textbf{\codeloraevo{}} & \textbf{74.1} & \textbf{0.866} & \textbf{0.846} \\
\bottomrule
\end{tabular}

\end{center}
\end{small}
\end{table}

\subsection{Evolution Track}
\label{sec:results:evolution}

Table~\ref{tab:per_commit_results} reports evolution-track results on commit-derived prefixes.
Relative to the static track (Table~\ref{tab:main_results}), commit-derived tasks are substantially harder: \UseMacro{method-pretrained} CR EM drops from \UseMacro{cr-em-pretrained}\% to \UseMacro{cd-cr-em-pretrained}\%.
Context injection helps here---\UseMacro{method-rag} reaches \UseMacro{cd-cr-em-rag}\% / \UseMacro{cd-ir-em-rag}\% and \UseMacro{method-drc} \UseMacro{cd-cr-em-drc}\% / \UseMacro{cd-ir-em-drc}\% EM, $5$--$8$\,pp above \UseMacro{method-pretrained}---but both plateau roughly $23$--$27$\,pp below \codeloraevo{}, and neither closes the gap to any trained adapter.
Among fine-tuned methods, \UseMacro{method-slora} reaches \UseMacro{cd-cr-em-slora}\% / \UseMacro{cd-ir-em-slora}\% EM; \UseMacro{method-plora} reaches 64.2\% IR EM (the only applicable split).
\codelorastatic{}, included as a within-framework reference on the same commit-derived inputs, scores \UseMacro{cd-cr-em-codelorastatic}\% / \UseMacro{cd-ir-em-codelorastatic}\%, which is close to \UseMacro{method-slora} on CR and markedly below its static-track performance (\UseMacro{cr-em-codelorastatic}\% / \UseMacro{ir-em-codelorastatic}\%). %
\codeloraevo{} is the strongest method on both splits (\UseMacro{cd-cr-em-codeloraevo}\% CR, \UseMacro{cd-ir-em-codeloraevo}\% IR EM), +\UseMacro{cd-cr-em-delta-codeloraevo}\,pp over \UseMacro{method-slora} on CR and matching the \UseMacro{method-plora} upper bound on IR (64.5\% vs.\ 64.2\%) without per-repository training; its gaps over \UseMacro{method-slora} and \codelorastatic{} are significant on both splits and on the OOD holdout (Appendix~\ref{app:significance}).
Appendix Figure~\ref{fig:commit_pct_trend} (\S\ref{app:commit_pct_trend}) shows that this lead holds in every part of the commit history, widening from ${\sim}3.5$\,pp on early commits to ${\sim}6$\,pp on late ones.

\subsection{Out-of-Distribution Generalization}
\label{sec:results:ood}

The OOD set comprises \UseMacro{num-repos-ood} repositories created strictly after the training cutoff (\texttt{2025-04-01}) and used for evaluation only.
Table~\ref{tab:ood_results} reports results on the temporal holdout under the same commit-derived protocol as Table~\ref{tab:per_commit_results}.
\codeloraevo{} achieves the highest EM (\UseMacro{ood-em-codeloraevo}\%), ahead of \codelorastatic{} (\UseMacro{ood-em-codelorastatic}\%) and \UseMacro{method-slora} (\UseMacro{ood-em-slora}\%).
OOD assertion targets are systematically shorter than in-distribution ones (median 7 vs.\ 12--13 characters; Appendix~\ref{app:ood_caveats}), which inflates EM for every method, so we compare only within Table~\ref{tab:ood_results}.
On that basis, \codeloraevo{} leads the next-best fine-tuned adapter by ${\sim}$1.8\,pp EM---narrower than the evolution-track gap (${\sim}$5\,pp CR EM, Table~\ref{tab:per_commit_results}) but positive and consistent across EditSim and CodeBLEU.

\section{Conclusion}
\label{sec:conclusion}

We introduced \codelora{}, a hypernetwork framework that generates repository-specific LoRA adapters, effectively injecting repository knowledge with zero inference-time token overhead,
and \repopeftbench{}, a benchmark of \UseMacro{num-repos-total} Python repositories suitable for evaluating repository-level PEFT methods.
\codelorastatic{} maps a repository snapshot to an adapter and reaches \UseMacro{cr-em-codelorastatic}\% CR / \UseMacro{ir-em-codelorastatic}\% IR EM on the static track;
\codeloraevo{} updates its adapter through a GRU hidden state on each code diff and reaches \UseMacro{cd-cr-em-codeloraevo}\% CR / \UseMacro{cd-ir-em-codeloraevo}\% IR EM on the evolution track.
On repositories created after the training cutoff, \codeloraevo{} remains the strongest method, although its margin over the next-best fine-tuned adapter narrows to ${\sim}$1.8\,pp EM.
Within the setting we study---one task, one language, and one frozen backbone---these results show that parametric adaptation outperforms the evaluated retrieve-and-prepend and dependency-resolved baselines, and that tracking software evolution matters once evaluation follows a repository's commit history.
We envision \codelora{} as a building block for stronger, repository-customized, and less costly AI coding assistants.

\section*{Limitations}

\paragraph{Scope of evaluation.}
We evaluate only on Python repositories, a single frozen LLM backbone, and one downstream task (assertion completion that requires challenging repository-level reasoning).
The \codelora{} architecture is language- and task-agnostic by construction, and can be expanded by collecting more training data and scaling up the training of the hypernetwork.
Due to the limited compute budget, we leave extending the empirical evidence to additional languages, backbones, and downstream tasks to future work.

\paragraph{Coverage of the baseline space.}
The retrieval and hybrid configurations we evaluate cover only the most standard configurations of context injection.
Retrieval quality is sensitive to how context is presented to the backbone (Appendix~\ref{app:context_calibration}), and stronger retrievers, rerankers, or agentic search may narrow the gap.
We also hold the backbone fixed across all methods to isolate the repository-injection mechanism, and thus do not compare against long-context LLMs; we view this direction as complementary, since parametric adaptation can be applied on top of a long-context backbone while still avoiding re-encoding the repository per query.

\paragraph{Fixed-size repository representation.}
\codelora{} conditions on a fixed-size pooled embedding of the repository or diff.
Performance does not degrade with repository size within \repopeftbench{} (up to \UseMacro{repo-tok-max} tokens; Appendix~\ref{app:repo_size}), but much larger codebases may require hierarchical or graph-structured encoders.

\paragraph{Memorization analysis.}
Appendix~\ref{app:leakage} evaluates the potential memorization effects of training answers, assertion templates, and repositories.
It scores held-out targets and assertion templates that never appear in training, and the test repositories least similar to the training set, including the cross-repository split and the post-cutoff repositories.
The advantage of \codelora{} persists on these unseen answers, templates, and repositories.
While a perturbation-based memorization test is not done, these experiments already show that \codelora{} generalizes to unseen repositories and patterns.

\paragraph{Syntactic-level metrics.}
Exact match misses functional equivalence; EditSim and CodeBLEU only partly mitigate this.
Evaluation by execution could provide a more semantic measure, but requires setting up reliable runtime environments for each project, which we leave to future work.

\paragraph{Model size.}
The LoRA-generation hypernetwork dominates the trainable parameter count---${\sim}$340M for \codelorastatic{} and ${\sim}$713M for \codeloraevo{}---and is itself a function of the backbone's projection dimensions.
The evolution-track finding is therefore most directly supported at the 1.5B-parameter scale; whether recurrent aggregation over commit diffs remains necessary (or sufficient) once the backbone is much larger is an open question.

\paragraph{Reproducibility.}
Code, model checkpoints, and \repopeftbench{} are available at the anonymized links in the abstract, and hyperparameters are in Appendix~\ref{app:implementation}; all experiments run on a single H100 80\,GB GPU.

\paragraph{Potential risks.}
\repopeftbench{} is constructed exclusively from public permissively licensed Python repositories (Appendix~\ref{app:dataset_construction}), so the dataset itself does not introduce new personal data, harmful content, or proprietary code into circulation, and we redistribute each repository under its original license terms with attribution preserved.
The downstream artifact---a code language model conditioned on a repository-specific LoRA---inherits the well-understood risks of code LLMs more broadly: it can be steered to emit insecure, incorrect, or licensed-code-resembling completions, and repository conditioning may increase attribution risk if a user feeds in a private repository and the generated assertions surface verbatim from training repositories.
Production deployment should therefore apply standard mitigations, such as license-aware filtering of generated code, human review of generated test assertions before commit, and rejection of completions matching long verbatim training spans.

\section*{Acknowledgments}
We thank Saarang Agarwal, Kyunghyun Cho, Bihui Jin, Jiale Amber Wang, Wentao Zhang, Yifan Zong
and the anonymous reviewers for their comments and feedback.
This work is enabled in part by support provided by Compute Ontario (computeontario.ca) and the Digital Research Alliance of Canada (alliancecan.ca).
This work is partially supported by the Natural Sciences and Engineering Research Council of Canada (NSERC) under funding reference number RGPIN-2024-04909 and RGPIN-2024-05178.

\bibliography{bib}

\newpage
\clearpage
\appendix

\section{Use of LLMs}
\label{app:llms}
We used an LLM-based writing assistant to polish grammar. All ideas, analyses, experiments, and scientific claims are our own, and we take full responsibility for the content of this work.

\section{Dataset Details}
\label{app:dataset}
\label{app:dataset_construction}

This section documents detailed construction process and statistics of \repopeftbench{}, organized as the data flows from raw GitHub repositories to the QnA splits actually consumed by the methods in Tables~\ref{tab:main_results}--\ref{tab:ood_results}:
task motivation (\S\ref{app:dataset:motivation}),
repository selection and licensing (\S\ref{app:dataset:selection}),
construction pipeline (\S\ref{app:dataset:pipeline}),
the splits used at training and evaluation (\S\ref{app:dataset:splits}),
composition by assertion family and target type (\S\ref{app:dataset_composition}),
token-length distributions (\S\ref{app:dataset:tokens}),
per-repository breakdown (\S\ref{app:per_repo}),
and the privacy / content review (\S\ref{app:dataset:privacy}).

\subsection{Motivation and Task}
\label{app:dataset:motivation}

\paragraph{Why a repository-conditioned assertion task.}
Our assertion-completion task is directly inspired by the \emph{code execution} task of LiveCodeBench~\cite{jain2025livecodebench}, which probes whether a model can predict the runtime value produced by a piece of code at a designated point of evaluation.
Treating an assertion target as the ``answer'' a developer wrote down for what a piece of code \emph{should} evaluate to at exactly that line, the prediction objective inherits the same semantics---compute, in the model's head, what this expression would resolve to in this concrete context---while replacing LiveCodeBench's hand-curated, single-function snippets with naturally occurring assertions extracted at scale from real test suites.
This reframing keeps the cognitive load of the original task (multi-step, type-aware, value-level reasoning over surrounding code) and additionally couples each prediction to a full repository's API surface, naming conventions, fixtures, and domain vocabulary---turning code execution into an explicit \emph{repository-conditioned} reasoning probe.

\paragraph{Why a new dataset.}
Existing repository-level benchmarks (RepoBench~\cite{liu2023repobenchbenchmarkingrepositorylevelcode}, CrossCodeEval~\cite{ding2024crosscodeeval}, RepoHyper~\cite{phan2024repohypersearchexpandrefinesemanticgraphs}, R2C2-Coder~\cite{deng2024r2c2coder}) ship only the slices their task consumes---a target file and a handful of retrieval-selected snippets---and discard the rest of the codebase and the Git history at release time.
This is fine for input-side methods but precludes any method that must ingest the \emph{whole} repository as parameters or as a streaming state.
We therefore release each repository in \repopeftbench{} \emph{whole}: every non-test source file (for the repository representation), every test file (for assertion QnAs), and every first-parent production commit (for the evolution track's diff sequences).

\subsection{Repository Selection and Licensing}
\label{app:dataset:selection}

\paragraph{In-distribution selection.}
The GitHub Search API was queried with \texttt{language:python license:mit stars:>=300 pushed:>=2023-01-01} together with a \texttt{pytest}/\texttt{unittest} usage filter; matching repositories were ranked by star count and downloaded in two passes (the upper pool with $\geq 1000$ stars and a mid-range pool with $300$--$1000$ stars), yielding the $512$ in-distribution repositories used for training and CR/IR evaluation.

\paragraph{Temporal OOD holdout.}
To probe generalization beyond the training scrape, we mined an additional set of repositories with the same language, testing, activity, size, and non-fork filters but \emph{without} the $\geq 300$-star constraint---star-count ranges were searched from $6$ upward so that enough candidates exist among repositories created strictly after \texttt{2025-04-01}.
Permissive licenses (MIT and Apache-2.0) were both considered during mining; $92$ repositories passed fork-chain and pytest checks and yield valid assertion pairs.
Together with the in-distribution corpus, these form the $604$ repositories in \repopeftbench{}.
Because the in-distribution query hard-filtered on \texttt{license:mit}, all $512$ in-distribution repositories are MIT-licensed; the OOD holdout may include Apache-2.0 repositories where that was the upstream license.
We retain a copy of each repository's \texttt{LICENSE} file alongside the source tree in the released dataset, and the dataset release itself is distributed under the same MIT terms with attribution to the upstream maintainers preserved.

\paragraph{Intended use and consistency with upstream terms.}
Using the source contents of MIT-licensed public repositories for research on code language models is consistent with the upstream license, which explicitly permits use, modification, and redistribution provided that the copyright notice is included.
\repopeftbench{} and the released \codelora{} checkpoints are intended exclusively for non-commercial research on repository-level adaptation of code LMs; downstream commercial or product deployment is out of scope for this release and would require an independent re-licensing review of each contributing repository.
Derivatives produced from the dataset (e.g., embeddings, generated LoRAs, predictions) inherit the same research-use scope.

\subsection{Construction Pipeline}
\label{app:dataset:pipeline}

\paragraph{Test file identification.}
Files are classified as test files if they match any of:
\texttt{test\_*.py}, \texttt{*\_test.py}, or reside in directories named \texttt{tests/}, \texttt{test/}.
Identified test files are moved to a separate \texttt{TEST\_HYPERNET/} directory within each repository, preserving relative paths.

\paragraph{Structured prefix construction.}
Each QnA prefix is constructed as follows:
\begin{enumerate}[nosep]
\item All import statements from the test file.
\item The enclosing class definition (if the test is a method).
\item Helper methods (\texttt{setUp}, \texttt{tearDown}, fixtures).
\item The test function signature and body up to the assertion cut point.
\end{enumerate}
This structured approach preserves the most informative context while managing token budget.

\paragraph{Quality filters applied.}
\begin{itemize}[nosep]
\item Targets starting with comma (malformed AST segmentation).
\item Targets outside function bodies (module-level assertions).
\item Empty or whitespace-only targets.
\item Duplicate targets within the same test function.
\item Targets containing only punctuation or single characters.
\end{itemize}

\begin{figure}[t]
\begin{center}
\includegraphics[width=\columnwidth]{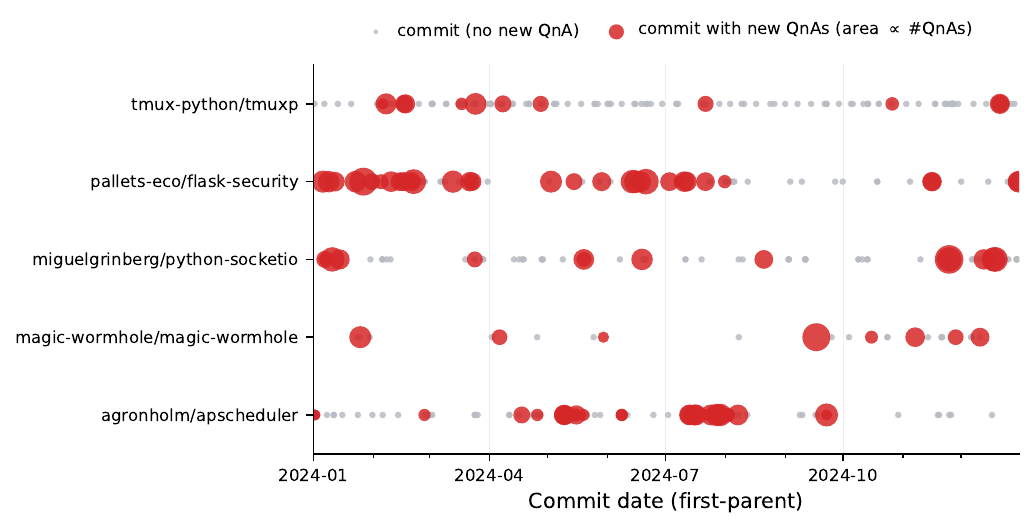}
\caption{Bursty commit pattern, illustrated using randomly selected 5 repositories out of the \UseMacro{num-repos-total} \repopeftbench{} repositories. Test-touching commits arrive irregularly; the median repository accumulates over 100 such commits, motivating per-commit (rather than one-shot) adaptation under software evolution.}
\label{fig:dataset_construction}
\end{center}
\end{figure}

\paragraph{Bursty commit pattern.}
Figure~\ref{fig:dataset_construction} shows the per-repository test-touching commit distribution that motivates the evolution track: test-touching commits arrive in irregular bursts rather than at uniform intervals, so a single static snapshot of any repository fails to capture the full history of assertion edits seen during active development.

\subsection{Splits Used in Experiments}
\label{app:dataset:splits}

Table~\ref{tab:dataset_actual} (in the main paper) reports the exact splits consumed by every number in Tables~\ref{tab:main_results}--\ref{tab:ood_results} (one row per split actually used at training or evaluation time); Table~\ref{tab:dataset_actual_full} below expands that overview with per-commit and per-repository densities, including the smart-cap output for \codeloraevo{} training.
For the evolution track we enforce a per-commit cap of ${\le}8$ QnAs at both training time (as part of the smart cap, which additionally bounds at ${\le}4$ QnAs per test file) and evaluation time: every evaluator scores the first ${\le}8$ QnAs per (repo, commit) group so that the EM / EditSim / CodeBLEU averages are not dominated by a few unusually large commits with hundreds of QnAs in a single test file.
The average density after the eval-time cap is ${\sim}6.8$ QnAs per commit (below the cap because many commits naturally have fewer than $8$ QnAs).

\begin{table*}[t]
\begin{small}
\begin{center}
\caption{Fine-grained statistics for every split actually consumed by the main tables. \emph{Static track}: one anchor snapshot per repository (rows feed Table~\ref{tab:main_results}). \emph{Evolution track}: multi-commit prefixes (rows feed Tables~\ref{tab:per_commit_results} and~\ref{tab:ood_results}); the smart cap (${\le}4$ QnAs per test file, ${\le}8$ per commit) is applied to \codeloraevo{} training rows so that no commit can dominate a backprop window.}\label{tab:dataset_actual_full}

\setlength{\tabcolsep}{4pt}
\begin{tabular}{l r r r r r r}
\toprule
Split & Repos & Commits & QnAs & Cmts / repo & QnAs / cmt & QnAs / repo \\
\midrule
\multicolumn{7}{@{}l}{\emph{Static track --- one anchor snapshot per repository (no per-commit cap)}} \\
Train             & 409 & 409 & 39{,}612 & 1.00 &  96.9 &  96.9 \\
CR Val            &  51 &  51 &  6{,}213 & 1.00 & 121.8 & 121.8 \\
CR Test           &  52 &  52 &  6{,}414 & 1.00 & 123.3 & 123.3 \\
IR Val            & 409 & 409 &  4{,}833 & 1.00 &  11.8 &  11.8 \\
IR Test           & 409 & 409 &  5{,}222 & 1.00 &  12.8 &  12.8 \\
OOD Test          &  92 &  92 &  9{,}942 & 1.00 & 108.1 & 108.1 \\
\midrule
\multicolumn{7}{@{}l}{\emph{Evolution track --- multi-commit; ${\le}8$ QnAs / commit at train (smart-cap, ${\le}4$/file) and eval}} \\
Train (\codelorastatic{}, anchor)        & 400 &         400 &  44{,}149 &   1.00  & 110.4 &    110.4 \\
Train (\codeloraevo{}, 8-cap)            & 400 &    45{,}516 & 215{,}129 & 113.79  &   4.73 &    537.8 \\
CR Val                                   &  49 &     8{,}614 &  58{,}944 & 175.80  &   6.84 & 1{,}203 \\
CR Test                                  &  51 &     6{,}618 &  44{,}732 & 129.76  &   6.76 &    877 \\
IR Val                                   & 389 &     5{,}710 &  38{,}783 &  14.68  &   6.79 &    99.7 \\
IR Test                                  & 389 &     6{,}179 &  42{,}061 &  15.88  &   6.81 &  108.1 \\
OOD Test                                 &  92 &     1{,}950 &  14{,}813 &  21.20  &   7.60 &  161.0 \\
\bottomrule
\end{tabular}

\end{center}
\end{small}
\end{table*}

\subsection{Composition by Assertion Family and Target Type}
\label{app:dataset_composition}

To characterize the assertion-completion task at the level of what the model actually predicts, Table~\ref{tab:dataset_details} breaks down each split by assertion family (which keyword triggers the test) and by target type (what the assertion expects).
The three splits are tightly aligned: bare \texttt{assert} accounts for ${\sim}$82--86\% of pairs and the target distribution (numeric/string literals, variables, function calls, complex expressions) varies by at most ${\sim}$2\,pp between train, CR test, and IR test.
This rules out distribution shift across splits as an explanation for the cross-repo gap, and confirms that improvements on CR test are genuine generalization rather than reweighting of easier target categories.

\begin{table}[t]
\begin{small}
\begin{center}
\caption{Composition of the static-track QnAs by assertion family (which keyword triggers the test) and target type (what the assertion expects), computed over the $62{,}294$ QnAs actually used at training and evaluation time (sum of static train, CR Val/Test, and IR Val/Test rows in Table~\ref{tab:dataset_actual}). Splits are tightly aligned: every target-type fraction differs by at most ${\sim}2$\,pp between train, CR test, and IR test.}\label{tab:dataset_details}

\begin{tabular}{lrrr}
\toprule
Property & Train & CR Test & IR Test \\
\midrule
Assertion types & & & \\
\quad \texttt{assert} & 82.5\% & 86.2\% & 82.2\% \\
\quad \texttt{self.assert*} & 13.5\% & 10.0\% & 13.6\% \\
\quad \texttt{pytest.*} & 4.1\% & 3.8\% & 4.3\% \\
\midrule
Target types & & & \\
\quad Numeric literal & 18.7\% & 19.9\% & 19.4\% \\
\quad String literal & 18.2\% & 18.2\% & 18.5\% \\
\quad Variable & 21.7\% & 21.9\% & 21.8\% \\
\quad Collection & 11.8\% & 10.2\% & 11.3\% \\
\quad Function call & 9.4\% & 10.2\% & 8.9\% \\
\quad Complex expression & 14.5\% & 14.0\% & 15.0\% \\
\quad Bool/None literal & 5.8\% & 5.5\% & 5.1\% \\
\bottomrule
\end{tabular}

\end{center}
\end{small}
\end{table}

\subsection{Token-Length Statistics}
\label{app:dataset:tokens}

\begin{figure}[t]
\begin{center}
\includegraphics[width=\columnwidth]{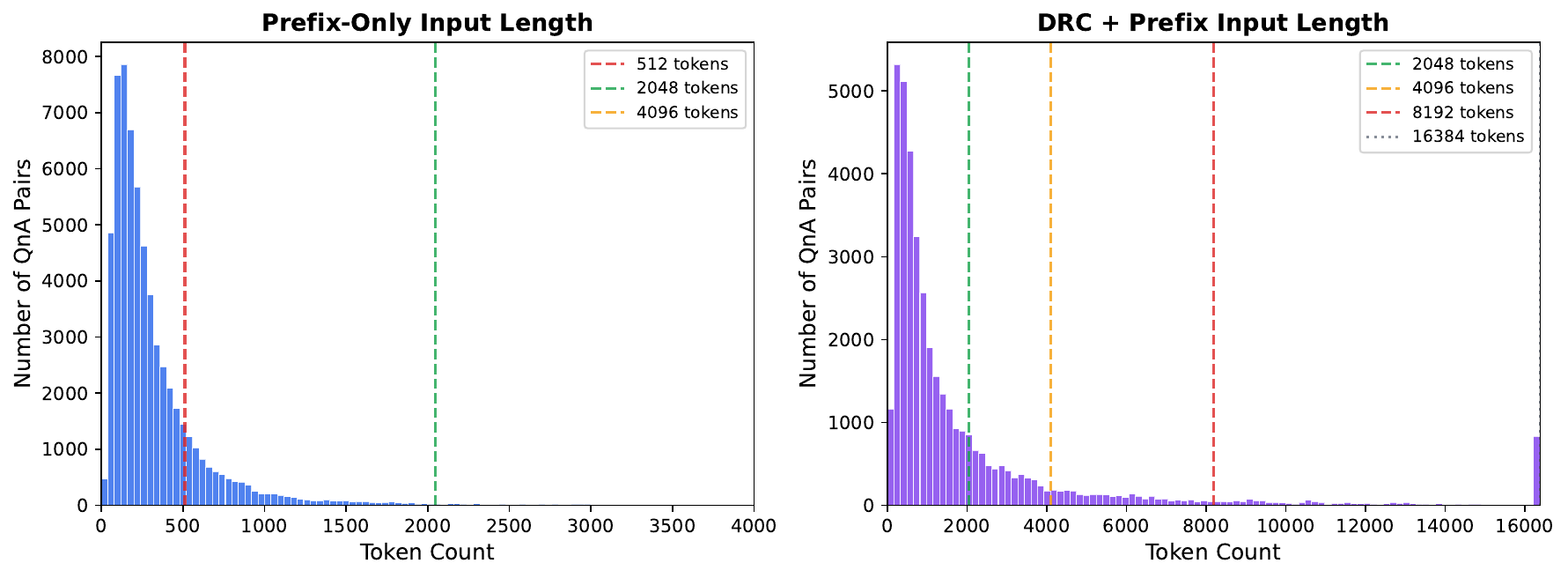}
\caption{Token length distributions for prefix-only (left) and DRC+prefix (right) input formats across all splits. Vertical dashed lines mark common context window sizes. Prefix-only inputs are compact (median $224$ tokens), while DRC+prefix inputs have a heavy right tail requiring larger context windows.}
\label{fig:input_dist}
\end{center}
\end{figure}

Table~\ref{tab:token_stats} reports token-length distributions for the four input components (repository, DRC context, structured prefix, target) over the $62{,}294$ static-track QnAs (Qwen2.5-Coder-1.5B tokenizer; same denominator as Table~\ref{tab:dataset_details}).
Repositories are large (median $165$K tokens), DRC context---when present---is moderate (median $517$ tokens) but heavy-tailed, prefixes are compact (median $224$ tokens), and targets are short (median $3$ tokens).
Figure~\ref{fig:input_dist} plots the prefix-only and DRC+prefix length distributions side by side and marks common context-window sizes, illustrating why the DRC baseline needs a much larger input budget than the prefix alone.

\begin{table*}[t]
\begin{small}
\begin{center}
\caption{Token length statistics across the $62{,}294$ static-track QnAs (Qwen2.5-Coder-1.5B tokenizer). Repo size is the total token count of all Python source files per repository (repeated per pair). DRC statistics are over the $64.1\%$ of pairs with resolvable dependency context.}\label{tab:token_stats}

\setlength{\tabcolsep}{6pt}
\renewcommand{\arraystretch}{1.1}
\begin{tabular}{lrrrrrrr}
\toprule
\textbf{Component} & \textbf{Mean} & \textbf{Med.} & \textbf{Std} & \textbf{p75} & \textbf{p95} & \textbf{p99} & \textbf{Max} \\
\midrule
Repo size       & 284{,}509 & 165{,}376 & 363{,}914 & 311{,}729 & 1{,}028{,}703 & 1{,}865{,}509 & 2{,}994{,}853 \\
DRC context$^\dag$ & 1{,}900 & 517 & 6{,}243 & 1{,}634 & 7{,}849 & 20{,}826 & 574{,}001 \\
Prefix          & 360 & 224 & 566 & 396 & 992 & 2{,}588 & 27{,}171 \\
Target          & 4.8 & 3.0 & 10.2 & 5.0 & 14 & 43 & 290 \\
\bottomrule
\multicolumn{8}{l}{\footnotesize $^\dag$ Computed over 39{,}902 pairs (64.1\%) with resolvable dependency context.}
\end{tabular}

\end{center}
\end{small}
\end{table*}

\subsection{Per-Repository Performance Breakdown}
\label{app:per_repo}

To support repository-by-repository scrutiny of every method, we release a per-repository table covering all $409$ IR-test repositories with EM, EditSim, CodeBLEU, and example counts for pretrained, FFT, sLoRA, per-repo LoRA, and \codelorastatic{}.
The supplementary materials contain the full table; aggregate distributions and the data-sparsity scatter for per-repo LoRA are summarized in Figures~\ref{fig:per_repo_boxplot} and~\ref{fig:plora_data_sparsity}.

\subsection{Privacy and Content Review}
\label{app:dataset:privacy}

The dataset contains only non-test source files and test files from public open-source projects with permissive licenses, copied verbatim from the upstream repositories.
No private repositories, user accounts, commit messages, issue bodies, or PR discussions are included; identifying information is therefore limited to whatever the upstream maintainers chose to embed in public Python source (e.g., author docstrings, copyright headers in \texttt{LICENSE} files, contact emails inside module-level docstrings of well-known libraries).
We did not perform automated PII scrubbing because (i) the dataset is a redistribution of already-public, license-permitted source, and (ii) any aggressive scrubbing would alter the very identifiers (function names, fixture names, class names) that the benchmark task requires the model to predict.
We did not observe offensive content in random spot checks of the dataset, which is consistent with the high-star permissive-license selection criterion; users who identify problematic content in any of the released repositories may file an issue against the dataset repository for removal.

\section{Additional Ablation Studies}
\label{app:additional_results}

\subsection{RAG with Different $k$}
\label{app:rag_k}

We sweep the number of retrieved chunks $k$ and the chunk size under the calibrated configuration of \S\ref{app:context_calibration} (Table~\ref{tab:rag_k}).
$k$ chunks of $c$ tokens give a retrieval budget of at most $k\cdot c$ tokens per query.
The main tables use the budget-matched setting $k{=}3$ with 512-token chunks (at most $1{,}536$ tokens), which is comparable to the context \UseMacro{method-drc} adds (median $517$, 75th percentile $1{,}634$, mean $1{,}900$ tokens on the static track; Table~\ref{tab:token_stats}); on the evolution track the retrieved context is compressed to the same $1{,}536$-token budget.
Larger budgets help pretrained RAG only marginally: $k{=}10$ at 512-token chunks (up to $5{,}120$ tokens) adds $0.6$\,pp CR and $0.9$\,pp IR EM, and across all five configurations EM varies by at most $1.3$\,pp.
At an equal budget, 512-token chunks are slightly better than 256-token chunks ($47.8$ vs.\ $46.9$ CR EM at $2{,}560$ tokens).
Retrieval also helps the trained baselines (Table~\ref{tab:rag_k}, bottom): at 256-token chunks and $k{=}5$, \UseMacro{method-fft} gains $2.7$\,/\,$0.9$\,pp and \UseMacro{method-slora} $4.1$\,/\,$2.7$\,pp EM (CR\,/\,IR).
Even the largest budget we tried ($48.2$\,/\,$48.9$ EM) stays below \UseMacro{method-drc} ($49.1$\,/\,$50.0$) and far below \codelorastatic{} ($63.8$\,/\,$66.2$), so the gap to \codelora{} is not an artifact of the retrieval budget.

\begin{table*}[t]
\begin{small}
\begin{center}
\caption{RAG ablation over chunk size and $k$ on CR and IR test, under the calibrated configuration of \S\ref{app:context_calibration}. Top: pretrained + RAG (native repository layout); bottom: trained models + RAG at inference (plain layout). The main tables use the budget-matched settings: $k{=}3$ with 512-token chunks for \UseMacro{method-rag} (at most $1{,}536$ retrieved tokens) and $k{=}5$ with 256-token chunks for \UseMacro{method-fft-rag} (at most $1{,}280$).}\label{tab:rag_k}

\begin{tabular}{llccc@{\hskip 12pt}ccc}
\toprule
& & \multicolumn{3}{c}{CR Test} & \multicolumn{3}{c}{IR Test} \\
\cmidrule(lr){3-5} \cmidrule(lr){6-8}
Chunk & $k$ & EM & EditSim & CB & EM & EditSim & CB \\
\midrule
\multicolumn{8}{l}{\emph{Pretrained + RAG}} \\
\midrule
512 & 3 & 47.6 & 0.624 & 0.655 & 48.0 & 0.632 & 0.661 \\
512 & 5 & 47.8 & 0.624 & 0.656 & 48.8 & 0.639 & 0.665 \\
512 & 10 & 48.2 & 0.629 & 0.658 & 48.9 & 0.639 & 0.663 \\
256 & 5 & 47.6 & 0.623 & 0.654 & 48.4 & 0.637 & 0.662 \\
256 & 10 & 46.9 & 0.618 & 0.648 & 47.7 & 0.630 & 0.657 \\
\midrule
\multicolumn{8}{l}{\emph{Trained + RAG}} \\
\midrule
256 & 5 (FFT) & 54.1 & 0.703 & 0.689 & 56.8 & 0.730 & 0.713 \\
256 & 5 (sLoRA) & 51.5 & 0.689 & 0.674 & 53.1 & 0.706 & 0.692 \\
\bottomrule
\end{tabular}

\end{center}
\end{small}
\end{table*}

\section{Additional Analyses}
\label{app:additional_analyses}

This section reports the analyses that validate the comparisons in Tables~\ref{tab:main_results}--\ref{tab:ood_results}: how the context-injection baselines are calibrated to the backbone (\S\ref{app:context_calibration}), the exact-match rule and its strict variant (\S\ref{app:em_rule}), statistical significance of the headline gaps on the static track, the evolution track, and the OOD holdout (\S\ref{app:significance}), the split between locally-solvable and repository-knowledge-required instances together with breakdowns by target length and target type (\S\ref{app:local_solvability}), behavior as a function of repository size (\S\ref{app:repo_size}), and memorization probes at the level of answers, assertion templates, and repository similarity (\S\ref{app:leakage}).

\subsection{Calibrating the Context-Injection Baselines}
\label{app:context_calibration}

Context-injection baselines are sensitive to how retrieved code is presented to the backbone, so we configure both of them for Qwen2.5-Coder before drawing any parametric-versus-context conclusion.
When retrieved code precedes the test prefix, Qwen2.5-Coder can switch onto its fill-in-the-middle code path and emit its own control sentinels (\texttt{<|fim\_suffix|>}, \texttt{<|file\_sep|>}, \texttt{<|repo\_name|>}) instead of continuing the assertion.
Both baselines therefore keep the test prefix un-truncated and ban those sentinels at decode time (\texttt{bad\_words\_ids}), with the retriever, index, $k{=}3$, and context budget fixed.
On top of this we consider two layouts for the retrieved code: plain prepending, and the backbone's native repository-completion layout, in which retrieved files are delimited by \texttt{<|repo\_name|>} and \texttt{<|file\_sep|>}.
On the static track retrieval uses 512-token windows; on the evolution track, where the index is rebuilt per commit, chunks are aligned to AST boundaries.

We tune each context baseline in its own favor and report the stronger layout per method.
For RAG the native layout is stronger on the static track ($44.6$\,/\,$46.1$ plain vs.\ \UseMacro{cr-em-rag}\,/\,\UseMacro{ir-em-rag} native, CR\,/\,IR); for DRC plain prepending is stronger (\UseMacro{cr-em-drc}\,/\,\UseMacro{ir-em-drc} vs.\ $48.2$\,/\,$48.6$), since import-resolved definitions are already file-scoped and gain nothing from file delimiters.
This is the configuration used throughout the main tables.
Under it, every context baseline improves over \UseMacro{method-pretrained}, by $1.2$--$3.4$\,pp on the static track, $5.0$--$8.1$\,pp on the evolution track, and $11.0$--$12.1$\,pp on OOD, yet even the \emph{strongest} one trails \codelorastatic{} by $14.7$\,/\,$16.2$\,pp (CR\,/\,IR) on the static track and \codeloraevo{} by $23.5$\,/\,$27.1$\,pp on the evolution track.

\subsection{Exact-Match Rule}
\label{app:em_rule}

The only relaxation in our EM metric is trimming overgeneration before comparison, which is standard for left-to-right code completion and is applied identically to every method (\texttt{evaluation/metrics.py}).
Concretely, we strip Qwen FIM sentinels, drop trailing comments, and keep the leading segment of the generation at its natural boundary: the first line, and---matching the short assertion targets---up to the first comma or first token.
Both strings then receive identical normalization (strip surrounding and trailing whitespace and \texttt{\{\,:\,,\,\}}, collapse internal whitespace) and must be exactly equal for $\text{EM}{=}1$.
There is no fuzzy, aliased, or semantic matching.
Because only the \emph{start} of the generation is kept, a method is credited only when it produces the answer first, never when a correct answer appears after distractor text.

This rule cannot flatter \codelora{}; it rescues the baselines.
Table~\ref{tab:strict_em} compares it against a strict variant (exact string equality, no truncation).
\codelorastatic{} is \emph{identical} under both metrics (\UseMacro{cr-em-codelorastatic}\,/\,\UseMacro{ir-em-codelorastatic}) because its completions terminate cleanly at the target, so none of its correct answers come from truncated overgeneration.
The fine-tuned baselines, which run past the short target, are the ones the relaxation saves: \UseMacro{method-fft} falls from $51.4$ to $14.8$ and \UseMacro{method-slora} from $47.4$ to $7.5$ on CR test.
The method ranking is unchanged under strict scoring and \codelora{}'s margin is strictly larger, so no conclusion in this paper depends on the relaxation.

\begin{table*}[t]
\begin{small}
\begin{center}
\caption{Relaxed vs.\ strict exact match on the static track (\S\ref{app:em_rule}). \codelorastatic{} is unaffected by the relaxation; the fine-tuned baselines depend on it heavily, so the relaxed rule narrows rather than widens our reported lead.}\label{tab:strict_em}

\begin{tabular}{lcc@{\hskip 14pt}cc}
\toprule
& \multicolumn{2}{c}{Cross-Repo (CR Test)} & \multicolumn{2}{c}{In-Repo (IR Test)} \\
\cmidrule(lr){2-3} \cmidrule(lr){4-5}
Method & relaxed EM & strict EM & relaxed EM & strict EM \\
\midrule
\UseMacro{method-pretrained} & 45.7 & 44.4 & 46.8 & 45.2 \\
\UseMacro{method-fft}        & 51.4 & 14.8 & 55.9 & 18.4 \\
\UseMacro{method-slora}      & 47.4 &  7.5 & 50.4 &  9.7 \\
\codelorastatic{}            & \textbf{63.8} & \textbf{63.8} & \textbf{66.2} & \textbf{66.2} \\
\bottomrule
\end{tabular}

\end{center}
\end{small}
\end{table*}

\subsection{Statistical Significance}
\label{app:significance}

Table~\ref{tab:significance} attaches uncertainty to the headline static-track comparisons, computed against the calibrated context baselines of \S\ref{app:context_calibration} so that the tests match the numbers in Table~\ref{tab:main_results}.
We report the paired per-instance difference in EM with a 95\% bootstrap confidence interval ($5{,}000$ resamples).
On $n{=}6{,}414$ CR and $n{=}5{,}222$ IR instances every gap is far outside sampling error: the smallest, \codelorastatic{} over \UseMacro{method-fft-rag} on CR, is $+9.7$\,pp with a 95\% CI of $[+8.6, +10.7]$ and $978$ versus $358$ discordant pairs.
Because instances are clustered within repositories, we additionally report a repository-clustered bootstrap that resamples whole repositories; the intervals widen---most on CR test, where only \UseMacro{num-repos-cr-test} repositories are resampled---but every interval stays comfortably above zero.
A McNemar test (with continuity correction) on the discordant pairs gives $p < 10^{-47}$ for every comparison in the table.
The \UseMacro{method-plora} rows of Tables~\ref{tab:main_results} and~\ref{tab:per_commit_results} are deliberately excluded from these tests: they are an in-repo upper bound trained per repository and have no cross-repo counterpart, so they are not a like-for-like comparison.

Table~\ref{tab:significance_evo} applies the same three tests to \codeloraevo{} on the evolution track and the OOD holdout, against the strongest non-\codelora{} baseline there (\UseMacro{method-slora}) and against \codelorastatic{}.
The evolution-track gaps are again well outside sampling error, and even the narrow OOD margin over \UseMacro{method-slora} ($+1.8$\,pp) keeps its repository-clustered interval above zero ($[+0.8, +3.0]$ over $92$ repositories).
McNemar gives $p < 10^{-10}$ for every comparison in that table.

\begin{table*}[t]
\begin{small}
\begin{center}
\caption{Paired significance tests for the static-track headline gaps (\S\ref{app:significance}). $\Delta$EM is the per-instance paired difference in percentage points against the calibrated baselines of Table~\ref{tab:main_results}; CIs are $5{,}000$-resample bootstrap intervals, and the repository-clustered variant resamples whole repositories. Every comparison has McNemar $p < 10^{-47}$.}\label{tab:significance}

\begin{tabular}{@{}l@{\hskip 8pt}rcc@{\hskip 12pt}rcc@{}}
\toprule
& \multicolumn{3}{c}{\UseMacro{TH-cr}, $n{=}6{,}414$} & \multicolumn{3}{c}{\UseMacro{TH-ir}, $n{=}5{,}222$} \\
\cmidrule(lr){2-4} \cmidrule(lr){5-7}
\codelorastatic{} vs. & $\Delta$EM & 95\% CI & repo-clust.\ CI & $\Delta$EM & 95\% CI & repo-clust.\ CI \\
\midrule
\UseMacro{method-fft-rag}    & $+9.7$  & $[+8.6, +10.7]$  & $[+7.6, +11.7]$  & $+9.4$  & $[+8.2, +10.7]$  & $[+7.9, +10.9]$  \\
\UseMacro{method-fft}        & $+12.4$ & $[+11.4, +13.4]$ & $[+10.5, +14.4]$ & $+10.4$ & $[+9.2, +11.5]$  & $[+8.9, +12.0]$ \\
\UseMacro{method-drc}        & $+14.7$ & $[+13.6, +15.9]$ & $[+12.2, +17.4]$ & $+16.2$ & $[+14.9, +17.5]$ & $[+14.6, +18.0]$ \\
\UseMacro{method-rag}        & $+16.2$ & $[+15.0, +17.4]$ & $[+13.4, +19.0]$ & $+18.2$ & $[+16.9, +19.6]$ & $[+16.4, +20.1]$ \\
\UseMacro{method-slora}      & $+16.4$ & $[+15.3, +17.5]$ & $[+13.8, +19.0]$ & $+15.8$ & $[+14.6, +17.1]$ & $[+14.2, +17.5]$ \\
\textlora{}                  & $+18.0$ & $[+16.9, +19.2]$ & $[+15.4, +20.8]$ & $+19.6$ & $[+18.3, +20.8]$ & $[+18.0, +21.2]$ \\
\UseMacro{method-pretrained} & $+18.1$ & $[+17.0, +19.2]$ & $[+15.4, +20.9]$ & $+19.4$ & $[+18.2, +20.7]$ & $[+17.9, +21.1]$ \\
\bottomrule
\end{tabular}

\end{center}
\end{small}
\end{table*}

\begin{table*}[t]
\begin{small}
\begin{center}
\caption{Paired significance tests for \codeloraevo{} on the evolution track (Table~\ref{tab:per_commit_results}) and the OOD holdout (Table~\ref{tab:ood_results}), computed as in Table~\ref{tab:significance}. The repository-clustered intervals resample $51$ (CR), $389$ (IR), and $92$ (OOD) repositories. Every comparison has McNemar $p < 10^{-10}$.}\label{tab:significance_evo}

\resizebox{\textwidth}{!}{%
\begin{tabular}{@{}l@{\hskip 8pt}rcc@{\hskip 10pt}rcc@{\hskip 10pt}rcc@{}}
\toprule
& \multicolumn{3}{c}{Evolution \UseMacro{TH-cr}, $n{=}44{,}732$} & \multicolumn{3}{c}{Evolution \UseMacro{TH-ir}, $n{=}42{,}061$} & \multicolumn{3}{c}{OOD, $n{=}14{,}813$} \\
\cmidrule(lr){2-4} \cmidrule(lr){5-7} \cmidrule(lr){8-10}
\codeloraevo{} vs. & $\Delta$EM & 95\% CI & repo-clust.\ CI & $\Delta$EM & 95\% CI & repo-clust.\ CI & $\Delta$EM & 95\% CI & repo-clust.\ CI \\
\midrule
\UseMacro{method-slora} & $+5.2$ & $[+4.9, +5.6]$ & $[+3.6, +6.9]$ & $+3.2$ & $[+2.8, +3.5]$ & $[+2.3, +4.0]$ & $+1.8$ & $[+1.3, +2.4]$ & $[+0.8, +3.0]$ \\
\codelorastatic{}       & $+4.6$ & $[+4.3, +5.0]$ & $[+3.3, +6.3]$ & $+3.9$ & $[+3.6, +4.3]$ & $[+3.2, +4.7]$ & $+1.9$ & $[+1.4, +2.4]$ & $[+1.2, +2.9]$ \\
\bottomrule
\end{tabular}%
}

\end{center}
\end{small}
\end{table*}

\subsection{Locally-Solvable and Repository-Knowledge-Required Instances}
\label{app:local_solvability}

Assertion targets are short (median ${\sim}3$ tokens), which raises the question of how much of \repopeftbench{} is solvable from the local test prefix alone and whether the reported gains concentrate on instances that genuinely need the wider repository.
Taking ``the no-context pretrained backbone solves it'' as the operational definition of locally solvable, \UseMacro{cr-locally-solvable}\% of CR-test and \UseMacro{ir-locally-solvable}\% of IR-test instances are locally solvable; the majority is not.
Table~\ref{tab:local_solvability} splits every method by this partition, using the calibrated context baselines of \S\ref{app:context_calibration}.

The two subsets separate the methods cleanly.
On the repository-knowledge-required subset, context injection recovers only ${\sim}10$\% of instances (\UseMacro{method-rag} \UseMacro{cr-reporeq-rag}\,/\,\UseMacro{ir-reporeq-rag}, \UseMacro{method-drc} \UseMacro{cr-reporeq-drc}\,/\,\UseMacro{ir-reporeq-drc}) while parametric adaptation recovers ${\sim}38$--$42$\% (\codelorastatic{} \UseMacro{cr-reporeq-codelorastatic}\,/\,\UseMacro{ir-reporeq-codelorastatic}).
On the locally-solvable subset \codelorastatic{} and the context baselines all stay near ceiling ($92$--$96$\%), whereas the fine-tuned baselines regress sharply (\UseMacro{method-slora} loses ${\sim}34$\,pp).
\codelora{} therefore recovers repository-knowledge instances without eroding local competence, which is the combination neither fine-tuning nor retrieval achieves.

\paragraph{Target length.}
Table~\ref{tab:target_length} bins the same instances by target length.
Every method degrades as targets get longer, but the gains of \codelorastatic{} are not confined to the shortest targets: its margin over the strongest other method is largest on mid-length targets of $6$--$19$ characters ($+14.4$ and $+15.9$\,pp on CR, $+13.7$ and $+11.7$\,pp on IR).
It leads in every bin except the longest IR bin, where \UseMacro{method-fft} is level ($27.4$ vs.\ $27.2$); on targets of $20$ or more characters the CR margin narrows to $+4.0$\,pp.
The context baselines track the no-context backbone in every bin, gaining at most ${\sim}4$\,pp.

\paragraph{Target type.}
Table~\ref{tab:target_type} bins the same instances by the target-type taxonomy of Table~\ref{tab:dataset_details}.
\codelorastatic{} leads on the types that depend on the repository: on variables, which name objects defined elsewhere in the code, it reaches $77.3$\,/\,$80.9$ EM (CR\,/\,IR) against at most $58.4$\,/\,$60.3$ for any other method, and it also leads on collections and complex expressions on both splits.
Where another method leads, the margin is at most $4.6$\,pp (\UseMacro{method-drc} on CR numeric literals, $74.9$ vs.\ $70.3$).
Numeric literals are largely solvable from the prefix ($71.4$\,/\,$74.9$ for the no-context backbone); there the fine-tuned baselines lose $20$--$30$\,pp while \codelorastatic{} stays within $1.1$\,pp of the backbone, consistent with the locally-solvable split above.

\begin{table*}[t]
\begin{small}
\begin{center}
\caption{Static-track EM (\%) split by whether the no-context backbone solves the instance (\S\ref{app:local_solvability}). \UseMacro{method-pretrained} is $0$ / $100$ by construction. The advantage of parametric adaptation is concentrated on the repository-knowledge-required subset, where context injection recovers only ${\sim}10$\% of instances.}\label{tab:local_solvability}

\begin{tabular}{lcc@{\hskip 14pt}cc}
\toprule
& \multicolumn{2}{c}{Cross-Repo (CR Test)} & \multicolumn{2}{c}{In-Repo (IR Test)} \\
\cmidrule(lr){2-3} \cmidrule(lr){4-5}
Method & repo-required & locally solvable & repo-required & locally solvable \\
& ($n{=}3{,}482$) & ($n{=}2{,}932$) & ($n{=}2{,}777$) & ($n{=}2{,}445$) \\
\midrule
\UseMacro{method-pretrained} & 0.0 & 100.0 & 0.0 & 100.0 \\
\UseMacro{method-rag}        & 10.1 & 92.2 & 9.7 & 91.5 \\
\UseMacro{method-drc}        & 9.6 & 96.0 & 9.5 & 96.1 \\
\UseMacro{method-fft}        & 33.7 & 72.5 & 40.4 & 73.5 \\
\UseMacro{method-slora}      & 32.1 & 65.6 & 36.5 & 66.2 \\
\codelorastatic{}            & \textbf{38.1} & 94.3 & \textbf{41.5} & 94.3 \\
\bottomrule
\end{tabular}

\end{center}
\end{small}
\end{table*}

\begin{table*}[t]
\begin{small}
\begin{center}
\caption{Static-track EM (\%) by target length in characters (\S\ref{app:local_solvability}). \UseMacro{method-rag} and \UseMacro{method-drc} use the calibrated configuration of Appendix~\ref{app:context_calibration}; each split's bins sum to the instance counts of Table~\ref{tab:main_results}, and the per-split averages reproduce its EM column. Best per row in bold.}\label{tab:target_length}

\resizebox{\textwidth}{!}{%
\begin{tabular}{lr@{\hskip 10pt}cccccc}
\toprule
Target length (chars) & $n$ & \UseMacro{method-pretrained} & \UseMacro{method-rag} & \UseMacro{method-drc} & \UseMacro{method-fft} & \UseMacro{method-slora} & \codelorastatic{} \\
\midrule
\multicolumn{8}{l}{\emph{\UseMacro{TH-cr}}} \\
\midrule
$<3$      & 1{,}489 & 61.9 & 63.8 & 65.7 & 58.7 & 52.0 & \textbf{74.8} \\
$3$--$5$  & 1{,}226 & 56.5 & 58.8 & 60.4 & 69.0 & 65.7 & \textbf{78.1} \\
$6$--$9$  & 1{,}076 & 47.6 & 48.9 & 51.1 & 58.2 & 56.1 & \textbf{72.6} \\
$10$--$19$ & 1{,}543 & 39.4 & 41.9 & 43.4 & 47.2 & 42.6 & \textbf{63.1} \\
$\geq 20$ & 1{,}080 & 18.3 & 19.4 & 19.3 & 20.7 & 18.4 & \textbf{24.7} \\
\midrule
\multicolumn{8}{l}{\emph{\UseMacro{TH-ir}}} \\
\midrule
$<3$      & 1{,}217 & 65.7 & 66.8 & 69.7 & 58.5 & 52.8 & \textbf{75.2} \\
$3$--$5$  & 1{,}109 & 54.6 & 56.3 & 58.7 & 72.0 & 67.9 & \textbf{78.8} \\
$6$--$9$  & 898     & 49.1 & 49.1 & 52.2 & 61.6 & 55.7 & \textbf{75.3} \\
$10$--$19$ & 1{,}218 & 38.5 & 39.5 & 41.2 & 52.5 & 47.4 & \textbf{64.2} \\
$\geq 20$ & 780     & 16.5 & 19.1 & 18.3 & \textbf{27.4} & 20.5 & 27.2 \\
\bottomrule
\end{tabular}%
}

\end{center}
\end{small}
\end{table*}

\begin{table*}[t]
\begin{small}
\begin{center}
\caption{Static-track EM (\%) by target type, using the taxonomy of Table~\ref{tab:dataset_details} (\S\ref{app:local_solvability}). Best per row in bold.}\label{tab:target_type}

\resizebox{\textwidth}{!}{%
\begin{tabular}{lr@{\hskip 10pt}cccccc}
\toprule
Target type & $n$ & \UseMacro{method-pretrained} & \UseMacro{method-rag} & \UseMacro{method-drc} & \UseMacro{method-fft} & \UseMacro{method-slora} & \codelorastatic{} \\
\midrule
\multicolumn{8}{l}{\emph{\UseMacro{TH-cr}}} \\
\midrule
Numeric literal & 1{,}274 & 71.4 & 73.6 & \textbf{74.9} & 51.0 & 41.3 & 70.3 \\
String literal & 1{,}167 & 36.8 & 42.1 & 43.9 & 53.3 & 54.2 & \textbf{57.2} \\
Bool literal & 131 & 67.2 & 69.5 & 74.0 & 87.0 & \textbf{90.8} & \textbf{90.8} \\
None literal & 227 & 81.5 & 84.6 & 85.5 & \textbf{97.4} & 95.6 & 96.9 \\
Variable & 1{,}405 & 56.3 & 56.7 & 58.4 & 44.8 & 39.2 & \textbf{77.3} \\
Collection & 657 & 17.5 & 20.9 & 20.7 & 36.4 & 31.8 & \textbf{42.6} \\
Function call & 651 & 20.4 & 17.8 & 20.3 & \textbf{34.1} & 31.8 & 32.0 \\
Complex expression & 902 & 31.3 & 32.3 & 33.5 & 66.6 & 64.4 & \textbf{68.4} \\
\midrule
\multicolumn{8}{l}{\emph{\UseMacro{TH-ir}}} \\
\midrule
Numeric literal & 1{,}012 & 74.9 & 75.3 & \textbf{77.2} & 54.7 & 47.1 & 74.5 \\
String literal & 967 & 37.1 & 39.0 & 43.4 & \textbf{59.5} & 56.4 & 58.1 \\
Bool literal & 110 & 54.5 & 63.6 & 64.5 & \textbf{85.5} & 83.6 & 84.5 \\
None literal & 155 & 81.9 & 82.6 & 84.5 & \textbf{98.7} & 97.4 & 98.1 \\
Variable & 1{,}140 & 58.2 & 59.0 & 60.3 & 50.8 & 40.8 & \textbf{80.9} \\
Collection & 588 & 15.1 & 18.0 & 19.7 & 38.6 & 32.3 & \textbf{39.6} \\
Function call & 466 & 22.1 & 21.2 & 22.7 & 45.5 & 41.0 & \textbf{45.9} \\
Complex expression & 784 & 36.4 & 37.4 & 38.4 & 66.8 & 66.5 & \textbf{67.5} \\
\bottomrule
\end{tabular}%
}

\end{center}
\end{small}
\end{table*}

\subsection{Scaling with Repository Size}
\label{app:repo_size}

\codelora{} conditions on a fixed-size pooled repository embedding, which raises the concern that the representation saturates once a codebase grows past some size.
We test this by binning every test instance by the total token count of its repository's non-test source and recomputing EM per bin (Table~\ref{tab:repo_size}).
The repositories span a wide range: median $165$K tokens, mean \UseMacro{repo-tok-mean}, p95 ${\approx}1.03$M, p99 ${\approx}1.87$M, and max \UseMacro{repo-tok-max} tokens, so the $>$300K bin already sits well past any single context window.

Among the static-track methods, \codelorastatic{} leads every bin on both splits, and its advantage does not shrink with size---on CR test its margin in the largest bin ($82.9$ vs.\ $69.4$ for \UseMacro{method-drc}, the strongest baseline in that bin) is within $0.3$\,pp of its widest margin in any bin.
Its EM does dip in the largest IR bin, but this is not embedding-specific: every method drops in lockstep there, which points to harder completions in those repositories rather than a saturating representation.
Per-repository EM correlates flat-to-slightly-positively with $\log_{10}$(repository size) for all methods, and \codelorastatic{} has the \emph{largest} positive correlation on both splits ($+0.24$ CR, $+0.08$ IR).
The recurrent representation behaves the same way: repeating the ablation for \codeloraevo{} on the evolution track gives $56.5$--$73.3$ (CR) and $62.8$--$65.5$ (IR) across bins with correlations of $+0.105$ and $-0.046$.
We do not claim the pooled representation scales indefinitely---hierarchical or graph-structured encoders are a natural next step for repositories far larger than those tested---but at the scales present in \repopeftbench{} it degrades neither absolutely nor relative to any baseline.

\begin{table*}[t]
\begin{small}
\begin{center}
\caption{EM (\%) by total repository source size (\S\ref{app:repo_size}). Bins are on the token count of all non-test Python files. \codelorastatic{} leads every bin on both splits and its per-repository EM does not degrade with repository size.}\label{tab:repo_size}

\begin{tabular}{@{}l@{\hskip 6pt}r@{\hskip 8pt}c@{\hskip 5pt}c@{\hskip 5pt}c@{\hskip 5pt}c@{\hskip 7pt}c@{\hskip 7pt}c@{}}
\toprule
Repo size & $n$ inst.\,/\,repos & \UseMacro{method-pretrained} & \UseMacro{method-drc} & \UseMacro{method-fft} & \UseMacro{method-slora} & \codelorastatic{} & \codeloraevo{}$^\dag$ \\
\midrule
\multicolumn{8}{l}{\emph{\UseMacro{TH-cr}}} \\
\midrule
$<$25K    &   534 / 10 & 36.7 & 38.8 & 51.3 & 47.0 & \textbf{57.9} & 64.7 \\
25--75K   & 1{,}720 / 16 & 43.5 & 46.7 & 45.8 & 41.2 & \textbf{58.5} & 56.5 \\
75--150K  & 2{,}293 / 15 & 47.4 & 50.3 & 52.0 & 47.7 & \textbf{65.8} & 64.1 \\
150--300K & 1{,}383 /  8 & 44.3 & 46.8 & 52.6 & 49.2 & \textbf{62.8} & 57.4 \\
$>$300K   &   484 /  3 & 59.7 & 69.4 & 65.1 & 63.4 & \textbf{82.9} & 73.3 \\
\multicolumn{2}{l}{Pearson $r$ vs.\ $\log_{10}$(size)} & $+.19$ & $+.23$ & $+.13$ & $+.14$ & $+.24$ & $+.11$ \\
\midrule
\multicolumn{8}{l}{\emph{\UseMacro{TH-ir}}} \\
\midrule
$<$25K    &   524 /  66 & 44.1 & 48.3 & 51.5 & 46.9 & \textbf{61.3} & 63.3 \\
25--75K   & 1{,}551 / 137 & 45.6 & 49.3 & 54.2 & 49.4 & \textbf{63.4} & 64.7 \\
75--150K  & 1{,}356 /  94 & 46.7 & 49.4 & 55.4 & 49.9 & \textbf{66.6} & 64.8 \\
150--300K & 1{,}021 /  66 & 50.0 & 54.1 & 61.4 & 54.5 & \textbf{72.2} & 65.5 \\
$>$300K   &   770 /  46 & 47.1 & 48.6 & 55.8 & 50.4 & \textbf{66.9} & 62.8 \\
\multicolumn{2}{l}{Pearson $r$ vs.\ $\log_{10}$(size)} & $+.04$ & $.00$ & $+.05$ & $+.03$ & $+.08$ & $-.05$ \\
\bottomrule
\multicolumn{8}{l}{\footnotesize $^\dag$ Scored on the evolution track (44{,}732 CR / 42{,}061 IR per-commit QnAs), so its instance counts} \\
\multicolumn{8}{l}{\footnotesize \phantom{$^\dag$} differ from the static-track columns; the size bins are identical.} \\
\end{tabular}

\end{center}
\end{small}
\end{table*}

\subsection{Memorization and Answer Leakage}
\label{app:leakage}

A ${\sim}340$M-parameter hypernetwork trained on hundreds of repositories could in principle memorize training patterns rather than learn a genuine adaptation mechanism.
The evaluation design already controls for this along a gradient of decreasing training exposure: IR repositories are seen in training but the scored completions are held out; the \UseMacro{num-repos-cr-test} CR repositories are absent from training entirely, so the hypernetwork encodes each for the first time at test; and the OOD repositories were created after the scrape cutoff, placing them outside both our training set and the backbone's pretraining corpus.
\codelora{} leads on all three, and the CR result in particular cannot be repository memorization.

To rule out memorization at the level of individual \emph{answers}, we collect the $20{,}861$ unique normalized targets in the static-track training split and label every CR-test instance by whether its answer string ever appears as a training target.
Half of CR test ($50.5$\%) has an answer that was never a training target and is therefore impossible to reproduce by recall.
Table~\ref{tab:leakage} reports EM on each subset together with a memorization probe: among the predictions a method gets \emph{right}, what fraction had a training-seen answer.
\UseMacro{method-pretrained} is the control here---it was never trained on our data, so its $63.7$\% is the overlap expected purely from recurring common values (\texttt{True}, \texttt{0}, \texttt{None}, and similar).
\codelorastatic{} scores $50.5$\% on the never-seen subset, above \UseMacro{method-fft} ($42.2$) and the backbone ($32.8$), so it generalizes to answer strings absent from training; and only $60.0$\% of its correct predictions have a training-seen answer, \emph{below} the never-trained control, whereas a memorizing model would sit above it.
IR test shows the same pattern, though there the seen-answer overlap is ${\approx}90$\% for every method because IR repositories were in training.
This probe bounds verbatim answer memorization; the two analyses below test the broader possibilities that the hypernetwork relies on assertion templates seen in training or on resemblance to training repositories.

\paragraph{Template memorization.}
We abstract each assertion statement (the last line of the prefix followed by the target) into a template by replacing non-keyword identifiers, numbers, and strings with placeholders (e.g., \texttt{assert ID.ID == NUM}), and count how often each template occurs among the $44{,}149$ static-track training statements ($7{,}666$ distinct templates).
Table~\ref{tab:template_probe} reports EM by template frequency.
On CR test \codelorastatic{} leads every bin, including the $20.3$\% of instances whose template never occurs in training ($39.6$ vs.\ $31.7$ for \UseMacro{method-fft}, the strongest other method there, and $26.7$ for the backbone).
Its margin over the strongest other method is largest on templates seen $100$ or more times ($+13.9$\,pp, $46.3$\% of CR test) and smaller on the rarer bins ($+3.2$ to $+7.9$\,pp), so familiarity with training templates accounts for part of the advantage.
On IR test only $5.5$\% of instances have an unseen template; on that subset \codelorastatic{} is level with \UseMacro{method-fft} ($38.5$ vs.\ $38.1$), and it leads every other bin.

\paragraph{Similarity to training repositories.}
For each test repository we compute the cosine similarity of its repository embedding to the nearest training repository and split the repositories into terciles (Table~\ref{tab:repo_similarity}).
On CR test \codelorastatic{} leads every tercile, and its margin over the strongest other method does not shrink on the least similar third ($+9.5$\,pp, vs.\ $+8.3$ and $+10.7$ on the middle and most similar thirds); across the $52$ repositories its gain over \UseMacro{method-fft-rag} is uncorrelated with similarity (Pearson $+0.04$), while its gain over the backbone correlates weakly with it ($+0.25$).
The OOD repositories span a wider similarity range ($0.70$--$0.97$, vs.\ $0.91$--$0.97$ on CR test); there the gain of \codeloraevo{} over \UseMacro{method-slora} is largest on the least similar third ($+2.6$\,pp, vs.\ $+1.8$ and $+0.8$) and uncorrelated with similarity across the $92$ repositories (Pearson $-0.02$).
Neither analysis perturbs the inputs, so together they bound, but do not exclude, reliance on transferable repository-level patterns.

\begin{table*}[t]
\begin{small}
\begin{center}
\caption{Answer-leakage analysis on the held-out CR-test split (\S\ref{app:leakage}). Answers are labeled by whether the normalized target string ever appears in the static-track training set. \codelorastatic{} is strongest on answers it cannot have copied, and its correct predictions are \emph{less} training-overlapping than the never-trained control's.}\label{tab:leakage}

\begin{tabular}{lccc}
\toprule
CR Test ($n{=}6{,}414$) & \specialcell{EM on never-seen\\answers (50.5\% of $n$)} & \specialcell{EM on seen\\answers} & \specialcell{of \emph{correct} preds,\\\% answer seen} \\
\midrule
\UseMacro{method-pretrained}$^\ddag$ & 32.8 & 58.9 & 63.7 \\
\UseMacro{method-fft}                & 42.2 & 60.9 & 58.6 \\
\UseMacro{method-slora}              & 39.3 & 55.8 & 58.2 \\
\codelorastatic{}                    & \textbf{50.5} & \textbf{77.4} & 60.0 \\
\bottomrule
\multicolumn{4}{l}{\footnotesize $^\ddag$ Never trained on \repopeftbench{}; serves as the no-memorization control.} \\
\end{tabular}

\end{center}
\end{small}
\end{table*}

\begin{table*}[t]
\begin{small}
\begin{center}
\caption{Static-track EM (\%) by how often an instance's assertion template (identifiers, numbers, and strings abstracted) occurs among the training statements (\S\ref{app:leakage}). Best per row in bold.}\label{tab:template_probe}

\resizebox{\textwidth}{!}{%
\begin{tabular}{lrr@{\hskip 10pt}ccccccc}
\toprule
Template count in training & $n$ & \% & \UseMacro{method-pretrained} & \UseMacro{method-rag} & \UseMacro{method-drc} & \UseMacro{method-fft} & \UseMacro{method-fft-rag} & \UseMacro{method-slora} & \codelorastatic{} \\
\midrule
\multicolumn{10}{l}{\emph{\UseMacro{TH-cr}}} \\
\midrule
unseen & 1{,}300 & 20.3 & 26.7 & 28.2 & 28.5 & 31.7 & 31.1 & 28.7 & \textbf{39.6} \\
$1$--$9$ & 745 & 11.6 & 36.4 & 38.4 & 37.7 & 45.2 & 48.7 & 41.2 & \textbf{51.9} \\
$10$--$99$ & 1{,}397 & 21.8 & 46.2 & 47.9 & 49.5 & 58.3 & 62.2 & 54.7 & \textbf{67.3} \\
$\geq 100$ & 2{,}972 & 46.3 & 56.2 & 58.2 & 60.7 & 58.4 & 61.8 & 53.8 & \textbf{75.7} \\
\midrule
\multicolumn{10}{l}{\emph{\UseMacro{TH-ir}}} \\
\midrule
unseen & 286 & 5.5 & 27.6 & 25.9 & 25.9 & 38.1 & 32.9 & 31.5 & \textbf{38.5} \\
$1$--$9$ & 1{,}201 & 23.0 & 31.4 & 33.0 & 33.5 & 44.8 & 43.8 & 39.5 & \textbf{52.1} \\
$10$--$99$ & 1{,}148 & 22.0 & 47.4 & 48.8 & 51.3 & 57.2 & 58.9 & 52.4 & \textbf{67.3} \\
$\geq 100$ & 2{,}587 & 49.5 & 55.9 & 57.1 & 59.8 & 62.4 & 64.6 & 56.7 & \textbf{75.4} \\
\bottomrule
\end{tabular}%
}

\end{center}
\end{small}
\end{table*}

\begin{table*}[t]
\begin{small}
\begin{center}
\caption{EM (\%) by similarity of each test repository to its nearest training repository, in terciles of repositories (\S\ref{app:leakage}). Top: CR test, static track, using the snapshot repository embeddings. Bottom: OOD holdout, evolution track, using each repository's state embedding at its earliest commit against the evolution-track training repositories. Best per row in bold.}\label{tab:repo_similarity}

\resizebox{\textwidth}{!}{%
\begin{tabular}{lrrc@{\hskip 10pt}ccccccc}
\toprule
\UseMacro{TH-cr} tercile & repos & $n$ & cosine & \UseMacro{method-pretrained} & \UseMacro{method-rag} & \UseMacro{method-drc} & \UseMacro{method-fft} & \UseMacro{method-fft-rag} & \UseMacro{method-slora} & \codelorastatic{} \\
\midrule
least similar & 17 & 1{,}253 & 0.91--0.94 & 45.5 & 48.4 & 48.6 & 48.3 & 50.0 & 42.6 & \textbf{59.5} \\
middle & 17 & 2{,}217 & 0.94--0.96 & 47.3 & 50.3 & 50.2 & 54.4 & 57.5 & 51.2 & \textbf{65.8} \\
most similar & 18 & 2{,}944 & 0.96--0.97 & 44.6 & 45.2 & 48.4 & 50.5 & 53.4 & 46.7 & \textbf{64.1} \\
\bottomrule
\end{tabular}%
}
\vspace{6pt}

\begin{tabular}{lrrc@{\hskip 10pt}ccc}
\toprule
OOD tercile & repos & $n$ & cosine & \UseMacro{method-slora} & \codelorastatic{} & \codeloraevo{} \\
\midrule
least similar & 30 & 5{,}543 & 0.70--0.90 & 69.5 & 69.6 & \textbf{72.1} \\
middle & 31 & 5{,}293 & 0.90--0.93 & 75.8 & 75.7 & \textbf{77.6} \\
most similar & 31 & 3{,}977 & 0.93--0.97 & 71.4 & 71.0 & \textbf{72.2} \\
\bottomrule
\end{tabular}

\end{center}
\end{small}
\end{table*}

\section{Implementation Details}
\label{app:implementation}

This section documents the dependency-resolved context (DRC) extraction algorithm and the exact hyperparameters used to train every method in Tables~\ref{tab:main_results}--\ref{tab:ood_results}.
All training and evaluation runs use a single H100 80\,GB GPU; total compute is summarized at the end of the section.

\subsection{Dependency-Resolved Context Construction}
\label{app:drc_construction}

DRC takes a test prefix and, via static import analysis, returns the function and class definitions reachable from its imports.
We describe the resolution strategy, the relevance-aware compression that fits results into the adaptive 8K-token budget, and the empirical coverage on \repopeftbench{}.

\paragraph{Import resolution strategy.}
For each import in the test prefix:
\begin{enumerate}[nosep]
\item Parse using AST with fallback regex for syntax errors.
\item Resolve the module to a file path, trying multiple source roots: repository root, \texttt{src/}, \texttt{lib/}, package directories with \texttt{\_\_init\_\_.py}.
\item For relative imports, resolve relative to the test file's location.
\item If the imported name is used in the test prefix, extract its definition (function or class) from the source file via AST.
\end{enumerate}

\paragraph{Coverage.}
DRC context is available for 70.3\% of CR-test pairs, 64.7\% of IR-test pairs, and approximately 64\% of training pairs.
When present, DRC adds a median of 517 tokens (mean 1,900, p95 7,850 tokens).
Pairs with no resolvable imports (e.g., testing third-party libraries or built-in functions only) receive no DRC augmentation and are trained and evaluated on the plain prefix.

\subsection{Detailed Architecture Diagrams}
\label{app:architecture}

Figure~\ref{fig:architecture_static} and Figure~\ref{fig:architecture_evo} expand the overview in Figure~\ref{fig:architecture} with step-by-step training and inference details for each usage scenario.

\begin{figure*}[t]
\begin{center}

\definecolor{encblue}{RGB}{55,119,168}
\definecolor{hypamb}{RGB}{185,125,40}
\definecolor{infgrn}{RGB}{65,125,85}
\definecolor{vecteal}{RGB}{20,115,115}
\definecolor{lossred}{RGB}{165,48,48}
\resizebox{\textwidth}{!}{%
\begin{tikzpicture}[
>=Stealth,
block/.style={rectangle,rounded corners=5pt,line width=0.85pt,
minimum height=1.10cm,align=center,font=\small\bfseries\sffamily,
draw=#1!75,fill=#1!8,text=#1!80!black},
block/.default=black,
filebox/.style={rectangle,rounded corners=3pt,draw=black!38,fill=white,
minimum width=1.70cm,minimum height=0.52cm,
font=\footnotesize\ttfamily,inner sep=3pt,align=center},
vec/.style={rectangle,rounded corners=4pt,draw=vecteal!65,fill=vecteal!9,
minimum height=0.58cm,font=\small\bfseries,inner sep=5pt,align=center,
text=vecteal!80!black},
lorabox/.style={rectangle,rounded corners=4pt,draw=hypamb!85,fill=hypamb!14,
minimum width=1.05cm,minimum height=0.68cm,line width=0.85pt,
font=\small\bfseries,text=hypamb!80!black},
eqbox/.style={rectangle,rounded corners=3pt,draw=hypamb!50,fill=hypamb!7,
inner sep=4pt,font=\footnotesize},
arr/.style={->,line width=0.9pt,color=black!55},
lorarr/.style={->,line width=1.20pt,color=hypamb!80},
outarr/.style={->,line width=1.20pt,color=infgrn!75},
gradarr/.style={->,dashed,line width=1.0pt,color=lossred!70,
dash pattern=on 4pt off 2.5pt},
sub/.style={font=\scriptsize\itshape\sffamily,text=black!45,align=center},
tiny/.style={font=\tiny\itshape\sffamily,text=black!40,align=center},
badge/.style={font=\footnotesize\sffamily\bfseries,rounded corners=3pt,
fill=#1!11,draw=#1!32,inner sep=4pt,text=#1!70!black},
]

\node[filebox](f1)at(0.90, 0.75){model.py};
\node[filebox](f2)at(0.90, 0.00){utils.py};
\node[filebox](f3)at(0.90,-0.75){config.py};
\node[font=\scriptsize,text=black!30](fdots)at(0.90,-1.25){$\vdots$};

\node[block=encblue,minimum width=2.0cm](emb)at(3.00,0.00)
{Frozen\\[-2pt]Embedder};
\node[sub]at(3.00,-0.88){Qwen3-Emb-0.6B};
\node[tiny]at(3.00,-1.20){precomputed offline};

\node[vec,minimum width=0.72cm](e1)at(5.20, 0.55){$\mathbf{f}_1$};
\node[vec,minimum width=0.72cm](e2)at(5.20, 0.00){$\mathbf{f}_2$};
\node[vec,minimum width=0.72cm](e3)at(5.20,-0.55){$\mathbf{f}_K$};
\node[font=\tiny,text=black!28]at(5.20,-0.27){$\vdots$};

\node[block=vecteal,minimum width=1.80cm,minimum height=1.0cm,
font=\footnotesize\bfseries\sffamily](agg)at(7.10,0.00)
{Weighted\\[-2pt]Mean+Max};
\node[sub]at(7.10,-0.78){importance-weighted};

\node[vec,minimum width=2.10cm,minimum height=0.76cm,
font=\small\bfseries](erep)at(9.30,0.00){$\mathbf{e}_{\mathrm{repo}}$};
\node[sub]at(9.30,-0.68){$\in\mathbb{R}^{2048}$};

\node[block=hypamb,minimum width=1.90cm](trunk)at(12.60,0.00)
{MLP\\[-2pt]Trunk};
\node[sub]at(12.60,-0.88){2-layer GELU, $H{=}512$};
\node[tiny]at(12.60,-1.20){L2-norm ${\cdot}\sqrt{H}$};

\node[vec,minimum width=0.72cm,draw=hypamb!70,fill=hypamb!11,
text=hypamb!80!black](hid)at(14.45,0.00){$\mathbf{h}$};

\node[block=hypamb,minimum width=1.50cm,minimum height=0.85cm,
font=\footnotesize\bfseries\sffamily](headA)at(16.20, 0.78)
{$\mathrm{Head}^A_t$};
\node[block=hypamb,minimum width=1.50cm,minimum height=0.85cm,
font=\footnotesize\bfseries\sffamily](headB)at(16.20,-0.78)
{$\mathrm{Head}^B_t$};
\node[sub]at(16.20,-1.54){$t\!\in\!\{\texttt{q,k,v,o,up,gate,dn}\}$};
\node[tiny]at(16.20,-1.86){$\tanh\!\cdot\!\exp(s_t)$};

\node[lorabox](loraA)at(18.20, 0.78){$\mathbf{A}_t$};
\node[lorabox](loraB)at(18.20,-0.78){$\mathbf{B}_t$};

\draw[decorate,decoration={brace,amplitude=4pt,raise=3pt},
hypamb!60!black,line width=0.6pt]
([xshift=2pt]loraA.north east)--([xshift=2pt]loraB.south east)
node[midway,right=8pt,font=\scriptsize\bfseries\sffamily,
text=hypamb!70!black,align=left]
{$\times7$ types,\\[-2pt]shared $\times28$ layers};

\foreach \f in {f1,f2,f3}{\draw[arr](\f.east)--(emb.west|-\f.east);}
\draw[arr](emb.east)--++(0.22,0)|-(e1.west);
\draw[arr](emb.east)--(e2.west);
\draw[arr](emb.east)--++(0.22,0)|-(e3.west);
\foreach \e in {e1,e2,e3}{\draw[arr](\e.east)--(agg.west|-\e.east);}
\draw[arr](agg.east)--(erep.west);
\draw[arr](erep.east)--(trunk.west);
\draw[arr](trunk.east)--(hid.west);
\draw[arr](hid.east)--++(0.20,0)|-(headA.west);
\draw[arr](hid.east)--++(0.20,0)|-(headB.west);
\draw[arr,hypamb!65](headA.east)--(loraA.west);
\draw[arr,hypamb!65](headB.east)--(loraB.west);

\node[filebox,minimum width=2.70cm,fill=infgrn!5,draw=infgrn!60,
font=\footnotesize\ttfamily](inp)at(10.00,-4.20)
{assert res\;==\;\colorbox{yellow!45}{\textbf{?}}};
\node[sub]at(10.00,-4.80){test prefix};

\node[block=infgrn,minimum width=3.20cm,minimum height=1.90cm]
(llm)at(13.80,-4.20){};
\node[font=\small\bfseries\sffamily,text=infgrn!75!black]
at(13.80,-3.68){Frozen LLM};
\node[sub]at(13.80,-4.00){Qwen2.5-Coder-1.5B};
\node[eqbox]at(13.80,-4.48)
{$\mathbf{W}'\!=\!\mathbf{W}\!+\!\tfrac{\alpha}{r}\mathbf{B}_t\mathbf{A}_t$};

\node[filebox,minimum width=2.40cm,fill=infgrn!12,draw=infgrn!72,
font=\footnotesize\ttfamily\bfseries](out)at(17.80,-4.20){expected\_val};
\node[sub]at(17.80,-4.80){predicted target};

\draw[arr](inp.east)--(llm.west);
\draw[outarr](llm.east)--(out.west);

\draw[lorarr]
(loraA.south)
-- (18.20, 0.44)
-- (21.50, 0.44)
-- (21.50,-2.10)
-- (13.95,-2.10)
-- (13.95,-3.25);

\draw[lorarr]
(loraB.south)
-- (18.20,-1.12)
-- (21.50,-1.12)
-- (21.50,-2.55)
-- (13.65,-2.55)
-- (13.65,-3.25);

\draw[gradarr]
(out.east)
-- (22.20,-4.20)
-- (22.20, 2.50)
-- (12.60, 2.50)
-- (trunk.north);

\node[font=\small\bfseries\sffamily,text=lossred!72,
anchor=south,inner sep=3pt]at(17.40,2.50)
{$\nabla_{\!\theta}\,\mathcal{L}_{\mathrm{LM}}$};

\begin{scope}[on background layer]
\node[fit=(f1)(f3)(fdots)(emb)(e1)(e3)(agg)(erep),
rounded corners=9pt,fill=encblue!4,draw=encblue!20,line width=0.65pt,
inner xsep=10pt,inner ysep=12pt](bg1){};
\node[fit=(trunk)(hid)(headA)(headB)(loraA)(loraB),
rounded corners=9pt,fill=hypamb!4,draw=hypamb!20,line width=0.65pt,
inner xsep=10pt,inner ysep=14pt](bg2){};
\node[fit=(inp)(llm)(out),
rounded corners=9pt,fill=infgrn!4,draw=infgrn!20,line width=0.65pt,
inner xsep=12pt,inner ysep=12pt](bg3){};
\end{scope}

\node[badge=encblue]at([yshift=6pt]bg1.north)
{\textbf{1}\enspace Repository Encoding \emph{(offline)}};
\node[badge=hypamb]at([yshift=6pt]bg2.north)
{\textbf{2}\enspace Hypernetwork (Code2LoRAHead)};
\node[badge=infgrn]at([yshift=6pt]bg3.north)
{\textbf{3}\enspace Adapted Inference};

\end{tikzpicture}%
}%

\caption{Detailed \codelorastatic{} architecture.
\textbf{(1)}~Repository-level context is encoded by a frozen embedding model (Qwen3-Embedding-0.6B) and aggregated into a 2048-dim repository embedding $\mathbf{e}_{\text{repo}}$; the result is stored in the dataset and consumed verbatim at training time---gradients never flow back through the embedder.
\textbf{(2)}~A shared MLP trunk (2-layer GELU, hidden $H{=}512$) maps $\mathbf{e}_{\text{repo}}$ to a hidden representation $\mathbf{h}$ (L2-normalized, rescaled by $\sqrt{H}$); separate $\text{Head}^A_m$, $\text{Head}^B_m$ heads emit $\mathbf{A}_m, \mathbf{B}_m$ for each of the 7 projection types via $\tanh\cdot\exp(s_m)$ scaling with a clamped learnable log-scale $s_m$. The same $(\mathbf{A}_m, \mathbf{B}_m)$ pair is shared across all 28 transformer layers.
\textbf{(3)}~Generated LoRA weights are injected into the frozen LLM via $\mathbf{W}' = \mathbf{W} + \tfrac{\alpha}{r}\mathbf{B}_m\mathbf{A}_m$.
Only the hypernetwork parameters $\theta$ are trained via the language-modeling loss (dashed red); the LLM and embedder stay frozen.}
\label{fig:architecture_static}
\end{center}
\end{figure*}

\begin{figure*}[t]
\begin{center}

\definecolor{encblue}{RGB}{70,130,180}     %
\definecolor{infgreen}{RGB}{95,145,100}    %
\definecolor{embedteal}{RGB}{0,128,128}    %
\definecolor{initviolet}{RGB}{130,90,160}  %
\definecolor{grurust}{RGB}{175,100,70}     %
\definecolor{hypamber}{RGB}{205,145,60}    %
\definecolor{lossred}{RGB}{180,60,60}      %
\resizebox{\textwidth}{!}{%
\begin{tikzpicture}[
>=Stealth,
component/.style={
rectangle, rounded corners=5pt, line width=0.9pt,
minimum height=1.1cm, align=center, font=\small\bfseries,
draw=#1!80, fill=#1!8, text=black!85
},
component/.default=black,
smallcomp/.style={
rectangle, rounded corners=4pt, line width=0.7pt,
minimum height=0.8cm, align=center, font=\footnotesize\bfseries,
draw=#1!75, fill=#1!10, text=black!80
},
smallcomp/.default=black,
vecbox/.style={
rectangle, rounded corners=3pt, draw=#1!70, fill=#1!12,
minimum height=0.52cm, font=\footnotesize, inner sep=3pt, align=center
},
vecbox/.default=embedteal,
loramat/.style={
rectangle, rounded corners=4pt, draw=hypamber!90, fill=hypamber!16,
minimum width=1.0cm, minimum height=0.65cm, line width=0.9pt,
font=\small\bfseries, text=hypamber!75!black
},
diffbox/.style={
rectangle, rounded corners=2pt, draw=black!50, fill=encblue!6,
minimum width=1.5cm, minimum height=0.45cm, font=\footnotesize\ttfamily,
inner sep=2pt, align=center
},
arr/.style={->, line width=1pt, color=black!60},
thinarr/.style={->, line width=0.6pt, color=black!45},
fatarr/.style={->, line width=1.4pt, color=#1!75},
fatarr/.default=black,
gradarr/.style={->, dashed, line width=1pt, color=lossred!70},
stglabel/.style={
font=\footnotesize\sffamily\bfseries, rounded corners=2pt,
fill=#1!14, draw=#1!40, inner sep=3pt, text=#1!70!black
},
sublabel/.style={font=\scriptsize\itshape, text=black!50},
tinylabel/.style={font=\tiny, text=black!45},
]

\node[diffbox, fill=initviolet!8, draw=initviolet!45] (rs0) at (0, 2.40)
{\texttt{repo}\,$_{t=0}$};
\node[tinylabel, text=initviolet!65, anchor=south] at (0, 2.70)
{repo state @ commit 0};

\node[vecbox=initviolet, minimum width=1.05cm,
draw=initviolet!75, fill=initviolet!12,
font=\footnotesize\bfseries] (erep) at (2.0, 2.40)
{$\mathbf{e}_{\text{repo}}^{(0)}$};
\node[tinylabel, text=initviolet!55] at (2.0, 1.90) {$\in\mathbb{R}^{2048}$};

\node[component=initviolet, minimum width=2.0cm, minimum height=0.95cm]
(init) at (4.2, 2.40) {Repo-State\\[-1pt]Initializer};
\node[sublabel] at (4.2, 1.72)
{Linear\,$\to$\,GELU\,$\to$\,LayerNorm};

\node[vecbox=initviolet, minimum width=1.0cm,
font=\small\bfseries] (h0) at (6.5, 2.40) {$\mathbf{h}_0$};
\node[tinylabel, text=initviolet!55, anchor=west]
at ([xshift=-1.5pt]h0.east) {$\in\mathbb{R}^{H}$};

\draw[arr, initviolet!55] (rs0.east) -- (erep.west);
\draw[arr, initviolet!55] (erep.east) -- (init.west);
\draw[fatarr=initviolet] (init.east) -- (h0.west);

\node[diffbox] (d1) at (8.5, 3.20) {\texttt{diff}\,$\Delta_1$};
\node[diffbox] (d2) at (8.5, 2.55) {\texttt{diff}\,$\Delta_2$};
\node[font=\tiny, text=black!30] at (8.5, 2.10) {$\vdots$};
\node[diffbox] (dT) at (8.5, 1.60) {\texttt{diff}\,$\Delta_T$};
\node[tinylabel, anchor=south] at (8.2, 3.55) {per-commit diffs};

\node[component=encblue, minimum width=2.0cm] (enc) at (11.1, 2.4)
{Frozen\\[-1pt]Embedder};
\node[sublabel] at (11.1, 1.55) {Qwen3-Emb-0.6B};
\node[font=\tiny\itshape, text=encblue!55] at (11.1, 1.20)
{precomputed offline};

\node[vecbox=embedteal, minimum width=0.85cm] (e1) at (13.4, 3.20) {$\mathbf{e}_1$};
\node[vecbox=embedteal, minimum width=0.85cm] (e2) at (13.4, 2.55) {$\mathbf{e}_2$};
\node[font=\tiny, text=black!30] at (13.4, 2.10) {$\vdots$};
\node[vecbox=embedteal, minimum width=0.85cm] (eT) at (13.4, 1.60) {$\mathbf{e}_T$};
\node[tinylabel, anchor=west] at ([xshift=2pt]eT.east) {$\in\mathbb{R}^{2048}$};

\foreach \d in {d1,d2,dT} { \draw[thinarr] (\d.east) -- (enc.west |- \d.east); }
\draw[arr] (enc.east) -- ++(0.15,0) |- (e1.west);
\draw[arr] (enc.east) -- ++(0.15,0) |- (e2.west);
\draw[arr] (enc.east) -- ++(0.15,0) |- (eT.west);

\node[component=grurust, minimum width=1.45cm, minimum height=1.25cm]
(gru1) at (2.5, -0.40) {GRU};
\node[component=grurust, minimum width=1.45cm, minimum height=1.25cm]
(gru2) at (5.5, -0.40) {GRU};
\node[font=\large, text=grurust!55] (grudots) at (8.0, -0.40) {$\cdots$};
\node[component=grurust, minimum width=1.45cm, minimum height=1.25cm]
(gruT) at (10.5, -0.40) {GRU};

\node[vecbox=grurust, minimum width=0.7cm, font=\footnotesize\bfseries]
(h1) at (4.0, -0.40) {$\mathbf{h}_1$};
\node[vecbox=grurust, minimum width=0.7cm, font=\footnotesize\bfseries]
(h2) at (6.95, -0.40) {$\mathbf{h}_2$};
\node[vecbox=grurust, minimum width=0.95cm, font=\small\bfseries,
draw=grurust!90, fill=grurust!18, line width=1pt]
(hT) at (12.2, -0.40) {$\mathbf{h}_T$};
\node[tinylabel, anchor=west] at ([xshift=3pt]hT.east) {$\in\mathbb{R}^{2048}$};

\draw[arr, grurust!65] (gru1.east) -- (h1.west);
\draw[arr, grurust!65] (h1.east) -- (gru2.west);
\draw[arr, grurust!65] (gru2.east) -- (h2.west);
\draw[arr, grurust!65] (h2.east) -- (grudots.west);
\draw[arr, grurust!65] (grudots.east) -- (gruT.west);
\draw[fatarr=grurust] (gruT.east) -- (hT.west);

\coordinate (echY) at (0, 0.8);
\draw[arr, encblue!60]
(e1.east) -- (e1.east |- echY)
-- (gru1.north |- echY)
-- (gru1.north);
\draw[arr, encblue!60]
(e2.east) -- (e2.east |- echY)
-- (gru2.north |- echY)
-- (gru2.north);
\draw[arr, encblue!60]
(eT.east) -- (eT.east |- echY)
-- (gruT.north |- echY)
-- (gruT.north);

\coordinate (h0chY) at (0, 1.3);
\draw[fatarr=initviolet, line width=1.4pt]
(h0.south) -- (h0.south |- h0chY)
-- ([xshift=-0.55cm]gru1.west |- h0chY)
-- ([xshift=-0.55cm]gru1.west)
-- (gru1.west);

\node[sublabel, text=grurust!55, fill=white, inner sep=0.5pt] at (8.0, 0.45)
{each step: input projection (Linear\,$+$\,LayerNorm) $\to$ GRU recurrence};

\draw[decorate, decoration={brace, mirror, amplitude=4pt, raise=4pt},
grurust!55, line width=0.6pt]
(gru1.south west) -- (gruT.south east)
node[midway, below=5pt, font=\scriptsize\sffamily, text=grurust!55]
{truncated BPTT (detach every $K{=}16$ steps)};

\node[smallcomp=hypamber, minimum width=1.4cm]
(lnT) at (0.5, -3.7) {LayerNorm};
\node[sublabel] at (0.5, -4.35) {$\mathbf{ctx} = \mathrm{LN}(\mathbf{h}_T)$};

\node[component=hypamber, minimum width=1.8cm, minimum height=1.05cm]
(trunk) at (2.95, -3.7) {MLP\\[-1pt]Trunk};
\node[sublabel] at (2.95, -4.40) {2-layer GELU, $H{=}1024$};
\node[font=\tiny\itshape, text=hypamber!60] at (2.95, -4.70)
{L2-norm\,$\cdot\,\sqrt{H}$};

\node[smallcomp=hypamber, minimum width=1.5cm, minimum height=0.85cm]
(headA) at (5.35, -3.15) {$\text{Head}^A_t$};
\node[smallcomp=hypamber, minimum width=1.5cm, minimum height=0.85cm]
(headB) at (5.35, -4.25) {$\text{Head}^B_t$};
\node[sublabel] at (5.35, -4.85) {$t\!\in\!\{\texttt{q,k,v,o,up,gate,dn}\}$};
\node[font=\tiny\itshape, text=hypamber!60] at (5.35, -5.15)
{$\tanh\cdot \exp(s_t)$};

\node[loramat] (loraA) at (7.40, -3.15) {$\mathbf{A}_t$};
\node[loramat] (loraB) at (7.40, -4.25) {$\mathbf{B}_t$};

\node[font=\scriptsize\bfseries, text=hypamber!70!black,
align=center] (sharedlbl) at (7.40, -5.50)
{$\times 7$ types, shared $\times 28$ layers};

\draw[arr, hypamber!65] (lnT.east) -- (trunk.west);
\draw[arr, hypamber!65] (trunk.east) -- ++(0.20,0) |- (headA.west);
\draw[arr, hypamber!65] (trunk.east) -- ++(0.20,0) |- (headB.west);
\draw[arr, hypamber!65] (headA.east) -- (loraA.west);
\draw[arr, hypamber!65] (headB.east) -- (loraB.west);

\node[component=infgreen, minimum width=2.95cm, minimum height=2.2cm]
(llm) at (11.7, -3.7) {};
\node[font=\small\bfseries, text=infgreen!70!black] at (11.7, -3.0)
{Frozen LLM};
\node[sublabel] at (11.7, -3.32) {Qwen2.5-Coder-1.5B};
\node[rectangle, rounded corners=2pt, draw=hypamber!55, fill=hypamber!8,
inner sep=3pt, font=\footnotesize] (inj) at (11.7, -3.92)
{$\mathbf{W}'\!=\!\mathbf{W}\!+\!\tfrac{\alpha}{r}\mathbf{B}_t\mathbf{A}_t$};

\node[diffbox, minimum width=2.6cm, fill=infgreen!6, draw=infgreen!75,
font=\footnotesize\ttfamily] (inp) at (11.7, -5.70)
{assert res ==\,\colorbox{yellow!35}{\textbf{?}}};
\node[sublabel] at (11.7, -6.15) {test prefix};

\node[component=lossred, minimum width=1.5cm, minimum height=0.9cm,
font=\small\bfseries] (loss) at (14.55, -3.7) {$\mathcal{L}_{\text{CE}}$};
\node[sublabel, text=lossred!60] at (14.55, -4.30) {target tokens};

\draw[fatarr=hypamber, line width=1.3pt]
(loraA.east) -- ++(0.55,0) |- ([yshift=4pt]llm.west);
\draw[fatarr=hypamber, line width=1.3pt]
(loraB.east) -- ++(0.55,0) |- ([yshift=-4pt]llm.west);
\draw[arr] (inp.north) -- ([yshift=-3pt]llm.south);
\draw[arr, infgreen!70!black, line width=1.2pt] (llm.east) -- (loss.west);

\coordinate (rowgap)  at (hT.south |- 0,-2.10);
\coordinate (rowgapL) at (lnT.north |- 0,-2.10);
\draw[fatarr=grurust, line width=1.4pt]
(hT.south) -- (rowgap) -- (rowgapL) -- (lnT.north);
\node[font=\scriptsize\itshape, text=grurust!65, anchor=west, inner sep=1pt]
at ([xshift=3pt, yshift=-0.18cm]rowgapL)
{context $\mathbf{h}_T$};

\begin{scope}[on background layer]
\node[fit=(d1)(dT)(enc)(e1)(eT), rounded corners=8pt,
fill=encblue!3, draw=encblue!14,
inner sep=8pt, inner ysep=8pt] (bg1) {};
\node[fit=(rs0)(erep)(init)(h0), rounded corners=8pt,
fill=initviolet!4, draw=initviolet!16,
inner sep=8pt] (bg2) {};
\node[fit=(gru1)(gruT)(hT), rounded corners=8pt,
fill=grurust!3, draw=grurust!14,
inner sep=10pt, inner ysep=14pt] (bg3) {};
\node[fit=(lnT)(trunk)(headA)(headB)(loraA)(loraB)(sharedlbl),
rounded corners=8pt, fill=hypamber!3, draw=hypamber!14,
inner sep=10pt, inner ysep=12pt] (bg4) {};
\node[fit=(llm)(inp)(loss), rounded corners=8pt,
fill=infgreen!3, draw=infgreen!14,
inner sep=10pt] (bg5) {};

\draw[gradarr]
(loss.south) -- ++(0, -2.55)
-| ([yshift=-0.08cm]bg4.south);
\end{scope}

\node[font=\scriptsize, text=lossred!70, anchor=north west, inner sep=2pt,
align=left, text width=12cm] at (0.5, -6.78)
{$\nabla_\phi \mathcal{L}_{\text{CE}}$:\,
update $\phi = \{$repo-state init,\, GRU,\, Code2LoRAHead$\}$;\,
LLM and embedder remain frozen};

\node[stglabel=encblue]    at ([yshift=4pt]bg1.north)
{\textsf{\textbf{1}\;\;Offline Embedding}};
\node[stglabel=initviolet] at ([yshift=4pt]bg2.north)
{\textsf{\textbf{2}\;\;Repo-State Init}};
\node[stglabel=grurust]    at ([yshift=4pt]bg3.north)
{\textsf{\textbf{3}\;\;Repository GRU}};
\node[stglabel=hypamber]   at ([yshift=4pt]bg4.north)
{\textsf{\textbf{4}\;\;LoRA Head (shared w/ Fig.~\ref{fig:architecture_static})}};
\node[stglabel=infgreen]   at ([yshift=4pt]bg5.north)
{\textsf{\textbf{5}\;\;Adapted LLM}};

\node[rectangle, rounded corners=3pt, draw=black!25, fill=white,
inner sep=4pt, font=\scriptsize, anchor=north east] at (16.3, -6.45) {%
\begin{tabular}{@{}rl@{}}
\textcolor{encblue!75}{\rule{6pt}{6pt}}\,\textcolor{infgreen!75}{\rule{6pt}{6pt}} & frozen \\[1pt]
\textcolor{initviolet!75}{\rule{6pt}{6pt}}\,\textcolor{grurust!75}{\rule{6pt}{6pt}}\,\textcolor{hypamber!75}{\rule{6pt}{6pt}} & trainable ($\phi$) \\[1pt]
\textcolor{lossred!70}{-\,-\,-} & gradient flow \\
\end{tabular}
};

\end{tikzpicture}%
}%

\caption{Detailed \codeloraevo{} architecture and training procedure.
\textbf{(1)}~Per-commit production-code diffs $\Delta_t$ and the initial repository snapshot are encoded by the shared frozen embedder into 2048-dim vectors $\{\mathbf{e}_t\}_{t=1}^T$ and $\mathbf{e}_{\text{repo}}^{(0)}$; the resulting embeddings are stored in the dataset.
\textbf{(2)}~A small repo-state initializer (Linear\,$\to$\,GELU\,$\to$\,LayerNorm) maps the static snapshot $\mathbf{e}_{\text{repo}}^{(0)}$ to the initial hidden state $\mathbf{h}_0\!\in\!\mathbb{R}^{2048}$.
\textbf{(3)}~A 1-layer GRU walks the chronological diff sequence; each step projects $\mathbf{e}_t$ with a Linear\,+\,LayerNorm and applies the GRU recurrence to produce $\mathbf{h}_t$. Truncated BPTT detaches the hidden state every $K{=}16$ steps.
\textbf{(4)}~The final state $\mathbf{h}_T$ is fed (after LayerNorm) into \codeloraevo{}'s LoRA-generation projection head (analogous in design to \codelorastatic{}'s; Figure~\ref{fig:architecture_static}): a 2-layer GELU trunk with L2-norm rescaling, plus per-module-type $\text{Head}^A_m / \text{Head}^B_m$ output heads with $\tanh\cdot\exp(s_m)$ scaling. The resulting $(\mathbf{A}_m, \mathbf{B}_m)$ are shared across all 28 transformer layers per type.
\textbf{(5)}~Generated LoRAs are injected into the frozen LLM ($\mathbf{W}' = \mathbf{W} + \tfrac{\alpha}{r}\mathbf{B}_m\mathbf{A}_m$); training minimizes the cross-entropy loss on the assertion target. Gradients (dashed red) flow through the projection head, GRU, and repo-state initializer; the LLM and embedder stay frozen.}
\label{fig:architecture_evo}
\end{center}
\end{figure*}

\subsection{Training Details}
\label{app:training_details}

Table~\ref{tab:training_details} lists the optimizer, schedule, sequence length, batch size, and adapter configuration for every trained baseline (FFT, sLoRA, per-repo LoRA) and for \codelorastatic{} on the static track.
All methods share the same backbone (Qwen2.5-Coder-1.5B, bf16), the same optimizer (AdamW, cosine schedule, weight decay 0.01), and roughly the same effective compute budget; the methods differ in LR, sequence length, and (for adapter methods) LoRA rank, dropout, and module coverage.
\codelorastatic{} is trained with an 8K maximum sequence length; \codeloraevo{} truncates BPTT every 16 commits and uses a 4K sequence length per step (\S\ref{app:hyperparams_full}).

\begin{table*}[t]
\begin{small}
\begin{center}
\caption{Training hyperparameters. }
\label{tab:training_details}

\begin{tabular}{lcccc}
\toprule
& FFT & sLoRA & pLoRA & \codelorastatic{} \\
\midrule
LR & 2e-5 & 5e-5 & 2e-4 & 1e-4 \\
Epochs & 3 & 5 & 3 & 3 \\
Max seq len & 2,048 & 2,048 & 2,048 & 8,192 \\
Batch size & 4 & 4 & 4 & 1 \\
Grad accum & 8 & 8 & 4 & 8 \\
Effective batch & 32 & 32 & 16 & 8 \\
LoRA rank & --- & 16 & 16 & 16 \\
LoRA alpha & --- & 32 & 32 & 32 \\
LoRA dropout & --- & 0.1 & 0.05 & --- \\
Warmup ratio & 0.05 & 0.10 & 0.10 & 0.03 \\
Precision & bf16 & bf16 & bf16 & bf16 \\
Optimizer & AdamW & AdamW & AdamW & AdamW \\
LR schedule & cosine & cosine & cosine & cosine \\
\bottomrule
\end{tabular}

\end{center}
\end{small}
\end{table*}

\subsection{Compute Resources}
\label{app:compute}

All experiments were conducted on a single NVIDIA H100 80\,GB GPU per job.
Total GPU hours: FFT variants $\sim$6\,h, sLoRA variants $\sim$10\,h, \codelorastatic{} $\sim$17\,h, per-repo LoRA ($\sim$0.1\,h per repo $\times$ 409 repos) $\sim$41\,h, and evaluation jobs $\sim$30\,h.
\codeloraevo{} training requires an additional $\sim$24\,h on the commit-derived dataset.

\subsection{Hypernetwork Training Hyperparameters}
\label{app:hyperparams_full}

The \codelorastatic{} variant uses input dim $2{,}048$ (mean$+$max repository embedding), trunk hidden $H{=}512$, LoRA rank $r{=}16$, $\alpha{=}32$, and all seven attention/MLP projection types shared across all $28$ transformer layers.
\codeloraevo{} uses a 1-layer GRU with hidden size $2{,}048$ and a small \emph{repo-state initializer} (Linear\,$\to$\,GELU\,$\to$\,LayerNorm) that maps the initial 2048-dim repository embedding to $\mathbf{h}_0$; the LayerNorm-ed final state $\mathbf{h}_T$ feeds into a LoRA-generation head of the same design as \codelorastatic{}'s with a wider trunk (hidden $H{=}1024$).
The \codelorastatic{} reference reported on the evolution track (Tables~\ref{tab:per_commit_results} and~\ref{tab:ood_results}) uses the same $H{=}1024$ head, so it differs from \codeloraevo{} only by the recurrence.
Truncated BPTT detaches the hidden state every $K{=}16$ commits.
Both variants are trained for $3$ epochs with AdamW (cosine schedule, weight decay $0.01$): \codelorastatic{} at LR $1{\times}10^{-4}$ and max sequence length $8{,}192$; \codeloraevo{} at LR $5{\times}10^{-5}$ and max sequence length $4{,}096$.
Best checkpoint is selected by CR-val loss.

\section{OOD Evaluation Caveats}
\label{app:ood_caveats_top}
\label{app:ood_caveats}

Two confounds in Table~\ref{tab:ood_results} are worth surfacing.
\emph{(i) Prefix shape.}
Table~\ref{tab:ood_results} uses commit-derived prefixes (median ${\sim}$7.9\,KB), identical to Table~\ref{tab:per_commit_results} and \emph{not} the short static prefixes (${\sim}$0.9\,KB) of Table~\ref{tab:main_results}; OOD-vs-Table~\ref{tab:per_commit_results} deltas are therefore unconfounded by prefix shape, and only the underlying repositories differ.
\emph{(ii) Target length.}
OOD assertion targets are systematically shorter (median 7 chars) than CR/IR-test (12--13 chars), inflating exact-match credit on every OOD row uniformly; sLoRA's OOD EM (72.3\%) substantially exceeds its in-distribution EM (55.1/61.3\%) for this reason.
The within-table \codeloraevo{} vs.\ sLoRA gap on OOD is $+1.8$\,pp---narrower than the in-distribution gap ($+5.2/+3.2$\,pp, Table~\ref{tab:per_commit_results}) but positive and significant (Table~\ref{tab:significance_evo}), so \codeloraevo{} remains the best method on every split under matched inputs.
We interpret the narrower OOD margin as evidence that part of the streaming advantage is recovered from within-distribution edit patterns seen at training: the OOD repositories were created strictly after the scrape cutoff, so their early-life commit trajectories were never observed.

\section{Broader Analysis}
\label{app:analysis}

This section complements the main-paper analysis with the supporting figures and tables:
per-repository variance and data-sparsity scatter (\S\ref{app:per_repo_variance}),
the repository-count scaling curve (\S\ref{app:repo_scaling}),
the per-commit-position trend (\S\ref{app:commit_pct_trend}),
structural analysis of the generated LoRAs (\S\ref{app:lora_weight}),
the LiveCodeBench-style error taxonomy and qualitative examples (\S\ref{app:errors}, \S\ref{app:qualitative}),
DRC coverage broken out by availability (\S\ref{app:oracle_coverage}),
and the efficiency comparison (\S\ref{app:efficiency}).

\subsection{Per-Repository Performance and Data Sparsity}
\label{app:per_repo_variance}

\begin{figure}[t]
\begin{center}
\includegraphics[width=\columnwidth]{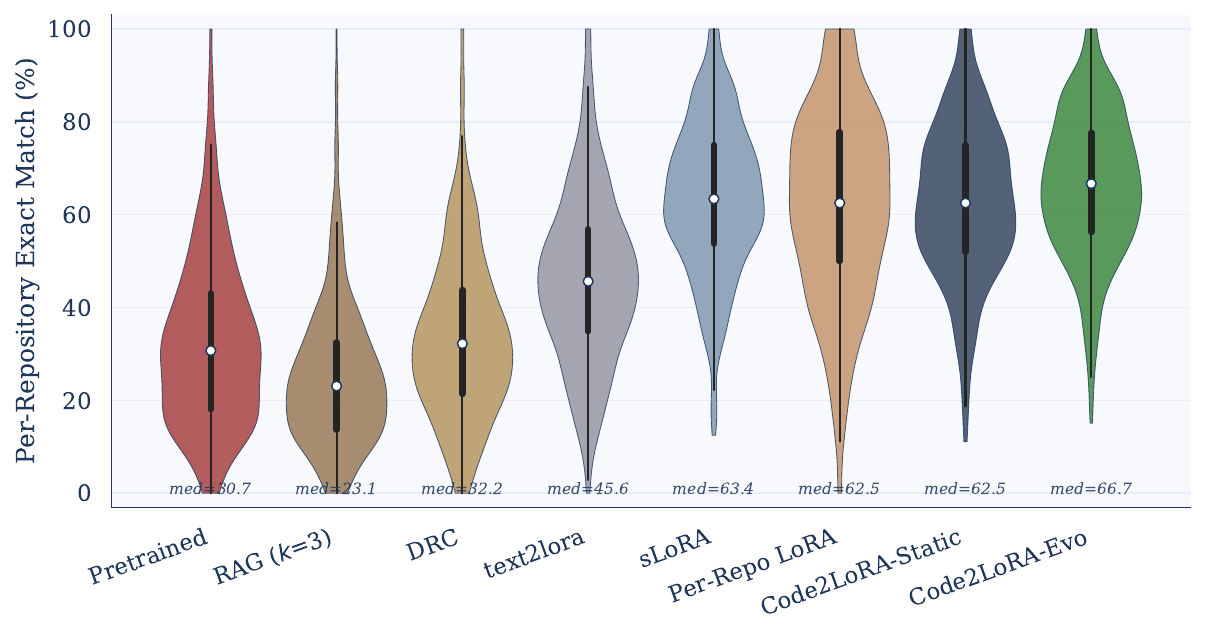}
\caption{Per-repository EM distribution on the IR-test split of \repopeftbench{} (Table~\ref{tab:per_commit_results} checkpoints; $n{=}389$ repositories common to all methods). Each violin shows the full distribution of per-repository EM for one method; the inner box reports the IQR and the white dot marks the median. \codelorastatic{} (median $62.5\%$, $\sigma{=}16.8$) and \codeloraevo{} (median $66.7\%$, $\sigma{=}15.8$) achieve consistently high performance with substantially lower variance than per-repo LoRA (median $62.5\%$, $\sigma{=}20.9$); per-repo LoRA falls below the pretrained baseline on $10.5\%$ of repositories versus only $1.3\%$ and $1.8\%$ for \codelorastatic{} and \codeloraevo{}, demonstrating the regularizing effect of cross-repository knowledge transfer.}
\label{fig:per_repo_boxplot}
\end{center}
\end{figure}

\begin{figure}[t]
\begin{center}
\includegraphics[width=\columnwidth]{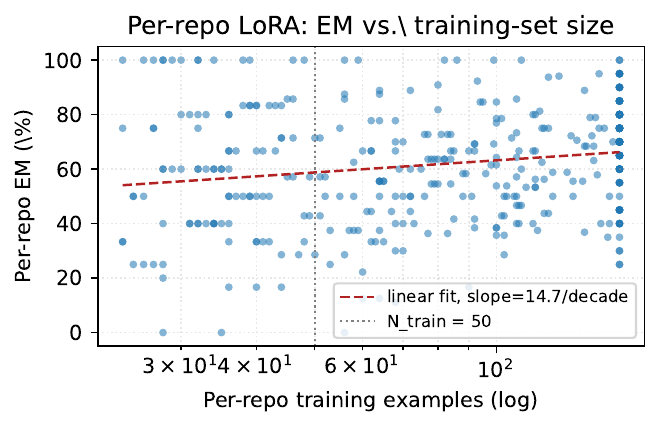}
\caption{Per-repo LoRA EM vs.\ training-set size on IR test. Repositories with fewer than 50 training pairs frequently underperform the IR-test pretrained baseline (46.8\%), while \codelorastatic{} maintains stable performance regardless of per-repo data availability.}
\label{fig:plora_data_sparsity}
\end{center}
\end{figure}

The aggregate IR-test EM in Table~\ref{tab:per_commit_results} hides substantial repository-to-repository variance for per-repo LoRA.
Across the $389$ repositories evaluated by every method, per-repo LoRA EM spans the full $[0, 100]\%$ range with a median of $62.5\%$ and a standard deviation of $20.9$; on $10.5\%$ of repositories ($41 / 389$) per-repo LoRA scores \emph{below} the pretrained baseline (per-repo median $30.7\%$).
The dominant driver is training-data availability: per-repo LoRA overfits to small in-repo datasets and frequently regresses below the unadapted backbone whenever the in-repo training pool is thin.
\codelorastatic{} sidesteps this failure mode through cross-repository knowledge transfer: the hypernetwork learns shared patterns from 409 repositories (39{,}612 examples) and regularizes the generated adapters, yielding the tighter per-repository EM distribution shown in Figure~\ref{fig:per_repo_boxplot} ($\sigma{=}16.8$ for \codelorastatic{} and $15.8$ for \codeloraevo{} vs.\ $20.9$ for per-repo LoRA; only $1.3\%$ and $1.8\%$ of repositories fall below pretrained, respectively) and the flatter EM-vs-data-size profile shown in Figure~\ref{fig:plora_data_sparsity}.

\subsection{Repository-Count Scaling}
\label{app:repo_scaling}

To understand whether the hypernetwork benefits from \emph{breadth} (more distinct repositories) or merely \emph{depth} (more pairs), we sweep the number of training repositories at $\{10, 25, 50, 100, 150, 200, 409, 500, 623\}$ while keeping the per-repo data budget and training schedule fixed. The settings beyond $409$ add $214$ repositories that are not part of \repopeftbench{}: they come from the same GitHub search as the in-distribution set (MIT-licensed, ${\geq}300$ stars) but were excluded from the benchmark because each yields fewer than $30$ assertion QnAs ($13$ on median); none of them overlaps the CR validation, CR test, or OOD repositories, and the $623$ setting uses all of them.

\begin{figure}[t]
\begin{center}
\includegraphics[width=\columnwidth]{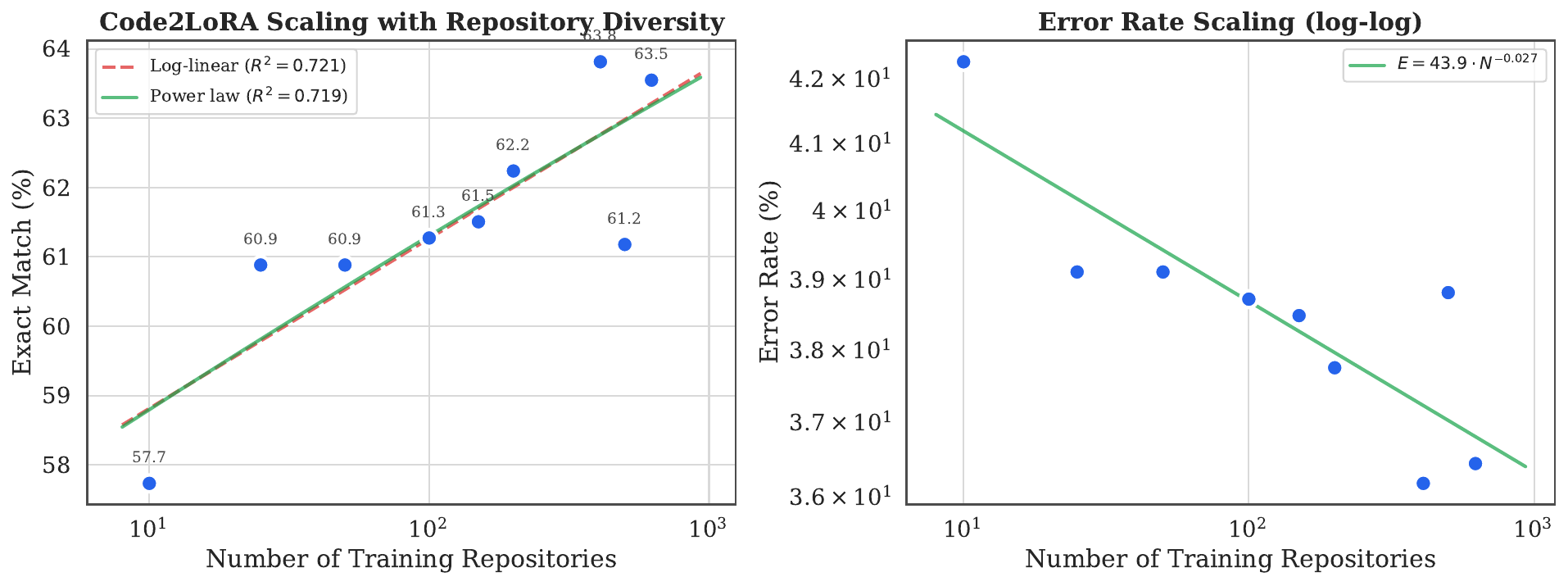}
\caption{CR-test EM as a function of training repository count. \codelorastatic{} benefits from repository diversity, with performance improving log-linearly.}
\label{fig:scaling_repos}
\end{center}
\end{figure}

Two findings emerge.
First, with only 10 repositories ($\sim$2\% of the full training set), \codelorastatic{} already reaches $57.7\%$ CR-test EM---above FFT trained on the full data ($51.4\%$, Table~\ref{tab:main_results}).
Second, CR-test EM scales log-linearly with repository count up to $\sim$200 repositories and is essentially flat between 200 and 623, suggesting that breadth saturates around a few hundred distinct codebases at the current backbone size.
Figure~\ref{fig:scaling_repos} plots the curve; Table~\ref{tab:ablation_data} reports the underlying numbers.

\begin{table}[t]
\begin{small}
\begin{center}
\caption{Effect of training-repository count on CR-test EM.}\label{tab:ablation_data}

\begin{tabular}{lcc}
\toprule
Training Repos & \% of Full & CR Test EM (\%) \\
\midrule
10 & 2\% & 57.7 \\
25 & 4\% & 60.9 \\
50 & 8\% & 60.9 \\
100 & 16\% & 61.3 \\
150 & 24\% & 61.5 \\
200 & 32\% & 62.2 \\
409 & 66\% & 63.8 \\
500 & 80\% & 61.2 \\
623 & 100\% & 63.5 \\
\bottomrule
\end{tabular}

\end{center}
\end{small}
\end{table}

\subsection{Per-Commit Position Trend}
\label{app:commit_pct_trend}

\begin{figure}[t]
\begin{center}
\includegraphics[width=\columnwidth]{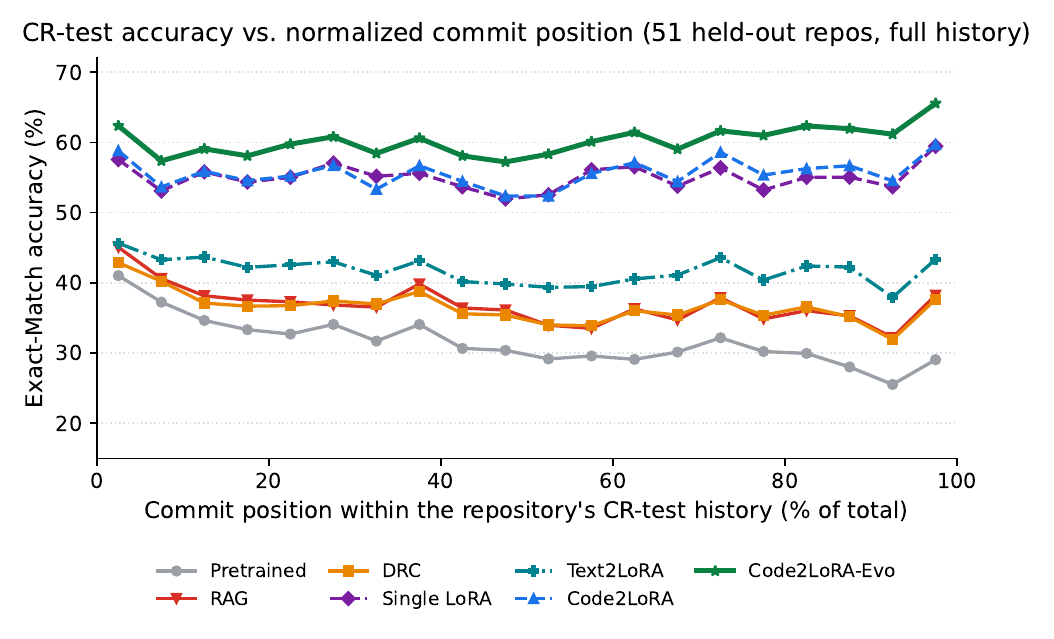}
\caption{CR-test exact-match vs.\ normalized commit position (51 held-out repositories, commit-derived prefixes). Each repository's timeline is scaled to 0--100\%; points are qna-weighted means per 5\% bin.}
\label{fig:commit_pct_trend}
\end{center}
\end{figure}

To verify that \codeloraevo{}'s evolution-track advantage is not driven by a few late-history commits, we plot CR-test EM as a function of each commit's normalized position within its repository's chronological history.
For every repository the timeline is rescaled to $[0\%, 100\%]$ (so $0\%$ is the first scored commit and $100\%$ the last), QnAs are bucketed into 5\%-wide bins, and each bin's score is the QnA-weighted mean EM across that bin.
This collapses short and long repository histories onto a single axis and visualizes the entire lifecycle of every repository rather than only its first commits.
Figure~\ref{fig:commit_pct_trend} shows that \codeloraevo{}'s lead persists across the entire history: it is the strongest method in every 5\% bin, $3.0$--$6.6$\,pp above the next-best method.
Over the timeline (linear fit from $0\%$ to $100\%$), the no-context backbone loses $9.9$\,pp, both context baselines $4.7$--$5.9$\,pp, and \textlora{} $2.8$\,pp, the snapshot adapters stay roughly flat (\UseMacro{method-slora} $+0.3$\,pp, \codelorastatic{} $+1.5$\,pp), and \codeloraevo{} gains the most ($+3.9$\,pp), so its margin widens toward the end of each history.

\subsection{Structure of the Generated LoRAs}
\label{app:lora_weight}

\begin{figure}[t]
\begin{center}
\includegraphics[width=\columnwidth]{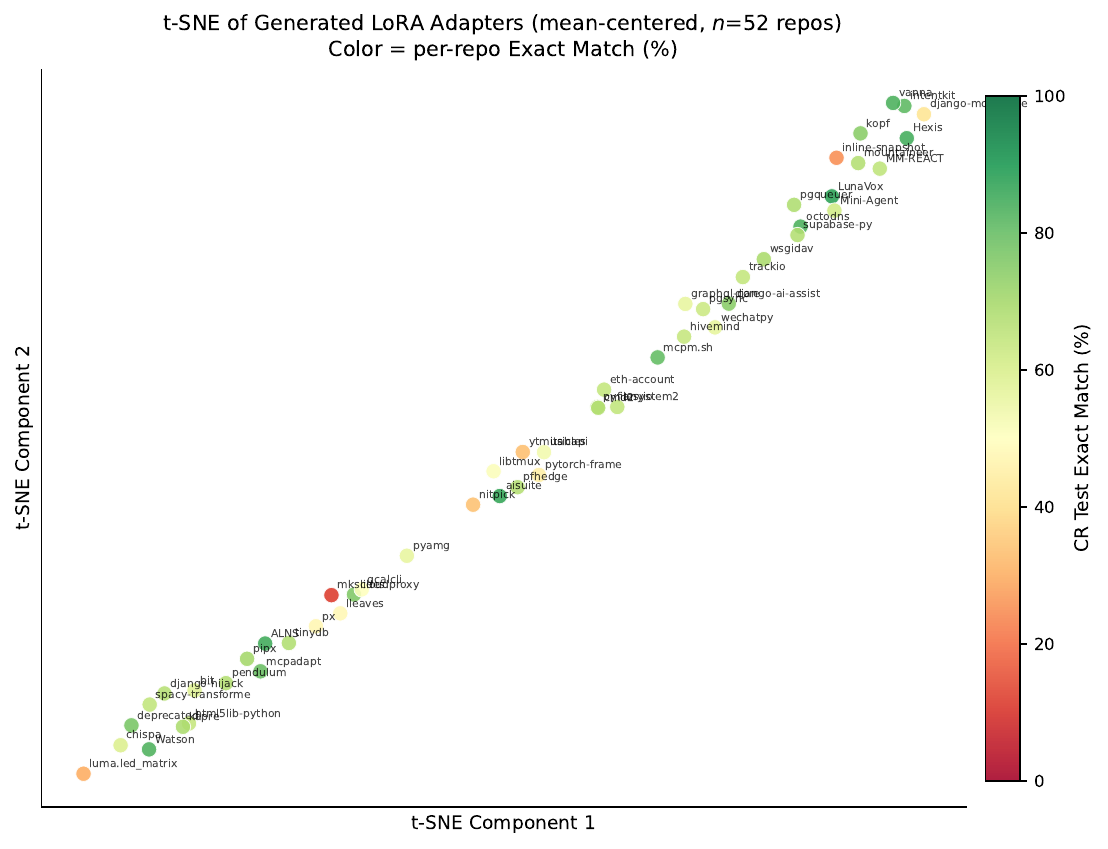}
\caption{t-SNE of generated LoRA adapters for 52 CR-test repositories (PCA pre-reduction to 50 dims, then t-SNE). Color indicates per-repo Exact Match (\%). Repositories with similar codebases tend to cluster together, and clusters show coherent EM ranges, demonstrating that the hypernetwork learns a smooth, semantically meaningful adapter manifold.}
\label{fig:lora_tsne}
\end{center}
\end{figure}

\begin{figure}[t]
\begin{center}
\includegraphics[width=\columnwidth]{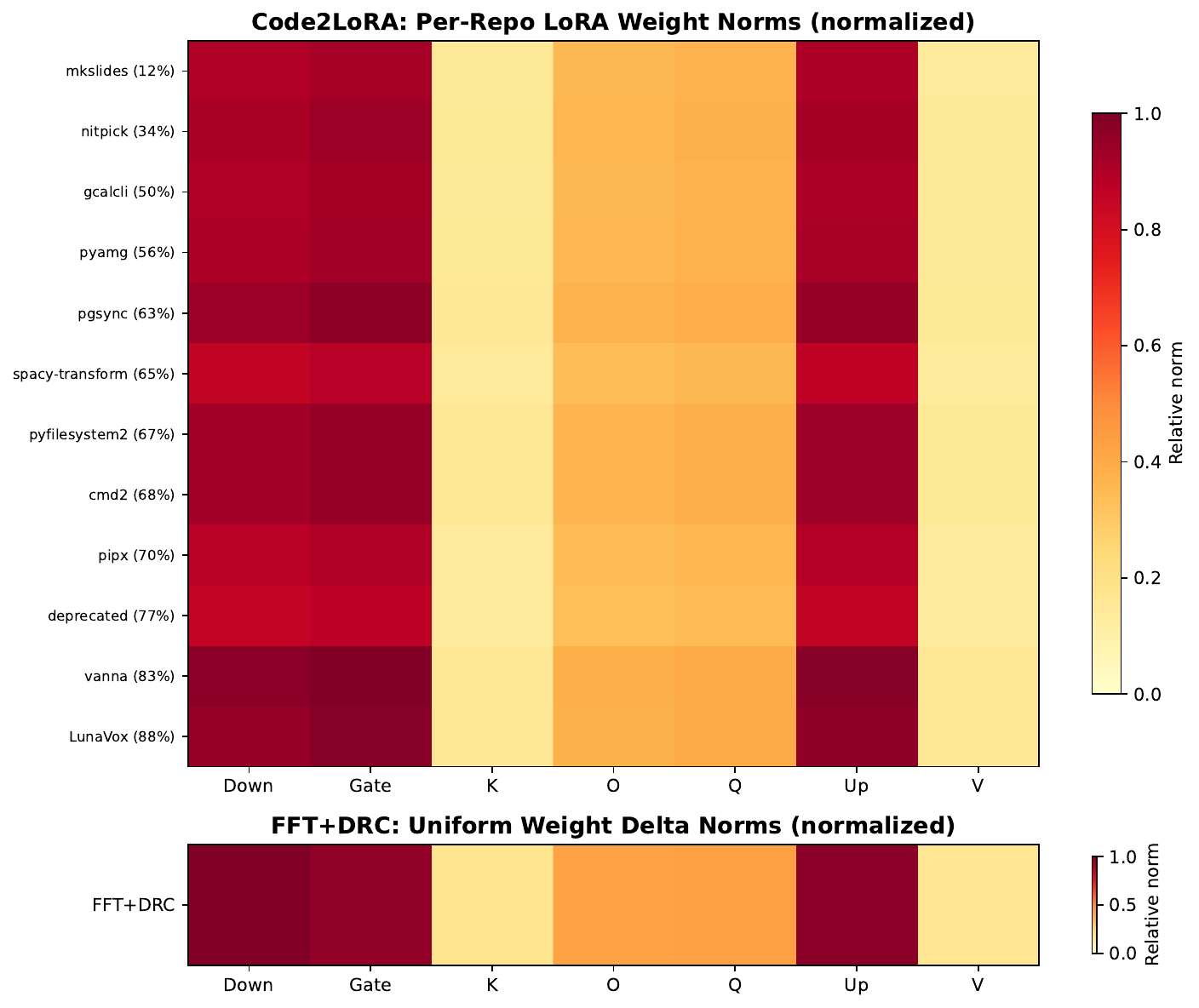}
\caption{Comparison of per-module weight norms. Top: \codelorastatic{} generates repo-specific LoRA adapters with varying weight distributions across module types. Bottom: full fine-tuning (an FFT checkpoint trained with dependency-resolved context) applies a near-uniform weight delta. \codelorastatic{}'s structured, repo-specific adaptations explain its stronger cross-repo performance.}
\label{fig:lora_comparison}
\end{center}
\end{figure}

A natural question is whether the hypernetwork emits genuinely repository-specific adapters or whether it converges to a single mean adapter that happens to behave well on average.
We probe this from two angles.
\emph{Diversity of adapters}: pairwise cosine similarities between the 52 mean-centered CR-test LoRAs (659K-dim flattened) span the full $[-1, +1]$ range with mean $0.01$ and standard deviation $0.94$, so the adapters are not a collapsed mean.
\emph{Semantic structure}: a t-SNE projection of those adapters (Figure~\ref{fig:lora_tsne}) shows that repositories with similar codebases cluster together and that clusters carry coherent EM ranges, indicating that the hypernetwork's adapter manifold is smooth and semantically organized rather than arbitrary.
\emph{Per-module concentration}: a comparison of per-module weight norms (Figure~\ref{fig:lora_comparison}) reveals that \codelorastatic{} concentrates updates on a repository-specific subset of modules (typically \texttt{gate} and \texttt{up} projections), whereas full fine-tuning (here an FFT checkpoint trained with dependency-resolved context) applies a near-uniform delta across all modules---a qualitative difference that helps explain \codelorastatic{}'s stronger cross-repo transfer.

\subsection{Error Analysis}
\label{app:errors}

\begin{figure}[t]
\begin{center}
\includegraphics[width=\columnwidth]{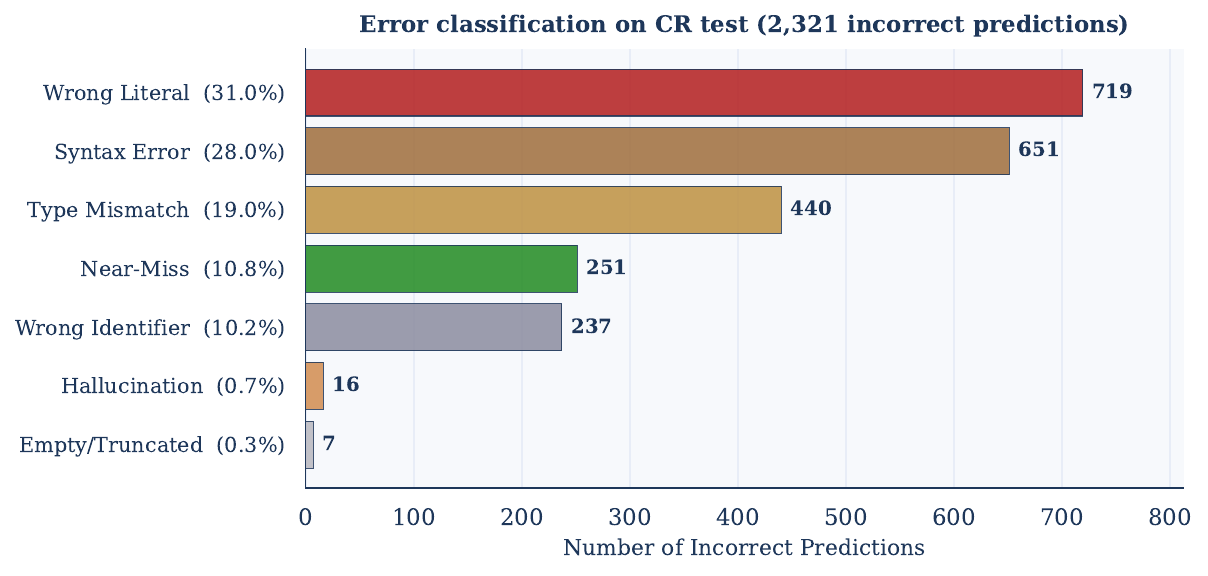}
\caption{Error classification of \codelorastatic{} failures on CR test (2{,}321 incorrect predictions), following a LiveCodeBench-inspired taxonomy. Wrong literal (31.0\%), syntax error (28.0\%), type mismatch (19.0\%), near-miss (10.8\%), wrong identifier (10.2\%); hallucinations and empty outputs are $<1\%$ each.}
\label{fig:error_breakdown}
\end{center}
\end{figure}

We classify all 2{,}321 incorrect CR-test predictions of \codelorastatic{} using a LiveCodeBench-inspired taxonomy~\cite{jain2025livecodebench}.
The breakdown in Figure~\ref{fig:error_breakdown} shows that no single failure mode dominates: \emph{wrong literal} ($31.0\%$) and \emph{syntax error} ($28.0\%$) together account for $\sim$60\% of errors, with the remainder split among \emph{type mismatch} ($19.0\%$), \emph{near-miss} ($10.8\%$), and \emph{wrong identifier} ($10.2\%$); hallucinations and empty outputs are each under $1\%$.
The wrong-literal class is dominated by numeric tests where the correct value depends on runtime state (e.g., expression-valued assertions); the near-miss class corresponds to syntactically valid completions that differ from the reference only in trailing punctuation or single tokens.

\subsection{Qualitative Examples}
\label{app:qualitative}

\definecolor{codegreen}{RGB}{34,139,34}
\definecolor{codered}{RGB}{180,30,30}
\definecolor{codebg}{RGB}{249,249,249}
\definecolor{prefixbg}{RGB}{243,246,252}

We complement the aggregate numbers with qualitative views of CR-test predictions. Figure~\ref{fig:qualitative} pairs two representative successes from \textit{inline-snapshot} and \textit{ALNS} where \codelorastatic{} recovers repo-specific identifiers and conventions that pretrained Qwen2.5-Coder and full fine-tuning miss. We then zoom in on a representative case with an expanded layout that shows the metadata header, full test prefix, retrieved repository context, and side-by-side per-method predictions: Figure~\ref{fig:qualitative_kopf_detailed} illustrates the \emph{context-quality bottleneck} case, where retrieval surfaces the relevant class definition but only the parametric methods complete the value-level reasoning step.

\definecolor{qexHead}{RGB}{30,55,90}
\definecolor{qexHeadBg}{RGB}{232,237,247}
\definecolor{qexCodeBg}{RGB}{247,249,252}
\definecolor{qexRetrBg}{RGB}{250,247,238}
\definecolor{qexPredBg}{RGB}{248,248,248}

\newtcolorbox{qexsection}[1]{
enhanced, breakable,
colback=qexHeadBg, colframe=qexHead!50, coltitle=qexHead,
boxrule=0.4pt, arc=2pt,
fonttitle=\bfseries\small, title={#1},
attach boxed title to top left={xshift=4pt, yshift=-2pt},
boxed title style={colback=qexHeadBg, colframe=qexHead!50, boxrule=0.3pt, arc=1pt},
top=8pt, bottom=4pt, left=4pt, right=4pt,
before skip=4pt, after skip=2pt,
}

\begin{figure*}[t]
\begin{center}
\begin{minipage}[t]{0.48\textwidth}
\noindent\textbf{(a)}\;\textit{inline-snapshot} (CR)\hfill\textcolor{codegreen}{\ding{51} Success}
\begin{lstlisting}[
language=Python,
basicstyle=\ttfamily\small,
backgroundcolor=\color{prefixbg},
frame=single, framerule=0.3pt, rulecolor=\color{black!15},
xleftmargin=4pt, xrightmargin=4pt,
aboveskip=3pt, belowskip=2pt,
showstringspaces=false, columns=fullflexible
]
s2 = s.run(reported_flag)
assert s2.source == (*@\colorbox{yellow!40}{\textbf{???}}@*)
\end{lstlisting}
\vspace{1pt}
\begin{tabular}{@{}l@{\;\;}l@{\;\;}c@{}}
{\scriptsize\textsc{Reference}} & \texttt{s2.source} & \\
{\scriptsize\textsc{\codelorastatic{}}} & \texttt{s2.source} & \textcolor{codegreen}{\ding{51}} \\
{\scriptsize\textsc{FFT}} & \texttt{s.source} & \textcolor{codered}{\ding{55}} \\
{\scriptsize\textsc{Pretrained}} & \texttt{s.source} & \textcolor{codered}{\ding{55}} \\
\end{tabular}\\[2pt]
{\scriptsize\itshape\color{black!50} \codelorastatic{} captures the \texttt{s2} naming pattern; baselines default to \texttt{s}.}
\end{minipage}\hfill
\begin{minipage}[t]{0.48\textwidth}
\noindent\textbf{(b)}\;\textit{ALNS} (CR)\hfill\textcolor{codegreen}{\ding{51} Success}
\begin{lstlisting}[
language=Python,
basicstyle=\ttfamily\small,
backgroundcolor=\color{prefixbg},
frame=single, framerule=0.3pt, rulecolor=\color{black!15},
xleftmargin=4pt, xrightmargin=4pt,
aboveskip=3pt, belowskip=2pt,
showstringspaces=false, columns=fullflexible
]
select.update(Zero(), 0, 0, 1)
assert_almost_equal(
select.destroy_weights[0], (*@\colorbox{yellow!40}{\textbf{???}}@*)
\end{lstlisting}
\vspace{1pt}
\begin{tabular}{@{}l@{\;\;}l@{\;\;}c@{}}
{\scriptsize\textsc{Reference}} & \texttt{expected[0])} & \\
{\scriptsize\textsc{\codelorastatic{}}} & \texttt{expected[0])} & \textcolor{codegreen}{\ding{51}} \\
{\scriptsize\textsc{FFT}} & \texttt{expected)} & \textcolor{codered}{\ding{55}} \\
{\scriptsize\textsc{Pretrained}} & \texttt{1)} & \textcolor{codered}{\ding{55}} \\
\end{tabular}\\[2pt]
{\scriptsize\itshape\color{black!50} The repo uses \texttt{expected[i]} arrays for ground truth.}
\end{minipage}

\caption{Qualitative examples from CR test. Each panel shows a test prefix with the completion target (\colorbox{yellow!30}{\textbf{???}}), ground-truth reference, and model predictions.
\textbf{(a)--(b)}: \codelorastatic{} correctly infers repo-specific identifiers and conventions that pretrained Qwen2.5-Coder and full fine-tuning miss.}
\label{fig:qualitative}
\end{center}
\end{figure*}

Beyond the two short panels of Figure~\ref{fig:qualitative}, we feature one additional commit-derived CR-test case in full detail below.
The case is drawn from the supplementary file \texttt{positive\_analysis.md} (10 cases total, 5 per category) and demonstrates the complementary \emph{context-quality bottleneck} phenomenon: RAG@3 / DRC retrieval surfaces the exact class definition that determines the assertion's outcome, yet pretrained, RAG, DRC, and sLoRA all fail to translate that prepended evidence into the correct prediction; only the hypernetwork variants complete the value-level reasoning step from the retrieved evidence.

\begin{figure*}[!htbp]
\begin{center}
\begin{tcolorbox}[
enhanced,
colback=white, colframe=qexHead, coltitle=white,
fonttitle=\bfseries,
title={Detailed qualitative example: \texttt{nolar/kopf} (Code2LoRA-exclusive, CR test, commit-derived)},
arc=3pt, boxrule=0.6pt,
top=3pt, bottom=3pt, left=6pt, right=6pt,
]

\begin{qexsection}{QnA metadata}
\footnotesize
\begin{tabular}{@{}l@{\;\;}l@{\hskip 14pt}l@{\;\;}l@{}}
\textsc{Repository}      & \texttt{nolar/kopf}                         & \textsc{Commit SHA}        & \texttt{d848601b0df0\ldots} \\
\textsc{Commit position} & $19.2\%$ (55 / 287)                         & \textsc{Python files}      & 131 \\
\textsc{Repo size}       & $\sim$423K chars / $\sim$120K tok           & \textsc{Assertion family}  & \texttt{pytest.raises} (exception class) \\
\textsc{Test location}   & \multicolumn{3}{l}{\texttt{tests/basic-structs/test\_resource.py:7:9}} \\
\end{tabular}
\end{qexsection}

\begin{qexsection}{Test prefix (model input)}
\begin{lstlisting}[
language=Python,
basicstyle=\ttfamily\footnotesize,
backgroundcolor=\color{qexCodeBg},
frame=none, xleftmargin=2pt, xrightmargin=2pt,
aboveskip=1pt, belowskip=1pt,
showstringspaces=false, columns=fullflexible,
]
import pytest
from kopf.structs.resources import Resource

...

def test_no_args():
with pytest.raises( (*@\colorbox{yellow!40}{\textbf{???}}@*)
\end{lstlisting}
\end{qexsection}

\begin{qexsection}{Retrieved repository context}
\textbf{\footnotesize DRC (import-resolved, 4K-token budget):}
\begin{lstlisting}[
language=Python,
basicstyle=\ttfamily\footnotesize,
backgroundcolor=\color{qexRetrBg},
frame=none, xleftmargin=2pt, xrightmargin=2pt,
aboveskip=1pt, belowskip=1pt,
showstringspaces=false, columns=fullflexible,
]
# kopf/structs/resources.py
class Resource(NamedTuple):
group: str
version: str
plural: str
@property
def name(self):
return f'{self.plural}.{self.group}'
\end{lstlisting}
\textbf{\footnotesize RAG@3 (top-3 retrieved 512-token chunks):} surfaces the identical \texttt{Resource} \texttt{NamedTuple} definition plus two unrelated chunks (truncated; full text in supplementary).
\end{qexsection}

\begin{qexsection}{Per-method predictions and exact-match outcome}
\footnotesize
\setlength{\tabcolsep}{6pt}
\begin{tabular}{@{}llc@{}}
\toprule
Method & Prediction & EM \\
\midrule
\textsc{Reference}                            & \texttt{TypeError)}    & \\
\midrule
Pretrained (Qwen2.5-Coder-1.5B)               & \texttt{ValueError)}   & \textcolor{codered}{\ding{55}} \\
RAG ($k\!=\!3$)                               & \texttt{ValueError)}   & \textcolor{codered}{\ding{55}} \\
Dependency-Resolved Context                   & \texttt{ValueError)}   & \textcolor{codered}{\ding{55}} \\
Single LoRA (sLoRA)                           & \texttt{ValueError)}   & \textcolor{codered}{\ding{55}} \\
\midrule
\codelorastatic{}                 & \texttt{TypeError)}    & \textcolor{codegreen}{\ding{51}} \\
\codeloraevo{}                                 & \texttt{TypeError)}    & \textcolor{codegreen}{\ding{51}} \\
\bottomrule
\end{tabular}
\end{qexsection}

\vspace{1pt}
{\footnotesize\itshape\color{black!55}
The retrieved context surfaces the exact \texttt{Resource} \texttt{NamedTuple} with three required fields, so the evidence to deduce that \texttt{Resource()} with zero arguments raises \texttt{TypeError} is in the prompt. Yet pretrained, RAG, DRC, and sLoRA all default to \texttt{ValueError}---the more common \texttt{pytest.raises} idiom---showing that input-side methods do not reliably execute the type-level reasoning hop even when the relevant evidence has been retrieved. Both \codelora{} variants predict \texttt{TypeError} because the repository's \texttt{NamedTuple}-vs-class conventions were distilled into the LoRA-generation step.
}

\end{tcolorbox}
\caption{Qualitative example of the QnA from the CR test set}
\label{fig:qualitative_kopf_detailed}
\end{center}
\end{figure*}

We further feature four detailed qualitative examples drawn from the commit-derived IR-test set (GRU dataset variant; the source HTML report \texttt{report\_gru\_ir\_test\_qnas.html} samples 300 QnAs across 18 methods).
Each figure shows the full test prefix, the actual DRC and RAG@3 contexts that were injected at evaluation time (trimmed to the most relevant signatures and class initializers; non-essential method bodies are elided with ``\texttt{...}''), and the per-method predictions for the five methods that the report tracks for Table~\ref{tab:per_commit_results}: \codelorastatic{}, \codeloraevo{}, RAG, DRC, and \textlora{}.
Figures~\ref{fig:qualitative_ir_fla_46} and~\ref{fig:qualitative_ir_pynguin_68} are easy cases where the local prefix already exposes the completion pattern and retrieval merely corroborates it.
Figure~\ref{fig:qualitative_ir_beartype_195} is a \emph{retrieval-precision} case: only DRC retrieves the discriminating \texttt{->\;bool} signature, RAG misses it and collapses onto the n-gram-likely \texttt{is 1}; the parametric \codelora{} variants succeed without context.
Figure~\ref{fig:qualitative_ir_apscheduler_245} is a \emph{retrieval-degeneracy} case: DRC retrieves the literal answer \texttt{JobOutcome.abandoned} in a docstring and RAG retrieves the \texttt{JobOutcome} enum class plus the dotted access pattern (one inference hop away from the answer), yet both methods collapse onto a FIM-token artifact at the very first generated token; only the methods that bake the repository signal into the parameters complete the assertion.

\begin{figure*}[!htbp]
\begin{center}
\begin{tcolorbox}[
enhanced,
colback=white, colframe=qexHead, coltitle=white,
fonttitle=\bfseries,
title={Detailed qualitative example: \texttt{fla-org/flash-linear-attention} (IR test, commit-derived)},
arc=3pt, boxrule=0.6pt,
top=3pt, bottom=3pt, left=6pt, right=6pt,
]

\begin{qexsection}{QnA metadata}
\footnotesize
\begin{tabular}{@{}l@{\;\;}l@{\hskip 14pt}l@{\;\;}l@{}}
\textsc{Repository}      & \texttt{fla-org/flash-linear-attention}     & \textsc{Commit SHA}        & \texttt{d62e316ea88b\ldots} \\
\textsc{Commit position} & 277 / 409 (training-window)                 & \textsc{In-repo split}     & \texttt{train} \\
\textsc{Assertion family} & \multicolumn{3}{l}{\texttt{assert\_close(...)} -- repository utility comparing two tensors with a tolerance ratio} \\
\textsc{Test location}   & \multicolumn{3}{l}{\texttt{tests/ops/test\_kda.py::test\_naive\_chunk}, line 73} \\
\end{tabular}
\end{qexsection}

\begin{qexsection}{Test prefix (model input, trimmed)}
\begin{lstlisting}[
language=Python,
basicstyle=\ttfamily\footnotesize,
backgroundcolor=\color{qexCodeBg},
frame=none, xleftmargin=2pt, xrightmargin=2pt,
aboveskip=1pt, belowskip=1pt,
showstringspaces=false, columns=fullflexible,
]
from fla.ops.kda.naive import naive_chunk_kda, naive_recurrent_kda
from fla.utils import IS_INTEL_ALCHEMIST, assert_close, device
...
def test_naive_chunk(B, T, H, D, scale, gate_logit_normalizer, dtype):
...
ref, ref_ht = naive_recurrent_kda(q=..., k=..., v=v.clone(),
g=g.clone(), beta=beta.clone(),
scale=scale, initial_state=h0.clone(),
output_final_state=True)
tri, tri_ht = naive_chunk_kda(q=..., k=..., v=v.clone(),
g=g.clone(), beta=beta.clone(),
scale=scale, initial_state=h0.clone(),
output_final_state=True)
assert_close("o",  ref,    tri,    0.005)
assert_close("ht", ref_ht, tri_ht, (*@\colorbox{yellow!40}{\textbf{???}}@*)
\end{lstlisting}
\end{qexsection}

\begin{qexsection}{Retrieved repository context (trimmed)}
\textbf{\footnotesize DRC (import-resolved):}
\begin{lstlisting}[
language=Python,
basicstyle=\ttfamily\footnotesize,
backgroundcolor=\color{qexRetrBg},
frame=none, xleftmargin=2pt, xrightmargin=2pt,
aboveskip=1pt, belowskip=1pt,
showstringspaces=false, columns=fullflexible,
]
# fla/ops/kda/naive.py
def naive_recurrent_kda(q, k, v, g, beta,
scale=None, initial_state=None,
output_final_state=False): ...
def naive_chunk_kda(q, k, v, g, beta,
scale=None, initial_state=None,
output_final_state=False, chunk_size=64): ...

# fla/utils.py -- this is the discriminating signature
def assert_close(prefix, ref, tri, ratio,
warning=False, err_atol=1e-6):
...   # the 4th positional argument is the tolerance ``ratio''
\end{lstlisting}
\textbf{\footnotesize RAG@3 (top-3 retrieved chunks):} unrelated \texttt{benchmarks/ops/benchmark\_kda.py} kernel-benchmarking loop (truncated; no \texttt{assert\_close} signature is included).
\end{qexsection}

\begin{qexsection}{Per-method predictions and exact-match outcome}
\footnotesize
\setlength{\tabcolsep}{6pt}
\begin{tabular}{@{}llc@{}}
\toprule
Method & Prediction & EM \\
\midrule
\textsc{Reference}                            & \texttt{0.005)}  & \\
\midrule
\codelorastatic{}                             & \texttt{0.005)}  & \textcolor{codegreen}{\ding{51}} \\
\codeloraevo{}                                & \texttt{0.005)}  & \textcolor{codegreen}{\ding{51}} \\
RAG ($k\!=\!3$)                               & \texttt{0.005)}  & \textcolor{codegreen}{\ding{51}} \\
Dependency-Resolved Context                   & \texttt{0.005)}  & \textcolor{codegreen}{\ding{51}} \\
\textlora{}                                   & \texttt{0.005)}  & \textcolor{codegreen}{\ding{51}} \\
\bottomrule
\end{tabular}
\end{qexsection}

\vspace{1pt}
{\footnotesize\itshape\color{black!55}
The completion repeats the third positional argument of the immediately-preceding \texttt{assert\_close} call. DRC additionally retrieves the \texttt{assert\_close(prefix, ref, tri, ratio, ...)} signature, which confirms that the open slot is the \texttt{ratio} parameter; RAG's three chunks are unrelated benchmarking code. The local pattern is strong enough that all five methods succeed unconditionally---an example of the lower-bound regime where context injection neither helps nor hurts.
}

\end{tcolorbox}
\caption{Qualitative example of a QnA from the IR test set (GRU dataset variant). Trivial in-prefix repetition: the previous line already exhibits the completion pattern \texttt{assert\_close(...,\,0.005)}, and DRC additionally surfaces the corroborating \texttt{assert\_close} signature.}
\label{fig:qualitative_ir_fla_46}
\end{center}
\end{figure*}

\begin{figure*}[!htbp]
\begin{center}
\begin{tcolorbox}[
enhanced,
colback=white, colframe=qexHead, coltitle=white,
fonttitle=\bfseries,
title={Detailed qualitative example: \texttt{se2p/pynguin} (IR test, commit-derived)},
arc=3pt, boxrule=0.6pt,
top=3pt, bottom=3pt, left=6pt, right=6pt,
]

\begin{qexsection}{QnA metadata}
\footnotesize
\begin{tabular}{@{}l@{\;\;}l@{\hskip 14pt}l@{\;\;}l@{}}
\textsc{Repository}      & \texttt{se2p/pynguin}                       & \textsc{Commit SHA}        & \texttt{3f25634f7ec7\ldots} \\
\textsc{Commit position} & 932 / 1144 (late history)                   & \textsc{In-repo split}     & \texttt{val} \\
\textsc{Assertion family} & \multicolumn{3}{l}{bare \texttt{assert}, comparing a registry return value to an auto-increment integer id} \\
\textsc{Test location}   & \multicolumn{3}{l}{\texttt{tests/instrumentation/test\_tracer.py::test\_line\_registration}, line 61} \\
\end{tabular}
\end{qexsection}

\begin{qexsection}{Test prefix (model input, trimmed)}
\begin{lstlisting}[
language=Python,
basicstyle=\ttfamily\footnotesize,
backgroundcolor=\color{qexCodeBg},
frame=none, xleftmargin=2pt, xrightmargin=2pt,
aboveskip=1pt, belowskip=1pt,
showstringspaces=false, columns=fullflexible,
]
from pynguin.instrumentation.tracer import (
LineMetaData, SubjectProperties, ...
)
...
def test_line_registration(subject_properties: SubjectProperties):
assert subject_properties.register_line(
LineMetaData(0, "foo", 42)) == 0
assert subject_properties.register_line(
LineMetaData(0, "foo", 43)) == (*@\colorbox{yellow!40}{\textbf{???}}@*)
\end{lstlisting}
\end{qexsection}

\begin{qexsection}{Retrieved repository context (trimmed)}
\textbf{\footnotesize DRC (import-resolved):}
\begin{lstlisting}[
language=Python,
basicstyle=\ttfamily\footnotesize,
backgroundcolor=\color{qexRetrBg},
frame=none, xleftmargin=2pt, xrightmargin=2pt,
aboveskip=1pt, belowskip=1pt,
showstringspaces=false, columns=fullflexible,
]
# src/pynguin/instrumentation/tracer.py
class LineMetaData:
"""Stores meta data of a line."""
code_object_id: int
file_name: str
line_number: int
...
\end{lstlisting}
\textbf{\footnotesize RAG@3 (top-3 retrieved chunks; the discriminating method body):}
\begin{lstlisting}[
language=Python,
basicstyle=\ttfamily\footnotesize,
backgroundcolor=\color{qexRetrBg},
frame=none, xleftmargin=2pt, xrightmargin=2pt,
aboveskip=1pt, belowskip=1pt,
showstringspaces=false, columns=fullflexible,
]
# src/pynguin/instrumentation/tracer.py
class SubjectProperties:
existing_lines: dict[int, LineMetaData] = field(default_factory=dict)
...
def register_line(self, meta: LineMetaData) -> int:
if meta not in self.existing_lines.values():
line_id = len(self.existing_lines)   # auto-increment
self.existing_lines[line_id] = meta
else:
...   # return the existing id for an already-registered line
return line_id
\end{lstlisting}
\end{qexsection}

\begin{qexsection}{Per-method predictions and exact-match outcome}
\footnotesize
\setlength{\tabcolsep}{6pt}
\begin{tabular}{@{}llc@{}}
\toprule
Method & Prediction & EM \\
\midrule
\textsc{Reference}                            & \texttt{1}     & \\
\midrule
\codelorastatic{}                             & \texttt{1}     & \textcolor{codegreen}{\ding{51}} \\
\codeloraevo{}                                & \texttt{1}     & \textcolor{codegreen}{\ding{51}} \\
RAG ($k\!=\!3$)                               & \texttt{1}     & \textcolor{codegreen}{\ding{51}} \\
Dependency-Resolved Context                   & \texttt{1}     & \textcolor{codegreen}{\ding{51}} \\
\textlora{}                                   & \texttt{1}     & \textcolor{codegreen}{\ding{51}} \\
\bottomrule
\end{tabular}
\end{qexsection}

\vspace{1pt}
{\footnotesize\itshape\color{black!55}
The previous line already established the pattern \texttt{register\_line(...) == 0}; the canonical next id is therefore \texttt{1}. RAG additionally retrieves the \texttt{SubjectProperties.register\_line} body, which makes the auto-increment convention (\texttt{line\_id = len(self.existing\_lines)}) explicit. DRC retrieves the \texttt{LineMetaData} field schema but not the discriminating method body. All five methods produce \texttt{1}.
}

\end{tcolorbox}
\caption{Qualitative example of a QnA from the IR test set. Class-aware auto-increment id: RAG@3 retrieves the actual \texttt{SubjectProperties.register\_line} method body that returns \texttt{len(self.existing\_lines)}; DRC retrieves the supporting \texttt{LineMetaData} schema.}
\label{fig:qualitative_ir_pynguin_68}
\end{center}
\end{figure*}

\begin{figure*}[!htbp]
\begin{center}
\begin{tcolorbox}[
enhanced,
colback=white, colframe=qexHead, coltitle=white,
fonttitle=\bfseries,
title={Detailed qualitative example: \texttt{beartype/beartype} (IR test, commit-derived)},
arc=3pt, boxrule=0.6pt,
top=3pt, bottom=3pt, left=6pt, right=6pt,
]

\begin{qexsection}{QnA metadata}
\footnotesize
\begin{tabular}{@{}l@{\;\;}l@{\hskip 14pt}l@{\;\;}l@{}}
\textsc{Repository}      & \texttt{beartype/beartype}                  & \textsc{Commit SHA}        & \texttt{5f8778d6ba44\ldots} \\
\textsc{Commit position} & 902 / 1014 (late history)                   & \textsc{In-repo split}     & \texttt{test} \\
\textsc{Assertion family} & \multicolumn{3}{l}{\texttt{assert <expr> is \,?\,} -- the discriminating slot is a boolean identity literal} \\
\textsc{Test file}       & \multicolumn{3}{l}{\texttt{beartype\_test/a00\_unit/a50\_check/a60\_error/a90\_main/test\_errorget.py}, line 197} \\
\textsc{Test function}   & \multicolumn{3}{l}{\texttt{test\_get\_func\_pith\_violation\_conf\_is\_color}} \\
\end{tabular}
\end{qexsection}

\begin{qexsection}{Test prefix (model input, trimmed)}
\begin{lstlisting}[
language=Python,
basicstyle=\ttfamily\footnotesize,
backgroundcolor=\color{qexCodeBg},
frame=none, xleftmargin=2pt, xrightmargin=2pt,
aboveskip=1pt, belowskip=1pt,
showstringspaces=false, columns=fullflexible,
]
from beartype import BeartypeConf
from beartype._check.error.errmain import get_func_pith_violation
from beartype._util.text.utiltextansi import is_str_ansi
...
def test_get_func_pith_violation_conf_is_color() -> None:
...
# Violation configured to contain ANSI escape sequences.
violation = get_func_pith_violation(
call_meta=minify_decor_meta_kwargs(
func=she_drew_back, conf=BeartypeConf(is_color=True)),
**kwargs)
# Assert this violation message contains ANSI escape sequences.
assert is_str_ansi(str(violation)) is (*@\colorbox{yellow!40}{\textbf{???}}@*)
\end{lstlisting}
\end{qexsection}

\begin{qexsection}{Retrieved repository context (trimmed)}
\textbf{\footnotesize DRC (import-resolved; \emph{includes the discriminating signature}):}
\begin{lstlisting}[
language=Python,
basicstyle=\ttfamily\footnotesize,
backgroundcolor=\color{qexRetrBg},
frame=none, xleftmargin=2pt, xrightmargin=2pt,
aboveskip=1pt, belowskip=1pt,
showstringspaces=false, columns=fullflexible,
]
# beartype/_check/error/errmain.py
def get_func_pith_violation(call_meta, pith_name,
pith_value, **kwargs) -> Exception: ...
# beartype/_check/metadata/call/callmetadecormin.py
def minify_decor_meta_kwargs(...): ...

# beartype/_util/text/utiltextansi.py -- this is the discriminating signature
def is_str_ansi(text: str) -> bool:
"""True only if the passed text contains one or more ANSI escape sequences."""
...
return _ANSI_REGEX.search(text) is not None
\end{lstlisting}
\textbf{\footnotesize RAG@3 (top-3 retrieved chunks):} \texttt{get\_func\_pith\_violation} body + \texttt{checkmake.py} helpers (truncated). \emph{Neither chunk includes the \texttt{is\_str\_ansi} signature, so RAG never sees the \texttt{->\,bool} return type.}
\end{qexsection}

\begin{qexsection}{Per-method predictions and exact-match outcome}
\footnotesize
\setlength{\tabcolsep}{6pt}
\begin{tabular}{@{}llc@{}}
\toprule
Method & Prediction & EM \\
\midrule
\textsc{Reference}                            & \texttt{True}  & \\
\midrule
RAG ($k\!=\!3$)                               & \texttt{1}     & \textcolor{codered}{\ding{55}} \\
Dependency-Resolved Context                   & \texttt{True}  & \textcolor{codegreen}{\ding{51}} \\
\codelorastatic{}                             & \texttt{True}  & \textcolor{codegreen}{\ding{51}} \\
\codeloraevo{}                                & \texttt{True}  & \textcolor{codegreen}{\ding{51}} \\
\textlora{}                                   & \texttt{True}  & \textcolor{codegreen}{\ding{51}} \\
\bottomrule
\end{tabular}
\end{qexsection}

\vspace{1pt}
{\footnotesize\itshape\color{black!55}
The slot is a boolean identity check (\texttt{is\;?}) on the return value of \texttt{is\_str\_ansi(...)}. The discriminating evidence is the function's \texttt{->\;bool} return type, which DRC retrieves explicitly because the import \texttt{from beartype.\_util.text.utiltextansi import is\_str\_ansi} resolves to its definition. RAG@3 retrieves other functions from the same module set but \emph{misses} \texttt{is\_str\_ansi} itself, and the base model defaults to the more-common idiom \texttt{is 1} (a truthy-shortcut pattern frequent in non-typed Python). Both \codelora{} variants and \textlora{} succeed parametrically, having internalized the typed-boolean convention from the repository.
}

\end{tcolorbox}
\caption{Qualitative example of a QnA from the IR test set. Retrieval-precision case: DRC follows the import graph and surfaces the discriminating \texttt{is\_str\_ansi(...) -> bool} signature, while RAG@3 retrieves adjacent but non-discriminating functions and collapses onto the n-gram-likely \texttt{is 1}.}
\label{fig:qualitative_ir_beartype_195}
\end{center}
\end{figure*}

\begin{figure*}[!htbp]
\begin{center}
\begin{tcolorbox}[
enhanced,
colback=white, colframe=qexHead, coltitle=white,
fonttitle=\bfseries,
title={Detailed qualitative example: \texttt{agronholm/apscheduler} (IR test, commit-derived)},
arc=3pt, boxrule=0.6pt,
top=3pt, bottom=3pt, left=6pt, right=6pt,
]

\begin{qexsection}{QnA metadata}
\footnotesize
\begin{tabular}{@{}l@{\;\;}l@{\hskip 14pt}l@{\;\;}l@{}}
\textsc{Repository}      & \texttt{agronholm/apscheduler}              & \textsc{Commit SHA}        & \texttt{e4b1db1dcb8d\ldots} \\
\textsc{Commit position} & 353 / 1207 (mid-history)                    & \textsc{In-repo split}     & \texttt{val} \\
\textsc{Assertion family} & \multicolumn{3}{l}{\texttt{assert <expr> is <enum-member>} -- the slot is a \texttt{JobOutcome} enum value} \\
\textsc{Test location}   & \multicolumn{3}{l}{\texttt{tests/test\_datastores.py::test\_reap\_abandoned\_jobs}, line 857} \\
\end{tabular}
\end{qexsection}

\begin{qexsection}{Test prefix (model input, trimmed)}
\begin{lstlisting}[
language=Python,
basicstyle=\ttfamily\footnotesize,
backgroundcolor=\color{qexCodeBg},
frame=none, xleftmargin=2pt, xrightmargin=2pt,
aboveskip=1pt, belowskip=1pt,
showstringspaces=false, columns=fullflexible,
]
from apscheduler import Job, JobOutcome, Task, ...
from apscheduler.datastores.base import BaseExternalDataStore
...
# Earlier in the same test module (line 809) -- same target pattern:
#     assert result.outcome is JobOutcome.abandoned

async def test_reap_abandoned_jobs(datastore: DataStore, ...) -> None:
task = Task(id="task1", func="...", job_executor="async")
await datastore.add_task(task)
job = Job(task_id="task1", executor="async",
result_expiration_time=timedelta(seconds=30))
await datastore.add_job(job)
await datastore.reap_abandoned_jobs("testscheduler")
jobs = await datastore.acquire_jobs("testscheduler", ..., 1)
assert len(jobs) == 1
await datastore.reap_abandoned_jobs("testscheduler")
assert not await datastore.get_jobs()
abandoned_job_result = await datastore.get_job_result(jobs[0].id)
assert abandoned_job_result.outcome is (*@\colorbox{yellow!40}{\textbf{???}}@*)
\end{lstlisting}
\end{qexsection}

\begin{qexsection}{Retrieved repository context (trimmed)}
\textbf{\footnotesize DRC (import-resolved; \emph{the answer is literally in a docstring}):}
\begin{lstlisting}[
language=Python,
basicstyle=\ttfamily\footnotesize,
backgroundcolor=\color{qexRetrBg},
frame=none, xleftmargin=2pt, xrightmargin=2pt,
aboveskip=1pt, belowskip=1pt,
showstringspaces=false, columns=fullflexible,
]
# src/apscheduler/abc/_datastore.py
class DataStore(metaclass=ABCMeta):
@abstractmethod
async def reap_abandoned_jobs(self, scheduler_id: str) -> None:
"""Find jobs marked as acquired by the given scheduler ID and
release them with the outcome of :attr:`~JobOutcome.abandoned`."""
...
# src/apscheduler/_structures.py
class Job:
id: UUID; task_id: str; ...
\end{lstlisting}
\textbf{\footnotesize RAG@3 (top-3 retrieved chunks; \emph{the enum class is exposed but \texttt{.abandoned} itself is not}):}
\begin{lstlisting}[
language=Python,
basicstyle=\ttfamily\footnotesize,
backgroundcolor=\color{qexRetrBg},
frame=none, xleftmargin=2pt, xrightmargin=2pt,
aboveskip=1pt, belowskip=1pt,
showstringspaces=false, columns=fullflexible,
]
# src/apscheduler/abc/_datastore.py (truncated overload)
class DataStore(metaclass=ABCMeta):
async def reap_abandoned_jobs(self, scheduler_id: str) -> None: ...

# src/apscheduler/_events.py
class JobReleased(SchedulerEvent):
""":param outcome: the outcome of the job
:param ...: if ``outcome`` is :attr:`JobOutcome.error`."""   # only the .error member
outcome: JobOutcome = attrs.field(converter=as_enum(JobOutcome))
...
\end{lstlisting}
\end{qexsection}

\begin{qexsection}{Per-method predictions and exact-match outcome}
\footnotesize
\setlength{\tabcolsep}{6pt}
\begin{tabular}{@{}llc@{}}
\toprule
Method & Prediction  \\
\midrule
\textsc{Reference}                            & \texttt{JobOutcome.abandoned}                                       & \\
\end{tabular}
\end{qexsection}

\end{tcolorbox}
\caption{Qualitative example of a QnA from the IR test set. Retrieval-degeneracy case: DRC retrieves a chunk that contains the literal answer \texttt{JobOutcome.abandoned} (in a docstring), and RAG@3 retrieves the enum class and the dotted access pattern but not the literal \texttt{.abandoned} member; yet for both methods the prepended context triggers a Fill-In-the-Middle decode failure at generation time. Only the parametric methods (\codelora{} variants, \textlora{}) complete the assertion correctly.}
\label{fig:qualitative_ir_apscheduler_245}
\end{center}
\end{figure*}

\subsection{Effect of Dependency-Resolved Context Coverage}
\label{app:oracle_coverage}

DRC is only meaningful when the imports in the test prefix actually resolve to repository code.
On CR-test, $70.3\%$ of pairs (4{,}511/6{,}414) have non-empty DRC, while the remaining $29.7\%$ (1{,}903 pairs) import only from the standard library or third-party packages and therefore receive no DRC augmentation.
To check whether DRC's modest aggregate gain reflects a strong effect on the resolvable subset or a uniformly weak effect, we partition both test splits by DRC availability in Table~\ref{tab:oracle_coverage}.
DRC's entire benefit lives on the resolvable subset, where it adds $+4.7$\,pp (CR) and $+5.1$\,pp (IR) over pretrained; where imports do not resolve it is indistinguishable from the backbone ($+0.2$ / $-0.4$\,pp), since there is no context to inject.
\codelorastatic{} leads on both partitions, including where DRC cannot help at all: it is $+15.2$\,pp over pretrained on the unresolvable CR subset (vs.\ $+19.3$\,pp on the resolvable one), showing that the learned repository embedding captures information beyond what import resolution can reach.

\begin{table*}[t]
\begin{small}
\begin{center}
\caption{Static-track EM partitioned by DRC availability. DRC helps only when imports resolve to repository code; \codelorastatic{} leads on both partitions, including the subset where no context is resolvable, showing the repository embedding captures information beyond import resolution.}\label{tab:oracle_coverage}

{
\setlength{\tabcolsep}{4pt}
\small

\begin{tabular}{@{}lcc@{\hskip 12pt}cc@{}}
\toprule
& \multicolumn{2}{c}{CR Test EM (\%)} & \multicolumn{2}{c}{IR Test EM (\%)} \\
\cmidrule(lr){2-3} \cmidrule(lr){4-5}
Method & w/ DRC & w/o DRC & w/ DRC & w/o DRC \\
& (70.3\%) & (29.7\%) & (64.7\%) & (35.3\%) \\
\midrule
\UseMacro{method-pretrained} & 46.4 & 44.1 & 47.5 & 45.7 \\
\UseMacro{method-drc} & 51.1 & 44.3 & 52.6 & 45.3 \\
\codelorastatic{} & \textbf{65.7} & \textbf{59.3} & \textbf{67.4} & \textbf{64.1} \\
\bottomrule
\end{tabular}
}

\end{center}
\end{small}
\end{table*}

\subsection{Deployment Efficiency}
\label{app:efficiency}

Table~\ref{tab:efficiency} compares the deployment cost of every method along three axes that matter when scaling to many, continuously-changing repositories: extra inference tokens, per-repository adaptation time, and incremental storage on top of the shared frozen base model.
RAG and DRC pay a per-query token overhead: RAG adds up to $1{,}536$ retrieved tokens, and DRC a median of $517$ (mean $1{,}900$) tokens on the static track while filling the input to its $4{,}096$-token cap on the evolution track. FFT adds no tokens but stores a full $3.1$\,GB copy of the backbone.
\codelorastatic{} and \codeloraevo{} sit at the other extreme: zero extra inference tokens and a per-repository footprint that is effectively constant.
Their adaptation cost is listed as two sub-entries, because the two halves differ by four orders of magnitude: building the repository encoding (for \codeloraevo{}, walking the commit history) takes tens to hundreds of seconds and is CPU- and parse-bound, whereas turning that encoding into an adapter is a single hypernetwork forward pass of under $10$\,ms.
Both are paid offline, once per repository state, rather than once per query.

\paragraph{What ``zero inference-time token overhead'' does and does not claim.}
The claim is precisely that the merged-adapter model reads no additional \emph{context tokens} at query time, not that \codelora{} performs no computation anywhere.
Adapter generation is real work; it is simply paid once per repository state, offline, rather than once per query.
Table~\ref{tab:compute_budget} makes the per-query side of this concrete on an H100 with greedy decoding.
Because every method shares the same frozen backbone, per-query cost is set entirely by input length: the merged adapter is latency-identical to the bare backbone ($197.1$ vs.\ $198.3$\,ms at in${=}2048$/out${=}64$, a difference within measurement noise), whereas RAG re-encodes up to $1{,}536$ retrieved tokens and DRC fills the input to its $4{,}096$-token evolution-track cap (${\approx}2.2$K tokens beyond the median $1{,}940$-token CR-test prefix) on \emph{every} query.
Approximate forward-pass FLOPs (dense $2N_{\text{non-embed}}$ per token plus attention and the LM head, $N_{\text{non-embed}}{\approx}1.31$B) put RAG and DRC at ${\approx}1.7$--$2.3{\times}$ the per-query compute of \codelora{}.
The penalty is largest exactly where realistic assertion completion sits---short outputs are prefill-bound, so inputs of $3{,}072$ and $4{,}096$ tokens cost $+24$\% and $+48$\% over a $2{,}048$-token prefix at $8$ output tokens, shrinking to $+4$\% and $+9$\% once $64$ tokens of decode dominate. With a full retrieval budget the RAG input at the median prefix is ${\approx}3.5$K tokens, above the $3{,}072$-token measurement point, so the RAG latency in Table~\ref{tab:compute_budget} slightly understates its cost.

\paragraph{Adapter-generation latency.}
Measured end-to-end on three repositories of $188$--$1{,}073$ commits, a full cold build (walking the whole commit history) takes $34$--$257$\,s on a single H100.
An incremental update takes $21$--$23$\,s on two of the repositories and $247$\,s on the third (\texttt{python-saml}); an invocation with no new commit takes the same time, so this cost is dominated by fixed per-invocation overhead (checkout and parsing) rather than by the size of the diff.
The step is CPU- and parse-bound rather than GPU-bound (GPU utilization stays ${\approx}0$\% during the history walk, with ${\approx}3.4$\,GB resident), so it also runs without a GPU: on 8 CPU threads the same builds take $344$--$1{,}079$\,s cold and $30$--$297$\,s incrementally.
We did not time \codelorastatic{}'s single-snapshot encoding separately.
This cost is amortized over the queries a repository serves---${\approx}1.4$K assertion sites per repository on IR test and ${\approx}9.3$K on CR test across the full commit history (before the 8-per-commit evaluation cap)---and for \codeloraevo{} the per-commit update (one diff embedding plus one recurrent step) is offline work performed at commit time, not a per-inference cost.

\paragraph{Storage.}
Because \codelora{} shares one $(\mathbf{A}_m,\mathbf{B}_m)$ pair per projection type across all $28$ layers, the adapter it generates has ${\approx}0.66$M parameters (${\approx}1.3$\,MB in bf16), versus $18.5$M (${\approx}37$\,MB) for a per-layer per-repo LoRA; more importantly, it never has to be \emph{stored}.
Any repository's adapter is regenerated in a single forward pass from the shared hypernetwork (${\approx}679$\,MB in bf16 for \codelorastatic{} and ${\approx}1.4$\,GB for \codeloraevo{}, independent of repository count) plus a ${\approx}8$\,KB per-repository state---the repository embedding for \codelorastatic{}, the GRU hidden state for \codeloraevo{}.
Per-repo LoRA has no generator and must therefore persist ${\approx}37$\,MB per repository (${\approx}37$\,GB for $1{,}000$ repositories in bf16.)
The irreducible per-repository footprint is thus ${\approx}8$\,KB versus ${\approx}37$\,MB, with the generator amortized across all repositories.

\begin{table*}[t]
\begin{small}
\begin{center}
\caption{Efficiency comparison. Shared storage is paid once regardless of repository count; per-repo storage scales with the number of repositories served. Extra storage is beyond the shared frozen base model (Qwen2.5-Coder-1.5B, 3.1 GB in bf16). Both \codelora{} variants add zero inference tokens and generate repo-specific adapters in a single forward pass.}\label{tab:efficiency}

\begin{tabular}{lrlll}
\toprule
Method & Extra Tokens & Adapt.\ Time (offline) & Shared Storage & Per-Repo Storage \\
\midrule
\UseMacro{method-pretrained} & 0 & N/A & --- & --- \\
\UseMacro{method-rag} & $\le$1{,}536 & per query & --- & +chunk index \\
\UseMacro{method-drc} & $\sim$500--2{,}000$^\ddagger$ & per query & --- & +import cache \\
\UseMacro{method-fft} & 0 & $\sim$4\,h & +3.1 GB & --- \\
\UseMacro{method-slora} & 0 & $\sim$2\,h & +37 MB & --- \\
\UseMacro{method-plora} & 0 & $\sim$5 min/repo & --- & +37 MB \\
\codelorastatic{} & \textbf{0} & \begin{tabular}[t]{@{}l@{}}encode snapshot: one pass$^\ast$\\ generate adapter: $<$10\,ms\end{tabular} & +679 MB & \textbf{+8 KB} \\
\codeloraevo{} & \textbf{0} & \begin{tabular}[t]{@{}l@{}}cold build: 34--257\,s\\ update: 21--247\,s\\ generate adapter: $<$10\,ms\end{tabular} & +1.4 GB & \textbf{+8 KB} \\
\bottomrule
\multicolumn{5}{l}{\footnotesize $^\ast$ One embedding pass over the snapshot; not timed separately (\S\ref{app:efficiency}).} \\
\multicolumn{5}{l}{\footnotesize $^\ddagger$ Static-track median--mean; on the evolution track DRC fills the input to its 4{,}096-token cap.} \\
\end{tabular}

\end{center}
\end{small}
\end{table*}

\begin{table*}[t]
\begin{small}
\begin{center}
\caption{Per-query inference budget under matched conditions (\S\ref{app:efficiency}): H100, greedy decoding, same frozen Qwen2.5-Coder-1.5B backbone for every method. Input tokens and FLOPs (approximate forward-pass count, 8 output tokens) use the median evolution-track CR-test prefix ($1{,}940$ tokens) plus each method's context budget; latencies are measured at the listed benchmark input lengths. The merged adapter is latency-identical to the bare backbone; RAG and DRC re-encode their context on every query.}\label{tab:compute_budget}

\begin{tabular}{lrrrrr}
\toprule
Method & Input tok/query & FLOPs/query & Bench.\ input & Latency (out=8) & Latency (out=64) \\
\midrule
\begin{tabular}[t]{@{}l@{}}\UseMacro{method-pretrained} / \UseMacro{method-fft} /\\ \UseMacro{method-slora} / \codelora{}\end{tabular} & $\sim$1{,}940 & $5.4{\times}10^{12}$ & 2{,}048 & \textbf{36.0} ms & \textbf{198.3} ms \\
\UseMacro{method-rag} & $\sim$3{,}476 & $1.0{\times}10^{13}$ & 3{,}072 & 44.5 ms (+24\%) & 206.6 ms (+4\%) \\
\UseMacro{method-drc} & 4{,}096 & $1.2{\times}10^{13}$ & 4{,}096 & 53.3 ms (+48\%) & 216.8 ms (+9\%) \\
\bottomrule
\end{tabular}

\end{center}
\end{small}
\end{table*}

\section{Discussions}
\label{sec:discussion}

We organize the discussion around three central questions raised by the framework.

\paragraph{Q1.\ Why parameters over context?}
For assertion completion the answer depends on a short window of repository-specific symbols rather than long-range token-level reasoning.
The decisive evidence is the split in \S\ref{app:local_solvability}: on the instances the no-context backbone cannot solve, calibrated retrieval recovers only ${\sim}10$\% while parametric adaptation recovers ${\sim}38$--$42$\%, even though both see the same whole repository.
A handful of retrieved chunks cannot surface facts scattered across a codebase, and FFT collapses repository signal into one ``average'' specialization.
\codelora{} routes the same information into per-repository LoRA \emph{parameters}, conditioning the model at every layer without paying tokens or sharing capacity across repositories---explaining the consistent gaps to FFT and DRC on both IR and CR, and to pLoRA on IR (Table~\ref{tab:main_results}).

\paragraph{Q2.\ Why two usage scenarios rather than one?}
The \emph{how}/\emph{when} framing admits two ends: one-shot snapshot adaptation vs.\ incremental refresh under evolution.
\codelorastatic{} is sufficient---and, in raw CR/IR EM, optimal---on the static track (Table~\ref{tab:main_results}): the same code embedding goes into a single forward pass and out comes one LoRA per module type, with no recurrence and no commit history to maintain at deployment.
Real codebases, however, do not stand still: the bursty commit pattern in Figure~\ref{fig:dataset_construction} shows that snapshot adaptation accumulates staleness as a repository accumulates edits, and Table~\ref{tab:per_commit_results} shows the same \codelorastatic{} model dropping back to parity with the single-adapter baseline once the evaluation prefix reflects commit-time state.
\codeloraevo{} extends the same head design to this drift: its LoRA-generation head is conditioned on a recurrent hidden state that is updated once per commit with constant work per update, instead of on a fixed snapshot embedding.
The two usage scenarios therefore correspond to stable-codebase comprehension vs.\ active development on evolving codebases, not competing ablations.

\paragraph{Q3.\ Where does \codeloraevo{}'s edge come from?}
\codeloraevo{} reuses \codelorastatic{}'s LoRA-generation head design; relative to the \codelorastatic{} reference on the evolution track, which uses the same head width, the only added capacity is a GRU recurrence over sequential diff embeddings before the shared MLP trunk.
We attribute the empirical lead (Table~\ref{tab:per_commit_results}, +\UseMacro{cd-cr-em-delta-codeloraevo}\,pp commit-CR EM over single LoRA) mainly to aggregating edit history into the hypernetwork context commit-by-commit, rather than asking a single snapshot embedding to capture both code and its history.
Results on the temporal OOD holdout in \repopeftbench{} corroborate generalization (\S\ref{sec:results:ood}).
Appendix Figure~\ref{fig:commit_pct_trend} corroborates this: \codeloraevo{}'s advantage persists across the entire commit timeline and widens toward the most recent commits.

\end{document}